\documentclass[a4paper, 11pt, oneside]{Thesis}
\graphicspath{Figures/}  

\usepackage[square, numbers, comma, sort&compress]{natbib}  
\usepackage{verbatim}  
\usepackage{vector}  
\usepackage{enumitem}
\usepackage{multicol}
\usepackage[T1]{fontenc}
\usepackage{siunitx}
\usepackage{dcolumn}

\usepackage{acronym}

\usepackage{hyperref}
\hypersetup
{
    pdfauthor={Aleksander Cis{\l}ak},
    pdfsubject={Master's Thesis},
    pdftitle={Full-text and Keyword Indexes for String Searching},
    pdfkeywords={approximate dictionary matching} {FM-index} {Hamming distance} {k-mismatches problem} {Master's thesis} {split index} {string matching} {string sketch} {text indexing},
    pdfborder={0 0 0},
    urlcolor=black,
    colorlinks=false
}

\begin{document}
\frontmatter      


\maketitle

\makesecondpage

\setstretch{1.3}  

\fancyhead{}  
\rhead{\thepage}  
\lhead{}  

\pagestyle{fancy}  

\Declaration{

\addtocontents{toc}{\vspace{1em}}  

I, Aleksander Cis{\l}ak, confirm that this master's thesis is my own work and I have documented all sources and material used.

\vspace{5em}
 
Signed:\\
\rule[1em]{25em}{0.5pt}  
 
Date:\\
\rule[1em]{25em}{0.5pt}  
}
\clearpage  

\addtotoc{Abstract}  
\abstract{
\addtocontents{toc}{\vspace{1em}}  

String searching consists in locating a substring in a longer text, and two strings can be approximately equal (various similarity measures such as the Hamming distance exist).
Strings can be defined very broadly, and they usually contain natural language and biological data (DNA, proteins), but they can also represent other kinds of data such as music or images.

One solution to string searching is to use online algorithms which do not preprocess the input text, however, this is often infeasible due to the massive sizes of modern data sets.
Alternatively, one can build an index, i.e.~a data structure which aims to speed up string matching queries.
The indexes are divided into full-text ones which operate on the whole input text and can answer arbitrary queries and keyword indexes which store a dictionary of individual words.
In this work, we present a literature review for both index categories as well as our contributions (which are mostly practice-oriented).

The first contribution is the FM-bloated index, which is a modification of the well-known FM-index (a compressed, full-text index) that trades space for speed.
In our approach, the count table and the occurrence lists store information about selected $q$-grams in addition to the individual characters.
Two variants are described, namely one using $O(n \log^2 n)$ bits of space with $O(m + \log m \log \log n)$ average query time, and one with linear space and $O(m \log \log n)$ average query time, where $n$ is the input text length and $m$ is the pattern length.
We experimentally show that a significant speedup can be achieved by operating on $q$-grams (albeit at the cost of very high space requirements, hence the name ``bloated'').

In the category of keyword indexes we present the so-called split index, which can efficiently solve the $k$-mismatches problem, especially for 1 error.
Our implementation in the C++ language is focused mostly on data compaction, which is beneficial for the search speed (by being cache friendly).
We compare our solution with other algorithms and we show that it is faster when the Hamming distance is used.
Query times in the order of 1 microsecond were reported for one mismatch for a few-megabyte natural language dictionary on a medium-end PC.

A minor contribution includes string sketches which aim to speed up approximate string comparison at the cost of additional space ($O(1)$ per string).
They can be used in the context of keyword indexes in order to deduce that two strings differ by at least $k$ mismatches with the use of fast bitwise operations rather than an explicit verification.
}

\clearpage  

\setstretch{1.3}  

\acknowledgements{
\addtocontents{toc}{\vspace{1em}}  

I would like to thank prof.~Szymon Grabowski for his constant support, advice, and mentorship.
He introduced me to the academia, and I would probably not pursue the scientific path if it were not for him.
His vast knowledge and ability to explain things are simply unmatched.

I would like to thank prof.~Burkhard Rost and Tatyana Goldberg for their helpful remarks and guidance in the field of bioinformatics.

I would like to thank the whole ``Gank Incoming'' Dota~2 team; the whole Radogoszcz football and airsoft pack; Ester C{\l}api{\'n}ska, who keeps my ego in check; {\L}ukasz Duda, whose decorum is still spoiled; Edwin Fung, for not letting the corporate giant consume him; {\L}ukasz Fuchs, who still promises the movie; Florentyna Gust, who always supported me in difficult times, for being a cat from Korea; Jacek Krasiukianis, for introducing me to the realm of street fighter games; Rafa{\l} Madaj, for biking tours; Jakub Przybylski, who recently switched focus from whisky to beer (while remaining faithful to the interdisciplinary field of malt engineering), for solving world problems together; and Wojciech Terepeta, who put up with my face being his first sight in the morning.

I am indebted to Ozzy Osbourne, Frank Sinatra, and Roger Waters for making the world slightly more interesting.
I would like to thank the developers of the free yUML software (\url{http://yuml.me/}) for making my life somewhat easier.

Many thanks also goes to my family as well as to all the giants who lent me their shoulders for a while.

}
\clearpage  

\pagestyle{fancy}  

\lhead{\emph{Contents}}  
\addtotoc{Contents}
\tableofcontents  

\pagestyle{empty}  

\null\vfill
\textit{``Why did the scarecrow win the Nobel Prize?\\Because he was out standing in his field.''}

\begin{flushright}
Unknown
\end{flushright}

\vfill\vfill\vfill\vfill\vfill\vfill\null
\clearpage  


\setstretch{1.3}  

\pagestyle{empty}  
\dedicatory{For everybody who have devoted their precious time to read this work in its entirety}

\addtocontents{toc}{\vspace{2em}}  

\mainmatter	  
\pagestyle{fancy}  


\chapter{Introduction}
\label{Chap:intro}
\lhead{\emph{Introduction}}

The Bible, which consists of the Old and the New Testament, is composed of roughly 800 thousand words (in the English language version)~\cite{Bib}.
Literary works of such stature were often regarded as good candidates for creating concordances --- listings of words that originated from the specific work.
Such collections usually included positions of the words, which allowed the reader to learn about their frequency and context.
Their assembly was a non-trivial task that required a lot of effort.
Under a rather favorable assumption that a friar (today also referred to as a research assistant) would be able to achieve a throughput of one word per minute, compilation (do not confuse with code generation) for the Bible would require over thirteen thousand man-hours, or roughly one and a half years of constant work.
This naturally ignores additional efforts, for instance printing and dissemination.

Such a listing is one of the earliest examples of a text-based data structure constructed with the purpose of faster searches at the cost of space and preprocessing.
Luckily, today we are capable of building and using various structures in a much shorter time.
With the aid of silicone, electrons, and capable human minds, we have managed to decrease the times from years to seconds (indexing) and from seconds to microseconds (searching).

\section{Applications}
\label{Sec:applications}

String searching has always been ubiquitous in everyday life, most probably since the very creation of the written word.
In the modern world, we encounter text on a regular basis --- on paper, glass, rubber, human skin, metal, cement, and since the 20th century also on electronic displays.
We perform various text-based operations almost all the time, often subconsciously.
This happens in trivial situations such as looking for interesting news on a website on a slow Sunday afternoon, or trying to locate information in the bus timetable on a cold Monday morning.
Many familiar text-related tasks can be finished faster thanks to computers, and powerful machines are also crucial to scientific research.
Specific areas are discussed in the following subsections.

\subsection{Natural language}

For years, the main application of computers to textual data was \ac{nl} processing, which goes back to the work of Alan Turing in the 1950s~\cite{turing}.
The goal was to understand the meaning as well as the context in which the language was used.
One of the first programs that could actually comprehend 	and act upon English sentences was Bobrow's STUDENT (1967), which solved simple mathematical problems~\cite[p.~19]{norvig}.
The first application to text processing where string searching algorithms could really shine was \emph{spell checking}, i.e. determining whether a word is written in a correct form.
It consists in testing whether a word is present in a \ac{nl} dictionary (a set of words).
Such a functionality is required since spelling errors appear relatively often due to a variety of reasons, ranging from writer ignorance to typing and transmission errors.
Research in this area started around 1957, and the first spell checker available as an application is believed to have appeared in 1971~\cite{peterson1980computer}.
Today, spell checking is universal, and it is performed by most programs which accept user input.
This includes dedicated text editors, programming tools, email clients, command-line interfaces, and web browsers.
More sophisticated approaches which try to take the context into account were also described~\cite{cherry1981writing, mitton1987spelling}, due to the fact that checking for dictionary membership is prone to errors (e.g.,~mistyping ``were'' for ``where''; Peterson~\cite{peterson1986note} reported that up to 16\% of errors might be undetected).
Another familiar scenario is searching for words in a textual document such as a book or an article, which allows for locating relevant fragments in a much shorter time than by skimming through the text.
Determining positions of certain keywords in order to learn their context (neighboring words) may be also useful for plagiarism detection (including the plagiarism of computer programs~\cite{parker1989computer}).

With the use of \emph{approximate} methods, similar words can be obtained from the \ac{nl} dictionary and correct spelling can be suggested (spelling correction is usually coupled with spell checking).
This may also include proper nouns, for example in the case of shopping catalogs (relevant products) or geographic information systems (specific locations, e.g.,~cities).
Such techniques are also useful for \emph{\aclu{ocr}}~(\acs{ocr}) where they serve as a verification mechanism~\cite{elliman1990review}.
Other applications are in security, where it is desirable to check whether a password is not too close to a word from a dictionary~\cite{manber1994algorithm} and in data cleaning, which consists in detecting errors and duplication in data that is stored in the database~\cite{chaudhuri2003robust}.
String matching is also employed for preventing the registration of fraudulent websites having similar addresses, the phenomenon known as ``typosquatting''~\cite{moore2010measuring}.

It may happen that the pattern that is searched for is not explicitly specified, as is the case when we use a web search engine (i.e.~we would like to find the entire website, but we specify only a few keywords), which is an example of \emph{information retrieval}.
For instance, index-based methods form an important component of the architecture of the Google engine~\cite{brin1998anatomy}.

\subsection{Bioinformatics}
\label{Sec:bioinformatics}

The biological data is commonly represented in a textual form, and for this reason it can be searched just like any other text.
Most popular representations include:

\begin{itemize}

\item
\textbf{DNA} --- the alphabet of four symbols corresponding to nucleobases: \texttt{A}, \texttt{C}, \texttt{G}, and \texttt{T}.
It can be extended with an additional character \texttt{N}, indicating that there might be any nucleobase at a specified position.
This is used, for instance, when the sequencing method could not determine the nucleobase with a desired certainty.
Sometimes, additional information such as the quality of the read, i.e.~the probability that a specific base was determined correctly, is also stored.

\item
\textbf{RNA} --- four nucleobases: \texttt{A}, \texttt{C}, \texttt{G}, and \texttt{U}.
Similarly to the DNA, additional information may be present.

\item
\textbf{Proteins} --- 20 symbols corresponding to different amino acids (uppercase letters from the English alphabet), with 2 additional symbols for amino acids occurring only in some species (\texttt{O} and \texttt{U}), and 4 placeholders (\texttt{B}, \texttt{J}, \texttt{X}, \texttt{Z}) for situations where the amino acid is ambiguous.
All 26 letters from the English alphabet are used.

\end{itemize} 
 
Computational information was an integral part of the field of bioinformatics from the very beginning, and at the end of the 1970s there was a substantial activity in the development of string (sequence) alignment algorithms (e.g.,~for RNA structure prediction)~\cite{ouzounis2003early}.
Alignment methods allow for finding evolutionary relationships between genes and proteins and thus construct phylogenetic trees.
Sequence similarity in proteins is important because it may imply structural as well as functional similarity.
Researchers use tools such as BLAST~\cite{altschul1990basic}, which try to match the string in question with similar ones from the database (of proteins or genomes).
Approximate methods play an important role here, because related sequences often differ from one another due to mutations in the genetic material.
These include point mutations, that is changes at a single position, as well as insertions and deletions (usually called indels).

Another research area that would not thrive without computers is genome sequencing.
This is caused by the fact that sequencing methods cannot read the whole genome, but rather they produce hundreds of gigabytes of strings (DNA reads, whose typical length is from tens to a thousand base pairs~\cite{liu2012comparison}) whose exact positions in the genome are not known.
Moreover, the reads often contain mistakes due to the imperfection of the sequencing itself.
The goal of the computers is to calculate the correct order using complicated statistical and string-based tools, with or without a reference genome (the latter being called \emph{de novo} sequencing).
This process is well illustrated by its name --- \emph{shotgun sequencing}, and it can be likened to shredding a piece of paper and reconstructing the pieces.
String searching is crucial here because it allows for finding repeated occurrences of certain patterns~\cite{langmead2009ultrafast, simpson2010efficient}.

\subsection{Other}

Other data can be also represented and manipulated in a textual form.
This includes music, where we would like to locate a specific melody, especially using approximate methods which account for slight variations or imperfections (e.g.,~singing out of pitch)~\cite[p.~77]{grabowskitext2011}.
Another field where approximate methods play a crucial role is \emph{signal processing}, especially in the case of audio signals, which can be processed by speech recognition algorithms.
Such a functionality is becoming more and more popular nowadays, due to the evolution of multimedia databases containing audiovisual data~\cite{navarro2001guided}.
String algorithms can be also used in \emph{intrusion detection systems}, where their goal is to identify malicious activities by matching data such as system state graphs, instruction sequences, or packets with those from the database~\cite{kumar1994pattern, tuck2004deterministic}.
String searching can be also applied for the detection of arbitrary two-dimensional shapes in images~\cite{bunke1993applications}, and yet another application is in compression algorithms, where it is desirable to find repetitive patterns in a similar way to sequence searching in biological data.
Due to the fact that almost any data can be represented in a textual form many other application areas exist, see, e.g.,~Navarro~\cite{navarro2001guided} for more information.

This diversity of data causes the string algorithms to be used in very different scenarios.
The pattern size can vary from a few letters (\ac{nl} words) to a few hundred (DNA reads), and the input text can be  of almost arbitrary size.
For instance, Google reported in 2015 that their web search index has reached over 100 thousand terabytes ($10^{17}$ bytes)~\cite{googlesearch}.
Massive data is also present in bioinformatics, where the size of the genome of a single organism is often measured in gigabytes; one of the largest animal genomes belong to the lungfish and the salamander, each occupying approximately 120\,Gbp~\cite{gregory2002genome} (i.e.~roughly 28\,GB, assuming that each base is coded with 2 bits).
As regards proteins, the UniProt protein database stores approximately 50 million sequences (each composed of roughly a few hundred symbols) in 2015 and continues to grow exponentially~\cite{uniprot}.
It was remarked recently that biological textual databases grow more quickly than the ability to understand them~\cite{belazzougui2015space}.

When it comes to data of such magnitude, it is feasible only to perform an index-based search (meaning that the data is preprocessed), which is the main focus of this thesis.
It seems most likely that the data sizes will continue to grow, and for this reason there is a clear need for the development of algorithms which are efficient in practice.

\section{Preliminaries}

This section presents an overview of data structures and algorithms which act as building blocks for the ones presented later, and it introduces the necessary terminology.
String searching, which is the main topic of this thesis, is described in the following chapter.

Throughout this thesis, data structures are usually approached from two angles: theoretical, which concentrates on the worst-case space and query time, and a practical one.
The latter focuses on performance in real-world scenarios, and it is often heuristically oriented and focused on cache utilization and reducing slow RAM access.
It is worth noting that state-of-the-art theoretical algorithms sometimes perform very poor in practice because of certain constant factors which are ignored in the analysis.
Moreover, they might not be even tested or implemented at all.
On the other hand, a practical evaluation depends heavily on the hardware (peculiarities of the CPU cache, instruction prefetching, etc), properties of the data sets used as input, and most importantly on the implementation.
Moffat and Gog~\cite{moffat2014string} provided an extensive analysis of experimentation in the field of string searching, and they pointed out various caveats.
These include for instance a bias towards certain repetitive patterns when the patterns are randomly sampled from the input text, or the advantage of smaller data sets which increase the probability that (at least some of) the data would fit into the cache.

The theoretical analysis of the algorithms is based on the big $O$ family of asymptotic notations, including $O$, $\Omega$, $\Theta$, and the relevant lower case counterparts (we assume that the reader is familiar with these tools and with complexity classes).
Unless stated otherwise, the complexity analysis refers to the worst-case scenario, and all logarithms are assumed to be base 2 (this might be also stated explicitly as $\log_2$).
When we state that the complexity or the average or the worst case is equal to some value, we mean the running time of the algorithm.
On the other hand, if the time or space is explicitly mentioned, the word ``complexity'' might be omitted.
Array, string, and vector indexes are always 0-based and they are assumed to be contiguous, and collection indexes are 1-based (e.g.,~a collection of strings $s_1, \ldots, s_n$).
We consider a standard hierarchical memory model with RAM and a faster CPU cache, and we take for granted that the data always fits into the main memory, i.e.~disk \ac{io} is ignored.
Moreover, we assume that the size of the data does not exceed $2^{32}$ bytes, which means that it is sufficient for each pointer or counter to occupy 32 bits (4 bytes).
Sizes that are given in kilobytes and megabytes are indicated with abbreviations KB and MB, which refer to standard computer science quantities, i.e. $2^{10}$ (rather than 1,000) and $2^{20}$.

\subsection{Sorting}

Sorting consists in ordering $n$ elements from a given set $S$ in such a way that the following holds: $\forall i \in [0, n - 1) : S[i] \leqslant S[i + 1]$, that is the smallest element is always in front.
In reverse sorting, the highest element is in front, and the inequality sign is reversed.
Popular sorting methods include the heapsort and the mergesort with $O(n \log n)$ worst-case time guarantees.
Another well-known algorithm is the quicksort with average time $O(n \log n)$ (although the worst case is equal to $O(n^2)$), which is known to be 2--3 times faster in practice than both heapsort and mergesort~\cite{skiena1998algorithm}.
There also exist algorithms linear in $n$ which can be used in certain scenarios, for instance the radix sort for integers with time complexity $O(wn)$ (or $O(n (w / \log n))$ for a byte-wide radix), where $w$ is the machine word size.

When it comes to sorting $n$ strings of average length $m$, a comparison sorting method would take $O(n \log n m)$ time (assuming that comparing two strings is linear in time).
Alternatively, we could obtain an $O(n m)$ time bound by sorting each letter column with a sorting method which is linear for a fixed alphabet (essentially performing a radix sort), e.g.,~using a counting sort.
Moreover, we can even achieve an $O(n)$ complexity by building a trie with lexicographically ordered children at each level and performing a pre-order, \ac{dfs} (see the following subsections for details).
When it comes to suffix sorting (i.e.~sorting all suffixes of the input text), dedicated methods which do not have a linear time guarantee are often used due to reduced space requirements or good practical performance~\cite{manzini2004engineering, puglisi2007taxonomy}.
Recently, linear methods which are efficient in practice have also been described~\cite{DBLP:journals/tois/Nong13}.

\subsection{Trees}

A tree contains multiple nodes that are connected with each other, with one (top-most) node designated as the \emph{root}.
Every node contains zero or more children, and a tree is an undirected graph where any two vertexes are connected by exactly one path (there are no cycles).
Further terminology which is relevant to trees is as follows~\cite[Sec.~B.5]{cormen}:

\begin{itemize}

\item
A \textbf{parent} is a neighbor of a \textbf{child} and it is located closer to the root (and vice versa).
\item
A \textbf{sibling} is a node which shares the same parent.
\item
\textbf{Leaves} are the nodes without children and in the graphical representation they are always shown at the bottom of the diagram. The leaves are also called external nodes, and internal nodes are all nodes other than leaves.
\item
\textbf{Descendants} are the nodes located anywhere in the subtree rooted by the current node, and \textbf{ancestors} are the nodes anywhere on the path from the root (inclusive) to the current node. \textbf{Proper} descendants and ancestors exclude the current node. If $n_1$ is an ancestor of $n_2$, then $n_2$ is a descendant of $n_1$, and vice versa.
\item
The \textbf{depth} of a node is the length of the path from this node to the root.
\item
The \textbf{height} of a tree is the longest path from the root to any leaf (i.e.~the depth of the deepest node).

\end{itemize}

The maximum number of children can be limited for each node.
Many structures are \emph{binary} trees, which means that every node has at most two children, and a generic term is a $k$-ary tree (or multiary for $k > 2$).
A full (complete) $k$-ary tree is a structure where every node has exactly 0 (in the case of leaves) or $k$ (in the case of internal nodes) children.
A perfect tree is a tree where all leaves have the same depth.
A historical note: apparently, a binary tree used to be called a \emph{bifurcating arborescence} in the early years of computer science~\cite[p.~363]{knuth1997tauhe}.
A \emph{balanced} (height-balanced, self-balancing) tree is a tree whose height is maintained with respect to its total size irrespective of possible updates and deletions.
The height of a balanced binary tree is logarithmic, i.e.~$O(\log n)$ ($\log_k n$ for a $k$-ary tree).
It is often desirable to maintain such a balance because otherwise a tree may lose its properties (e.g.,~worst-case search complexity).
This is caused by the fact that the time complexity of various algorithms is proportional to the height of the tree.

There exist many kinds of trees, and they are characterized by some additional properties which make them useful for a certain purpose.

\subsubsection{Binary search tree}

\begin{figure}[ht]
    \centering
    \includegraphics[scale=0.65]{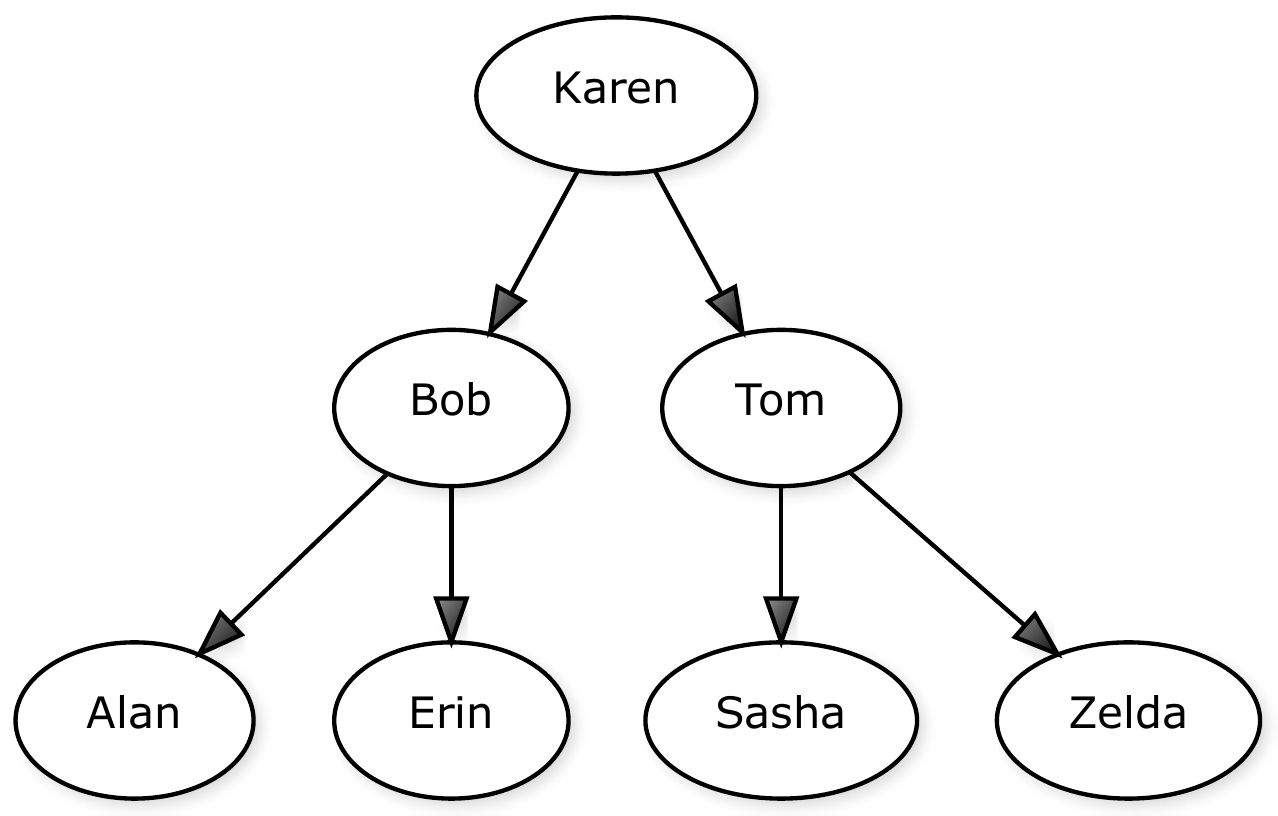}

    \caption[A \acf{bst} storing strings from the English alphabet.]{A \acf{bst} storing strings from the English alphabet. The value of the right child is always bigger than the value of the parent, and the value of the left child is always smaller than the value of the parent.}
    \label{Fig:BST}
\end{figure}

The \ac{bst} is used for determining a membership in the set or for storing key-value pairs.
Every node stores one value $V$; the value of its right child is always bigger than $V$, and the value of its left child is always smaller than $V$.
The lookup operation consists in traversing the tree towards the leaves until either the value is found or there are no more nodes to process, which indicates that the value is not present.
The \ac{bst} is often used to maintain a collection of numbers, however, the values can also be strings (they are ordered alphabetically), see Figure~\ref{Fig:BST}.
It is crucial that the \ac{bst} is balanced.
Otherwise, in the scenario where every node had exactly one child, its height would be linear (basically forming a linked list) and thus the complexity for the traversal would degrade from $O(\log n)$ to $O(n)$.
The occupied space is clearly linear (there is one node per value) and the preprocessing takes $O(n \log n)$ time because each insertion costs $O(\log n)$.

\subsubsection{Trie}
\label{Sec:trie}

\begin{figure}[ht]
    \centering
    \includegraphics[scale=0.65]{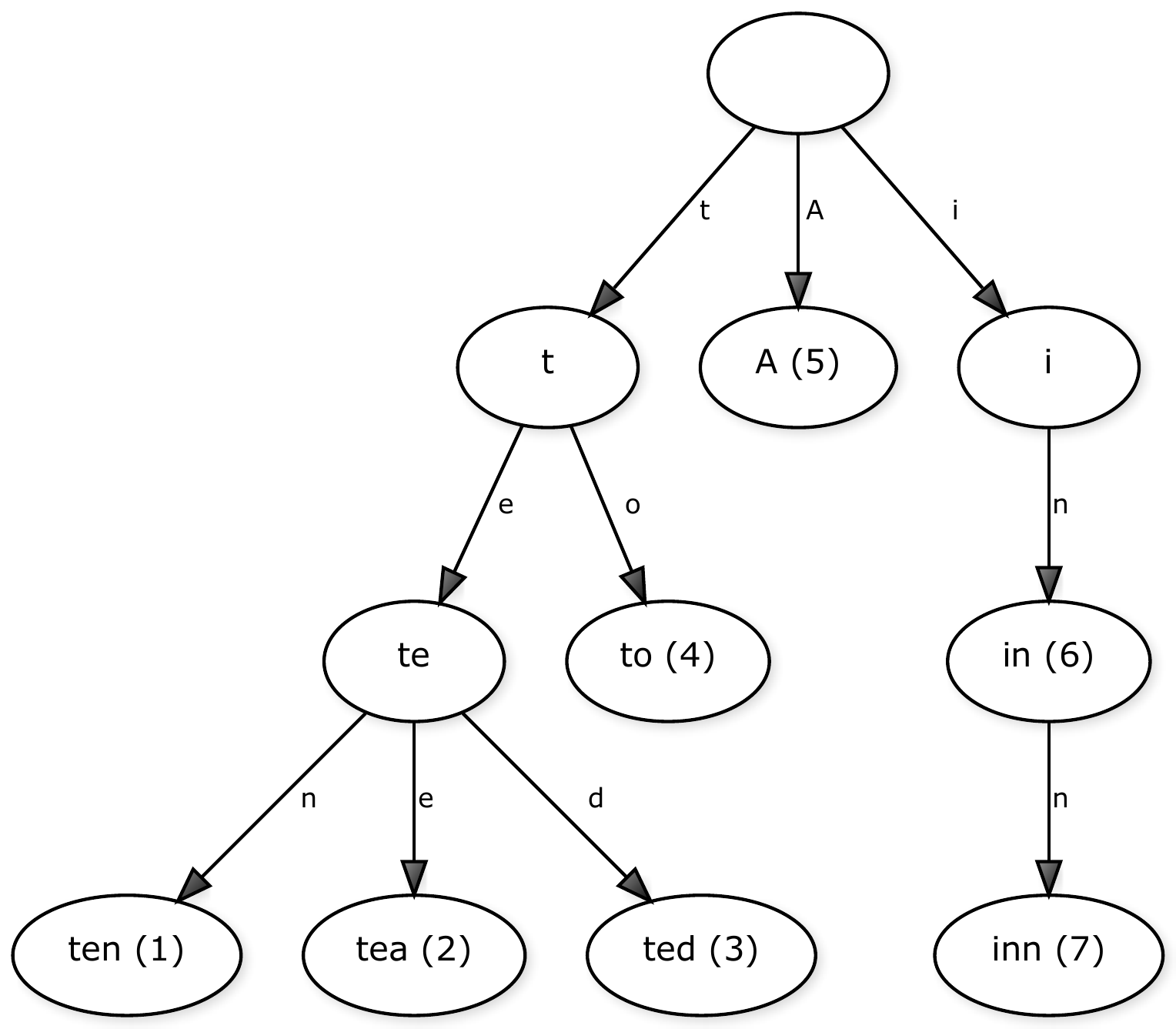}

    \caption[A trie which is one of the basic structures used in string searching.]{A trie, which is one of the basic structures used in string searching, constructed for strings from the set $\{\texttt{A}, \texttt{in}, \texttt{inn}, \texttt{tea}, \texttt{ted}, \texttt{ten}, \texttt{to}\}$. Each edge corresponds to one character, and the strings are stored implicitly (here shown for clarity). Additional information such as IDs (here shown inside the parentheses) is sometimes kept in the nodes.}
    \label{Fig:trie}
\end{figure}

The trie (digital tree)~\cite{fredkin1960trie} is a tree in which the position of a node (more specifically, a path from the root to the node) describes the associated value, see Figure~\ref{Fig:trie}.
The nodes often store IDs or flags which indicate whether a given node has a word, which is required because some nodes may be only intermediary and not associated with any value.
The values are often strings, and the paths may correspond to the prefixes of the input text.
A trie supports basic operations such as searching, insertion, and deletion.
For the lookup, we check whether each consecutive character from the query is present in the trie while moving towards the leaves, hence the search complexity is directly proportional to the length of the pattern.
In order to build a trie, we have to perform a full lookup for each word, thus the preprocessing complexity is equal to $O(n)$ for words of total length $n$.
The space is linear because there is at most one node per input character.

Various modifications of the regular trie exist --- an example could be the \emph{Patricia trie}~\cite{morrison1968patricia}, whose aim is to reduce the occupied space.
The idea is to merge every node which has no siblings with its parent, thus reducing the total number of nodes, and resulting edge labels include the characters from all edges that were merged (the complexities are unchanged).

\subsection{Hashing}
\label{Sec:hashing}

A \emph{hash function} $H$ transforms data of arbitrary size into the data of fixed size.
Typical output sizes include 64, 128, 256, and 512 bits.
The input can be in principle of any type, although hash functions are usually designed so that they work well for a particular kind of data, e.g.,~for strings or for integers.
Hash functions often have certain desirable properties such as a limited number of collisions, where for two chunks of data $d_1$ and $d_2$, the probability that $H(d_1) = H(d_2)$ should be relatively low (e.g., $H$ is called universal if~$Pr(H(d_1) = H(d_2)) \leqslant 1/n$ for an $n$-element hash table~\cite{carter1977universal}).
There exists a group of cryptographic hash functions, which offer certain guarantees regarding the number of collisions.
They also provide non-reversibility, which means that it is hard (in the mathematical sense, for example the problem may be NP-hard) to deduce the value of the input string from the hash value.
Such properties are provided at the price of reduced speed, and for this reason cryptographic hash functions are usually not used for string matching.

A \emph{perfect} hash function guarantees no collisions, e.g.,~the two-level FKS scheme with $O(n)$ space~\cite{fredman1984storing}.
All keys have to be usually known beforehand, although dynamic perfect hashing was also considered~\cite{dietzfelbinger1994dynamic}.
A \emph{minimal} perfect hash function (\acs{mphf}) uses every bucket in the hash table, i.e.~there is one value per bucket (the lower space bound for describing an \acs{mphf} is equal to roughly $1.44 n$ bits for $n$ elements~\cite{belazzougui2009hash}).
The complexity of a hash function is usually linear in the input length, although it is sometimes assumed that it takes constant time.

\begin{figure}[ht]
    \centering
    \includegraphics[scale=0.75]{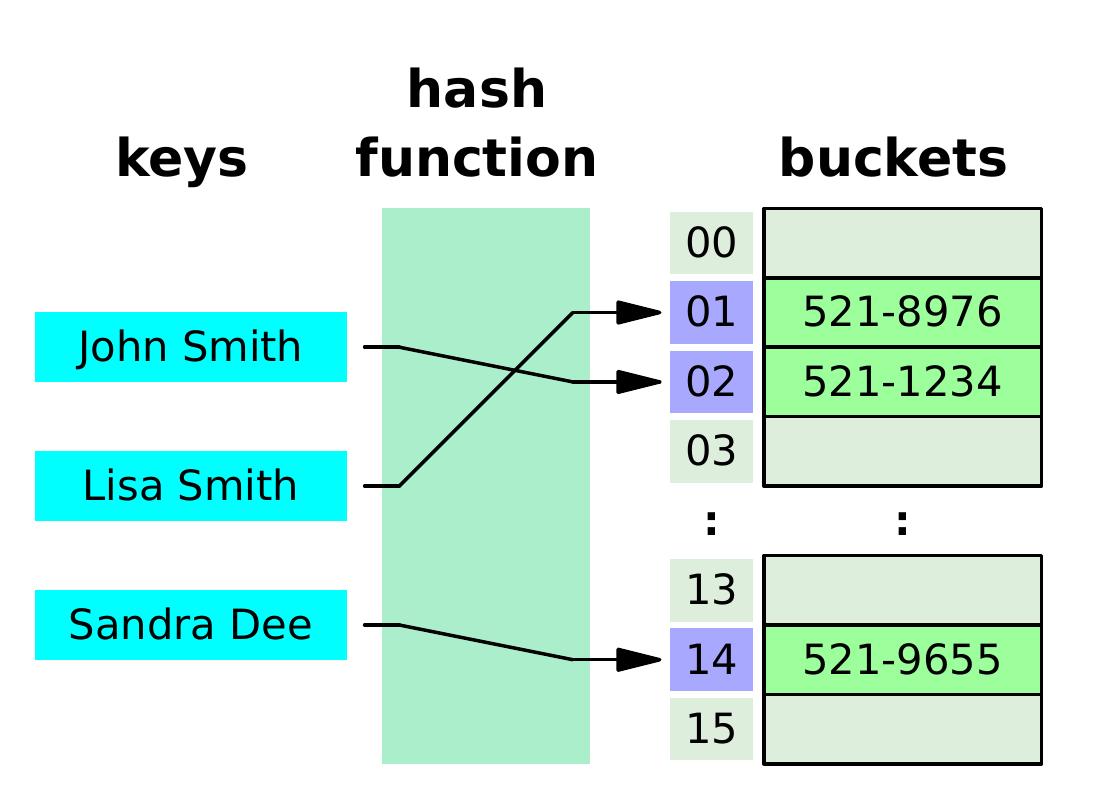}

    \caption[A hash table for strings.]{A hash table for strings; reproduced from Wikimedia Commons\protect\footnotemark.}
    \label{Fig:hash_table}
\end{figure}

A hash function is an integral part of a \emph{hash table} $H_T$, which is a data structure that associates the \emph{values} with \emph{buckets} based on the \emph{key}, i.e.~the hash of the value.
This can be represented with the following relation: $H_T[H(v)] = v$ for any value $v$.
Hash tables are often used in string searching because they allow for quick membership queries, see Figure~\ref{Fig:hash_table}.
The size of the hash table is usually much smaller than the number of all possible hash values, and it is often the case that a collision occurs (the same key is produced for two different values).
There exist various methods of resolving such collisions, and the most popular ones are as follows:

\begin{itemize}
\item
\textbf{Chaining} --- each bucket holds a list of all values which hashed to this bucket.

\item
\textbf{Probing} --- if a collision occurs, the value is inserted into the next unoccupied bucket. This may be linear probing, where the consecutive buckets are scanned linearly until an empty bucket is found, or quadratic probing, where the gaps between consecutive buckets are formed by the results of a quadratic polynomial.

\item
\textbf{Double hashing} --- gaps between consecutive buckets are determined by another hash function. A simple approach could be for instance to locate the next bucket index $i$ using the formula $i = H_1(v) + i H_2(v) \mod |H_T|$ for any two hash functions $H_1$ and $H_2$.

\end{itemize}

In order to resolve the collisions, the keys have to be usually stored as well.
The techniques which try to locate an empty bucket (as opposed to chaining) are referred to as \emph{open addressing}.
A key characteristic of the hash table is its \emph{load factor} (\acs{lf}), which is defined as the number of entries divided by the number of buckets.
Let us note that $0 \leqslant L_F \leqslant 1.0$ for open addressing (the performance degrades rapidly as $L_F \to 1.0$), however, in the case of chaining it holds that $0 \leqslant L_F \leqslant n$ for $n$ entries.

\setcounter{footnote}{0}

\subsection{Data structure comparison}
\label{Sec:ds_comparison}

In the previous subsections, we introduced data structures which are used by more sophisticated algorithms (described in the following chapters).
Still, they can be also used on their own for exact string searching.
In Figure~\ref{Tab:ds_comp} we present a comparison of their complexities together with a linear, direct access array.
It is to be noted that even though the worst case of a hash table lookup is linear (iterating over one bucket which stores all the elements), it is extremely unlikely, and any popular hash function offers reasonable guarantees against building such a degenerate hash table.

\footnotetext{Jorge Stolfi, available at \url{http://en.wikipedia.org/wiki/File:Hash_table_3_1_1_0_1_0_0_SP.svg}, CC A-SA 3.0.}

\begin{table}[ht]
\vspace{1em}
\centering
\begin{tabular}{c|ccc}
Data structure & Lookup & Preprocessing & Space \\
\hline
Array & $O(n)$ & $O(1)$ & $O(n)$ \\
Balanced BST & $O(\log n)$ & $O(n \log n)$ & $O(n)$ \\
Hash table & $O(m)$ avg, $O(n)$ worst-case & $O(n)$ & $O(n)$ \\
Trie & $O(m)$ & $O(n)$ & $O(n)$ \\
\end{tabular}
\vspace{4mm}
\caption[A comparison of the complexities of basic data structures which can be used for exact string searching.]{A comparison of the complexities of basic data structures which can be used for exact string searching. Here, we assume that string comparison takes constant time.}
\label{Tab:ds_comp}
\end{table}

\subsection{Compression}
\label{Sec:compression}

Compression consists in representing the data in an alternative (encoded) form with the purpose of reducing the size.
After compression, the data can be decompressed (decoded) in order to obtain the original representation.
Typical applications include reducing storage sizes and saving bandwidth during transmission.
Compression can be either lossless or lossy, depending on whether the result of decompression matches the original data.
The former is useful especially when it comes to multimedia (frequently used by domain-specific methods such as those based on human perception of images), where the lower quality may be acceptable or even indiscernible, and storing the data in an uncompressed form is often infeasible.
For instance, the original size of a 2-hour Full HD movie with 32 bits per pixel and 24 frames per second would amount to more than one terabyte.
Data that can be compressed is sometimes called \emph{redundant}.

One of the most popular compression methods is character substitution, where the selected symbols (bit $q$-grams) are replaced with ones that take less space.
A classic algorithm is called Huffman coding~\cite{huffman1952method}, and it offers an optimal substitution method.
Based on $q$-gram frequencies, it produces a codebook which maps more frequent characters to shorter codes, in such a way that every code is uniquely decodable (e.g.,~\texttt{00} and \texttt{1} are uniquely decodable, but \texttt{00} and \texttt{0} are not).
For real-world data, Huffman coding offers compression rates that are close to the entropy (see the following subsection), and it is often used as a component of more complex algorithms.
We refer the reader to Salomon's monograph~\cite{salomon2004data} for more information on data compression.

\subsubsection{Entropy}
\label{Sec:entropy}

We can easily determine the compression ratio by taking the size (a number of occupied bits) of the original data and dividing it by the size of the compressed data, $r = n / n_c$.
It may seem that the following should hold: $r \geqslant 1$, however, certain algorithms might actually increase the data size after compressing it when operating on an inconvenient data set (which is of course highly undesirable).
A related problem is how to determine the ``compressibility'' of the data, i.e.~the optimal compression ratio (the highest $r$).
This brings us to the notion of \emph{entropy}, sometimes also called Shannon's entropy after the name of the author~\cite{shannon1948mathematical}.
It describes the amount of information which is contained in a message, and in the case of strings it determines the average number of bits which is required in order to encode an input symbol under a specified alphabet and frequency distribution.
This means that the entropy describes a theoretical bound on data compression (one that cannot be exceeded by any algorithm).
Higher entropy means that it is more difficult to compress the data (e.g.,~when multiple symbols appear with equal frequency).
The formula is presented in Figure~\ref{Fig:entropy}.

\begin{figure}[h]

\[
E = -K \sum\limits^{n}_{i=1} p_i \log p_i.
\]

\caption[The formula for Shannon's entropy.]{The formula for Shannon's entropy, where $E$ is the entropy function, $p_i$ is the probability that symbol $i$ occurs, and $K$ is any constant.}
\label{Fig:entropy}
\end{figure}

A variation of entropy which is used in the context of strings is called a $k$-th order entropy.
It takes the context of $k$ preceding symbols into account and it allows for the use of different codes based on this context (e.g.,~ignoring a symbol $c_2$ which always appears after the symbol $c_1$).
Shannon's entropy corresponds to the case of $k = 0$ (denoted as $H_0$, or $H_k$ in general).
When we increase the $k$ value, we also increase the theoretical bound on compressibility, although the size of the data required for storing context information may at some point dominate the space~\cite{gog2011compressed}.

\subsection{Pigeonhole principle}

Let us consider a situation where we have $x$ buckets and $n$ items which are to be positioned inside those buckets.
The pigeonhole principle (often also called \emph{Dirichlet} principle) states that if $n > x$, then at least one of the buckets must store more than one item.
The name comes from an intuitive representation of the buckets as boxes and items as pigeons.
Despite its simplicity, this principle has been successfully applied to various mathematical problems.
It is also often used in computer science, for example to describe the number of collisions in a hash table.
Later we will see that the pigeonhole principle is also useful in the context of string searching, especially when it comes to string partitioning and approximate matching.

\section{Overview}

This thesis is organized as follows:
\begin{itemize}
\item
\textbf{Chapter~\ref{Chap:strings}} provides an overview of the field of string searching, deals with the underlying theory, introduces relevant notations, and discusses related work in the context of online search algorithms.
\item
\textbf{Chapter~\ref{Chap:full-text}} includes related work and discusses current state-of-the-art algorithms for full-text indexing as well as our contribution to this area.
\item
\textbf{Chapter~\ref{Chap:keyword}} does the same for keyword indexes.
\item
\textbf{Chapter~\ref{Chap:experimental}} describes the experimental setup and presents practical results.
\item
\textbf{Chapter~\ref{Chap:concs}} contains conclusions and pointers to the possible future work.
\vspace{\baselineskip}
\item
\textbf{Appendix~\ref{App:data_sets}} offers information regarding the data sets which were used for the experimental evaluation.
\item
\textbf{Appendix~\ref{App:str_comp}} discusses the complexity of exact string comparison.
\item
\textbf{Appendix~\ref{App:split_comp}} discusses the $q$-gram-based compression of the split index (Section~\ref{Sec:split_index}) in detail.
\item
\textbf{Appendix~\ref{App:str_sketches}} contains experimental results for string sketches (Section~\ref{Sec:str_sketches}) when used for the alphabet with uniform letter frequencies.
\item
\textbf{Appendix~\ref{App:letter_freq}} presents the frequencies of English alphabet letters.
\item
\textbf{Appendix~\ref{App:hashes}} contains Internet addresses where the reader can obtain the code for hash functions which were used to obtain experimental results for the split index (Section~\ref{Sec:split_index_res}).
\end{itemize}

\chapter{String Searching}
\label{Chap:strings}
\lhead{\emph{String Searching}}

This thesis deals with strings, which are sequences of symbols over a specified alphabet.
The string is usually denoted as \acs{S}, the alphabet as \acs{alph}, and the length (size) as \acs{n} or $|S|$ for strings, and \acs{alphSize} or $|\Sigma|$ for the alphabet.
An arbitrary string $S$ is sometimes called a word (which is not to be confused with the machine word, i.e.~a basic data unit in the processor), and it is defined over a given alphabet $\Sigma$, that is $S$ belongs to the set of all words specified over the said alphabet, $S \in \Sigma^*$.
Both strings and alphabets are assumed to be finite and well-defined, and alphabets are totally ordered.
A string with a specified value is written with the teletype font, as in \texttt{abcd}.
The brackets are usually used to indicate the character at a specified position and the index is 0-based, for instance if string $S$ = \texttt{text}, then $S[1] = \texttt{e}$.
A \emph{substring} (sometimes referred to as a factor) is written as $S[i_0, i_1]$ (an inclusive range); for the previous example, $S[1, 2] = \texttt{ex}$, and a single character is a substring of length 1 (usually denoted with $c$).
The last character is indicated with -1 and $P = P[0, -1]$.
$S_1 \subset S_2$ indicates that the string $S_1$ is a substring of $S_2$ (conversely, $S_1 \not\subset S_2$ indicates that $S_1$ is not a substring of $S_2$).
The subscripts are usually used to distinguish multiple strings, and two strings may be concatenated (merged into one), recorded as $S = S_1S_2$ or $S = S_1 + S_2$, in which case $|S| = |S_1| + |S_2|$.
Removing one substring from another is indicated with the subtraction sign, $S = S_1 - S_2$, provided that $S_2 \subset S_1$, and as a result $|S| = |S_1| - occ \cdot |S_2|$ for $occ$ occurrences of $S_2$ in $S_1$.
The equality sign indicates that the strings match exactly, which means that the following relation always holds: $S_1 = S_2 \to |S_1| = |S_2|$.

\emph{String searching}, or \emph{string matching}, refers to locating a substring (a pattern, $P$, or a query, $Q$, with length $m$) in a longer text $T$.
The textual data $T$ that is searched is called the \emph{input} (input string, input text, text, database), and its length is denoted by $n$ (e.g.,~$O(n)$ indicates that the complexity is linear with respect to the size of the original data).
The pattern is usually much smaller than the input, often in multiple orders of magnitude ($m \ll n$).
Based on the position of the pattern in the text, we write that $P$ occurs in $T$ with a shift $s$, i.e.~$P[0, m - 1] = T[s, s + m - 1] \land s \leqslant n - m$.
As mentioned before, the applications may vary (see Section~\ref{Sec:applications}), and the data itself can come from many different domains.
Still, string searching algorithms can operate on any text while being oblivious to the actual meaning of the data.
The field concerning the algorithms for string processing is sometimes called \emph{stringology}.

Two important notions are \emph{prefixes} and \emph{suffixes}, where the former is a substring $S[0, i]$ and the latter is a substring $S[i, |S| - 1]$ for any $0 \leqslant i < |S|$.
Let us observe at this point that every substring is a prefix of one of the suffixes of the original string, as well as a suffix of one of the prefixes.
This simple statement is a basis for many algorithms which are described in the following chapters.
A \emph{proper} prefix or suffix is not equal to the string itself.
The strings can be \emph{lexicographically} ordered, which means that they are sorted according to the ordering of the characters from the given alphabet (e.g.,~for the English alphabet, letter \texttt{a} comes before \texttt{b}, \texttt{b} comes before \texttt{c}, etc).
Formally, $S_1 < S_2$ for two strings of respective lengths $n_1$ and $n_2$, if $\exists s : \forall i \in [0, s) : S_1[i] = S_2[i] \land S_1[s] < S_2[s] \land 0 \leqslant s < \min(n1, n2) $ or $n_1 < n_2 \land S_1 = S_2[0, n_1-1]$~\cite[p.~304]{cormen}.
When it comes to strings, we often mention $q$-grams and $k$-mers, which are lists of contiguous characters (strings or substrings).
The former is usually used in general terms and the latter is used for biological data (especially DNA reads).
A 5-gram or 5-mer is a $q$-gram of length 5.
\section{Problem classification}
\label{Sec:problem_class}

The match between the pattern and the substring of the input text is determined according to the specified similarity measure, which allows us to divide the algorithms into two categories: \emph{exact} and \emph{approximate}.
The former refers to direct matching, where the length as well as all characters at the corresponding positions must be equal to each another.
This relation can be represented formally for two strings $S_1$ and $S_2$ of the same length $n$, which are equal if and only if $\forall i \in [0, n) : S_1[i] = S_2[i]$ (or simply, $S_1 = S_2$).
In the case of approximate matching, the similarity is measured with a specified \emph{distance}, also called an \emph{error metric}, between the two strings.
It is to be noted that the word approximation is not used here strictly in the mathematical sense, since approximate search is actually harder than the exact one when it comes to strings~\cite{navarro2001guided}.
In general, given two strings $S_1$ and $S_2$, the distance $D(S_1, S_2)$ is the minimum cost of \emph{edit} operations that would transform $S_1$ into $S_2$ or vice versa.
The edits are usually defined as a finite, well-defined set of rules $E = \{r : r(S) = S^\prime\}$, and each rule can be associated with a different cost.
When error metrics are used, the results of string matching are limited to those substrings which are close to the pattern.
This is defined by the threshold $k$, that is we report all substrings $s$ for which $D(s, P) \leqslant k$.
For metrics with fixed penalties for errors, $k$ is called the maximum allowed number of errors.
This value may depend both on the data set and the pattern length, for instance for spell checking a reasonable number of errors is higher for longer words.
It should hold that $k < m$, since otherwise the pattern could match any string, and $k = 0$ corresponds to the exact matching scenario.
See Subsection~\ref{Sec:error_metrics} for detailed descriptions of the most popular error metrics.

The problem of searching (also called a lookup) can vary depending on the kind of answer that is provided.
This includes the following operations:

\begin{itemize}

\item
\textbf{Match} --- determining the membership, i.e.~deciding whether $P \subset T$ (a decision problem).
When we consider the search complexity, we usually implicitly mean the $match$ query.
\item
\textbf{Count} --- stating how many times $P$ occurs in $T$.
This refers to the cardinality of the set containing all indexes $i$ s.t. $T[i, i+m-1]$ is equal to $P$.
Specific values of $i$ are ignored in this scenario.
The time complexity of the $count$ operation often depends on the number of occurrences (denoted with $occ$).
\item
\textbf{Locate} --- reporting all occurrences of $P$ in $T$, i.e.~returning all indexes $i$ s.t. $T[i, i+m-1]$ is equal to $P$.
\item
\textbf{Display} --- showing $k$ characters which are located before and after each match, that is for all aforementioned indexes $i$ we display substrings $T[i-k, i - 1]$ and $T[i+m, i+m+k-1]$. In the case of approximate matching, it might refer to showing all text substrings or keywords $s$ s.t. $D(s, P) \leqslant k$.

\end{itemize}

String searching algorithms can be also categorized based on whether the data is preprocessed.
One such classification adapted from Melichar et al.~\cite[p.~8]{melichar2005text} is presented in Table~\ref{Tab:algo_class}.
Offline searching is also called index-based searching because we preprocess the text and build a data structure which is called an index.
This is opposed to online searching, where no preprocessing of the input text takes place.
For detailed descriptions of the examples from these classes, consult Chapters~\ref{Chap:full-text}~and~\ref{Chap:keyword} (offline) and Section~\ref{Sec:online_searching} (online).

\begin{table}[ht]
\centering
\vspace{1em}
\begin{tabular}{cccc}
Text prepr. & Pattern prepr. & Algorithm type & Examples \\
\hline
No & No & Online & Naive, dynamic programming \\
No & Yes & Online & Pattern automata, rolling hash \\
Yes & No & Offline & Index-based methods \\
Yes & Yes & Offline & Signature methods \\
\end{tabular}
\vspace{4mm}
\caption{Algorithm classification based on whether the data is preprocessed.}
\label{Tab:algo_class}
\end{table}

\subsection{Error metrics}
\label{Sec:error_metrics}

The motivation behind error metrics is to minimize the score between the strings which are somehow related to each other.
Character differences that are more likely to occur should carry a lower penalty depending on the application area, for instance in the case of DNA certain mutations appear in the real world much more often than others.
The most popular metrics include:

\begin{itemize}
\item
\textbf{Hamming} distance --- relevant for two strings of equal length $n$, calculates the number of differing characters at corresponding positions (hence it is sometimes called the $k$-mismatches problem).
Throughout this thesis we denote the Hamming distance with $Ham$, and given that $|S_1| = |S_2| = n$, $Ham(S_1, S_2) = |E|$, where $E =\{i : i \in [0, n) \land S_1[i] \neq S_2[i]\}$, and $Ham(S_1, S_2) \leqslant n$.
Without preprocessing, calculating the Hamming distance takes $O(n)$ time.
Applications of the Hamming distance include, i.a., bioinformatics~\cite{kurtz2001reputer, landau2001algorithm}, biometrics~\cite{davida1998enabling}, cheminformatics~\cite{flower1998properties}, circuit design~\cite{girard1997reduction}, and web crawling~\cite{manku2007detecting}.
\item
\textbf{Levenshtein} distance --- measures the minimum number of \emph{edits}, here defined as insertions, deletions, and substitutions.
It was first described in the context of error correction for data transmission~\cite{levenshtein1966binary}.
It must hold that $Lev(S_1, S_2) \leqslant \max(|S_1|, |S_2|)$.
The calculation using the dynamic programming algorithm takes $O(|S_1| |S_2|)$ time using $O(\min(|S_1|, |S_2|))$ space (see Subsection~\ref{Sec:online_approx}).
Ukkonen \cite{ukkonen1985algorithms} recognized certain properties of the DP matrix and presented an algorithm with $O(k \min(|S_1|, |S_2|))$ time for $k$ errors, and an approximation algorithm in a near-linear time was also described~\cite{andoni2010polylogarithmic}.
Levenshtein distance is sometimes called simply the edit distance. 
When the distance for approximate matching is not explicitly specified, we assume the Levenshtein distance.
\item
Other \textbf{edit} distances.
These may allow only a subset of edit actions, e.g.,~\emph{\aclu{lcs}}~(\acs{lcs}) which is restricted to indels~\cite{needleman1970general} or the \emph{Episode} distance with deletions~\cite{das1997episode}.
Another approach is to introduce additional actions; examples include the \emph{Damerau--Levenshtein} distance which counts a transposition as one edit operation~\cite{bard2007spelling}, a distance which allows for matching one character with two and vice versa (specifically designed for \ac{ocr})~\cite{christodoulakis2008edit}, or a distance which has weights for substitutions based on the probability that a user may mistype one character for another~\cite{brill2000improved}.
\item
\textbf{Sequence alignment} --- there may exist gaps (other characters) in between substrings of $S_1$ and $S_2$.
Moreover, certain characters may match each other even though they are not strictly equal.
The gaps themselves (e.g.,~their lengths or positions~\cite{navarro2003fast, altschul1986optimal}) as well as the inequality between individual characters are quantified.
The score is calculated using a \emph{similarity matrix} which is constructed based on statistical properties of the elements from the domain in question (e.g.,~the BLOSUM62 matrix for the sequence alignment of proteins~\cite{eddy2004did}).
The problem can be also formulated as $(\delta, \alpha)$-matching, where the width of the gaps is at most $\alpha$ and for a set $P$ of positions of corresponding characters, it should hold that $\forall (i_1, i_2) \in P : |S_1[i_1] - S_2[i_2]| \leqslant \delta$~\cite{crochemore2002approximate}.
This means that absolute values of numerical differences between certain characters cannot exceed a specified threshold.
Sequence alignment is a generalization of the edit distance, and it can be also performed for multiple sequences, although this is known to be NP-complete~\cite{wang1994complexity}.
\item
\textbf{Regular expression} matching --- the patterns may contain certain metacharacters with various meanings.
These can specify ranges of characters which can match at certain positions or use additional constructs such as the wildcard symbol which matches 0 or more consecutive characters of any type.
\end{itemize}

\section{Online searching}
\label{Sec:online_searching}

In this section, we present selected algorithms for online string searching, and we divide them into exact and approximate ones.
Online algorithms do not preprocess the input text, however, the pattern may be preprocessed.
We assume that the preprocessing time complexity is equal to $O(1)$, and the time required for pattern preprocessing is subsumed under search complexity, which means that we consider a scenario where the patterns are not known beforehand.
Search time refers to the $match$ query.

\subsection{Exact}

Faro and Lecroq~\cite{faro2013exact} provided a survey on online algorithms for exact matching and remarked that over 80 algorithms have been proposed since the 1970s.
They categorized the algorithms into the following three groups:

\begin{itemize}
\item
Character comparisons
\item
Automata
\item
Bit parallelism
\end{itemize}

The \emph{naive} algorithm attempts to match every possible substring of $T$ of length $m$ with the pattern $P$.
This means that it iterates from left to right and checks whether $T[i,m - 1] = P$ for each $0 \leqslant i \leqslant n - m$.
Right to left iteration would be also possible and the algorithm would report the same results.
Time complexity is equal to $O(n m)$ in the worst case (although to $O(n)$ on average, see Appendix~\ref{App:str_comp} for more information), and there is no preprocessing or space overhead.

Even without text preprocessing, the performance of the naive algorithm can be improved significantly by taking advantage of the information provided by the mismatches between the text and the pattern.
Classical solutions for single-pattern matching include the \emph{\aclu{kmp}}\,(\acs{kmp})~\cite{knuth1977fast} and the \emph{\aclu{bm}}\,(\acs{bm})~\cite{boyer1977fast} algorithm.
The \ac{kmp} uses information regarding the characters that appear in the pattern in order to avoid repeated comparisons known from the naive approach.
It reduces the time complexity from $O(n m)$ to $O(n + m)$ in the worst case at the cost of $O(m)$ space.
When a mismatch occurs at position $i$ in the pattern ($P[i] \neq T[s + i]$), the algorithm shifts P by $i - l$, where $l$ is the length of the longest proper prefix of $P_S = P[0, i - 1]$ which is also a suffix of $P_S$ (instead of just 1 position), and it starts matching from the position $i=l$ instead of $i=0$.
Information regarding $P_S$ is precomputed and stored in a table of size $m$.
Let us observe that the algorithm does not skip any characters from the input string.
Interestingly, Baeza-Yates and Navarro~\cite{baeza2004text} reported that in practice the \ac{kmp} algorithm is roughly two times slower than the brute-force search (although this depends on the alphabet size).

The \ac{bm} algorithm, on the other hand, omits certain characters from the input.
It begins the matching from the end of the pattern and allows for forward jumps based on mismatches.
Thanks to the preprocessing, the size of each shift can be determined in constant time.
One of the two rules for jumping is called a bad character rule, which, given that $P[i] \neq T[s + i] \land T[s + i] = c$, aligns $T[s + i]$ with the rightmost occurrence of $c$ in $P$ ($P[j] = c$, where $j < i$), or shifts the pattern by $m$ if $c \notin P$.
The other rule is a complex good suffix rule, whose description we omit here, and which is also not a part of the \emph{\aclu{bmh}}\,(\acs{bmh})~\cite{horspool1980practical} algorithm which uses only the bad character rule --- this is because the good suffix rule requires extra cost to compute and it is often not practical.
The worst-case time complexity of the \ac{bm} algorithm is equal to $O(n m + \sigma)$, with $O(n / \min(m, \sigma) + m + \sigma)$ average (the same holds for \ac{bmh}), and the average number of comparisons is equal to roughly $3n$~\cite{cole1994tight}.
This can be improved to achieve a linear time in the worst case by introducing additional rules~\cite{galil1979improving}.

One of the algorithms developed later is the \emph{\aclu{rk}}\,(\acs{rk})~\cite{karp1987efficient} algorithm.
It starts with calculating the hash value of the pattern in the preprocessing stage, and then compares this hash with every substring of the text, sliding over it in a similar way to the naive algorithm.
Character-wise verification takes place only if two hashes are equal to each other.
The trick is to use a hash function which can be computed in constant time for the next substring given its output for the previous substring and the next character, a so-called \emph{rolling hash}~\cite{kornblum2006identifying}, viz.~$H(T[s, s + m - 1], T[s + m]) = H(T[s + 1, s + m])$.
A simple example would be to simply add the values of all characters, $H(S) = \sum\limits_{i=0}^{n-1} S[i]$.
There exist other functions such as the Rabin fingerprint~\cite{rabin1981fingerprinting}, which treats the characters as polynomial variables $c_i$, and the indeterminate $x$ is a fixed base, $R(S) = \sum\limits_{i = 1}^{n} c_i x^{i-1}$.
The \ac{rk} algorithm is suitable for multiple-pattern matching, since we can quickly compare the hash of the current substring with the hashes of all patterns using any efficient set data structure.
In this way, we obtain the average time complexity of $O(n + m)$ (assuming that hashing takes linear time), however, it is still equal to $O(n m)$ in the worst case when the hashes do match and verification is required.

Another approach is taken by the \emph{Aho--Corasick}~\cite{aho1975efficient} algorithm.
It builds a \ac{fsm}, i.e.~an automaton which has a finite number of states.
The structure of the automaton resembles a trie and it contains edges between certain nodes which represent the transitions.
It is constructed from the queries and attempts to match all queries at once when sliding over the text.
The transitions indicate the next possible pattern which can be still fully matched after a mismatch at a specified position occurs.
The search complexity is equal to $O(n \log \sigma + m + z)$, which means that it is linear with respect to the input length $n$, the length of all patterns ($m$, for building the automaton), and the number of occurrences $z$.

An example of a bit-parallel algorithm is the \emph{shift-or} algorithm by Baeza-Yates and Gonnet~\cite{baeza1992new}, which aims to speed up the comparisons.
The pattern length should be smaller than the machine word size, which is usually equal to 32 or 64 bits.
During the preprocessing, a mismatch mask $M$ is computed for each character $c$ from the alphabet, where $M[i] = 0$ if $P[i] = c$ and $M[i] = 1$ otherwise.
Moreover, we maintain a state mask $R$, initially set to all 1s, which holds information about the matches so far.
We proceed in a similar manner to the naive algorithm, trying to match the pattern with every substring, but instead of character-wise comparisons we use bit operations.
At each step, we shift the state mask to the left and \texttt{OR} it with the $M$ for the current character $T[i]$.
A match is reported if the most significant bit of $R$ is equal to 0.
Provided that $m \leqslant w$, the time complexity is equal to $O(n)$ and the masks occupy $O(\sigma)$ space.

Based on the practical evaluation, Faro and Lecroq~\cite{faro2013exact} reported that there is no superior algorithm and the effectiveness depends heavily on the size of the pattern and the size of the alphabet.
The differences in performance are substantial --- algorithms which are the fastest for short patterns are often among the slowest for long patterns, and vice versa.

\subsection{Approximate}
\label{Sec:online_approx}

In the following paragraphs, we use $\delta$ to denote the complexity of calculating the distance function between two strings (consult Subsection~\ref{Sec:error_metrics} for the description of the most popular metrics).
Navarro~\cite{navarro2001guided} presented an extensive survey regarding \emph{approximate} online matching, where he categorizes the algorithms into four categories which resemble the ones presented for the exact scenario:

\begin{itemize}
\item
Dynamic programming
\item
Automata
\item
Bit parallelism
\item
Filtering
\end{itemize}

The naive algorithm works in a similar manner to the one for exact searching, that is it compares the pattern with every possible substring of the input text.
It forms a generic idea which can be adapted depending on the edit distance which is used, and for this reason the time complexity is equal to $O(n \delta)$.

The oldest algorithms are based on the principle of \aclu{dp}~(\acs{dp}).
This means that they divide the problem into subproblems --- these are solved and their answers are stored in order to avoid recomputing these answers (i.e.~it is applicable when the subproblems overlap)~\cite[p.~359]{cormen}.
One of the most well-known examples is the \emph{\aclu{nw}}\,(\acs{nw})~\cite{needleman1970general} algorithm, which was originally designed to compare biological sequences.
Starting from the first character of both strings, it successively considers all possible actions (insertion, (mis)match, deletion) and constructs a matrix which holds all alignment scores.
It calculates the global alignment, and it can use a substitution matrix which specifies alignment scores (penalties).
The situation where the scores triplet is equal to $(-1, 1, -1$) (for gaps, matches, and mismatches, respectively) corresponds directly to the Levenshtein distance (consult Figure~\ref{Fig:NW_matrix}).
The \ac{nw} method can be invoked with the input text as one string and the pattern as the other.

A closely related variation of the NW algorithm is the \emph{\aclu{sw}}\,(\acs{sw})~\cite{smith1981identification} algorithm, which can also identify local (and not just global) alignments by not allowing negative scores.
This means that the alignment does not have to cover the entire length of the text, and it is therefore more suitable for locating a pattern as a substring.
Both algorithms can be adapted to other distance metrics by manipulating the scoring matrix, for example by assigning infinite costs in order to prohibit certain operations.
The time complexity of the \ac{nw} and \ac{sw} approaches is equal to $O(n m)$ and it possible to calculate them using $O(\min(n,m))$ space~\cite{hirschberg1975linear}.
Despite their simplicity, both methods are still popular for sequence alignment because they might be relatively fast in practice and they report the true answer to the problem, which is crucial when the quality of an alignment matters.

\begin{figure}[ht]
\centering
\vspace{1em}
\begin{tabular}{c|rrrrrr}
  &   & t & e & x & t \\
\hline
  & 0 & -1 & -2 & -3 & -4 \\
t & -1 & 1 & 0 & -1 & -2 \\
\cline{3-3}
a & -2 & 0 & 0 & -1 & -2 \\
\cline{4-4}
x & -3 & -1 & -1 & 1 & 0 \\
\cline{5-5}
i & -4 & -2 & -2 & 0 & 0 \\
\cline{6-6}
\end{tabular}
\vspace{4mm}
\caption[Calculating an alignment with Levenshtein distance using the \acf{nw} algorithm.]{Calculating an alignment with Levenshtein distance using the \acf{nw} algorithm. We follow the path from the top-left to the bottom-right corner selecting the highest possible score (underlined); the optimal global alignment is as follows: $\texttt{text} \to \texttt{taxi}$ (no gaps).}
\label{Fig:NW_matrix}
\end{figure}

Multiple other dynamic programming algorithms were proposed over the years, and they gradually tightened the theoretical bounds.
The difference lies mostly in their flexibility, that is a possibility of being adapted to other distance metrics, as well as practical performance.
Notable results for the edit distance include the \emph{Chang--Lampe}~\cite{chang1992theoretical} algorithm with $O(kn/\sqrt{\sigma})$ average time using $O(m)$ space, and the \emph{Cole--Hariharan}~\cite{cole2002approximate} algorithm with the worst-case time $O(n + m + k^c n / m)$ with $c=3$ for non-$k$-break periodic patterns and $c=4$ otherwise, taking $O(m + o_s)$ space, where $o_s$ refers to occurrences of certain substrings of the pattern in the text (the analysis is rather lengthy).
For the \ac{lcs} metric, Grabowski~\cite{grabowski14tab} provided the algorithm with $O(n m \log \log n / \log^2 n)$ time bound and linear space.

A significant achievement in the automata category is the \emph{Wu--Manber--Myers}~\cite{wu1996subquadratic} algorithm that uses the Four Russians technique, which consists in partitioning the matrix into fixed-size blocks, precomputing the values for each possible block, and then using a lookup table.
It implicitly constructs the automaton, where each state corresponds to the values in the \ac{dp} matrix.
They obtain an $O(kn / \log n)$ expected time bound using $O(n)$ space.
As regards bit parallelism, Myers~\cite{myers1999fast} presented the calculation of the DP matrix in $O(\lceil k/w \rceil n)$ average time.

An important category is formed by the filtering algorithms, which try to identify parts of the input text where it is not possible to match any substrings with the pattern.
After parts of the text are rejected, a non-filtering algorithm is used on the remaining parts.
Numerous filtering algorithms have been proposed, and one of the most significant is the \emph{Chang--Marr}~\cite{chang1994approximate} algorithm with the time bound of $O(n(k + \log_{\sigma} m) / m)$ for the error level $\alpha$ when it holds that $\alpha < 1 - e / \sqrt{\sigma}$ (for very large $\sigma$).

As regards the $k$-mismatches problem, a notable example is the \emph{Amir--Lewenstein--Porat}~\cite{amir2004faster} algorithm which can answer the locate query in $O(n \sqrt{k \log k })$ time.
This was refined to $O(n + n \sqrt{k/w} \log k)$ in the word RAM model, where $w = \Omega(\log n)$~\cite{fredriksson2009fast}.
Recently, Clifford et al.~\cite{clifford2015k} described an algorithm with search time complexity $O(n k^2 \log k / m + n\,\text{polylog}\,m)$.

\section{Offline searching}
\label{Sec:index-based}

An online search is often infeasible for real-world data, since the time required for one lookup might be measured in the order of seconds.
This is caused by the fact that any online method has to access at least $n/m$ characters from the input text~\cite[p.~155]{grabowskitext2011}, and it normally holds that $m \ll n$.
This thesis is focused on \emph{index-based} (offline) methods, where a data structure (an index, pl.~indexes or indices, we opt for the former term) is built based on the input text in order to speed up further searches, which is a classic example of data preprocessing.
This is justified even if the preprocessing time is long, since the same text is often queried with multiple patterns.
The indexes can be divided into two following categories: 

\begin{itemize}

\item
\textbf{Full-text} indexes
\item
\textbf{Keyword} (dictionary) indexes

\end{itemize}

The former means that we can search for any substring in the input text (string matching, text matching), whereas the latter operates on individual words (word matching, keyword matching, dictionary matching, matching in dictionaries).
Keyword indexes are usually appropriate where there exist well-defined boundaries between the keywords (which are often simply called words), for instance in the case of a natural language dictionary or individual DNA reads ($k$-mers).
It is worth noting that the number of distinct words is almost always smaller than the total number of words in a dictionary $\mathcal{D}$, all of which are taken from a document or a set of documents.
Heaps' law states that $|\mathcal{D}| = O(n^{\beta})$, where $n$ is the text size and $\beta$ is an empirical constant (usually in the interval $[0.4, 0.6]$)~\cite{heaps1978information}.
Full-text and keyword indexes are actually related to each other, because they are often based on similar concepts (e.g.,~the pigeonhole principle) and they may even use the other kind as the underlying data structure.

The indexes can be divided into \emph{static} and \emph{dynamic} ones, depending on whether updates are allowed after the initial construction.
Another category is formed by \emph{external} indexes --- these are optimized with respect to disk \ac{io}, and they aim to be efficient for the data which does not fit into the main memory.
We can also distinguish \emph{compressed} indexes (see Subsection~\ref{Sec:entropy} for more information on compression), which store the data in an encoded form.
One goal is to reduce storage requirements while still allowing fast searches, especially when compared to the scenario where a naive decompression of the whole index has to be performed.
On the other hand, it is also possible to achieve both space saving and a speedup with respect to the uncompressed index.
This can be achieved mostly due to reduced I/O and (rather surprisingly) fewer comparisons required for the compressed data~\cite[p.~131]{grabowskitext2011}.
Navarro and M{\"a}kinen~\cite{navarro2007compressed} note that the most successful indexes can nowadays obtain both almost optimal space and query time.
A compressed data structure usually also falls into the category of a \emph{succinct} data structure.
This is a rather loose term which is commonly applied to algorithms which employ efficient data representations with respect to space, often close to the theoretic bound.
Thanks to reduced storage requirements, succinct data structures can process texts which are an order of magnitude bigger than ones suitable for classical data structures~\cite{gog2014optimized}.
The term succinct may also suggest that the we are not required to decompress the entire structure in order to perform a lookup operation.
Moreover, certain indexes can be classified as \emph{self-indexes}, which means that they implicitly store the input string $S$.
In other words, it is possible to transform (decompress) the index back to $S$, and thus the index can essentially replace the text.

The main advantage of indexes when compared to the online scenario are fast queries, however, this naturally comes at a price.
Indexes might occupy a substantial amount of space (sometimes even orders of magnitude more than the input), they are expensive to construct, and it is often problematic to support functionality such as approximate matching and updates. 
Still, Navarro et al.~\cite{navarro2001indexing} point out that in spite of the existence of very fast (both from a practical and a theoretical point of view) online algorithms, the data size often renders online algorithms infeasible (which is even more relevant in the year 2015).

Index-based methods are explored in detail in the following chapters: full-text indexes in Chapter~\ref{Chap:full-text} and keyword indexes in Chapter~\ref{Chap:keyword}.
Experimental evaluation of our contributions can be found in Chapter~\ref{Chap:experimental}.

\chapter{Full-text Indexes}
\label{Chap:full-text}
\lhead{\emph{Full-text Indexes}}

Full-text indexes allow for searching for an arbitrary substring from the input text.
Formally, for a string $T$ of length $n$, having a set of $x$ substrings $\mathcal{S} = \{s_1, \ldots, s_x\}$ over a given alphabet $\Sigma$, $I(\mathcal{S})$ is a full-text index supporting matching with a specified distance $D$.
For any query pattern $P$, it returns all substrings $s$ from $T$ s.t. $D(P, s) \leqslant k$ (with $k=0$ for exact matching).
In the following sections, we describe data structures from this category, divided into exact (Section~\ref{Sec:full-exact}) and approximate ones (Section~\ref{Sec:full-approx}).
Our contribution in this field is presented in Subsection~\ref{Sec:fm-bloated}, which describes a variant of the well-known FM-index called FM-bloated.

\section{Exact}
\label{Sec:full-exact}

\subsection{Suffix tree}
\begin{figure}[ht]
    \centering
    \includegraphics[scale=0.65]{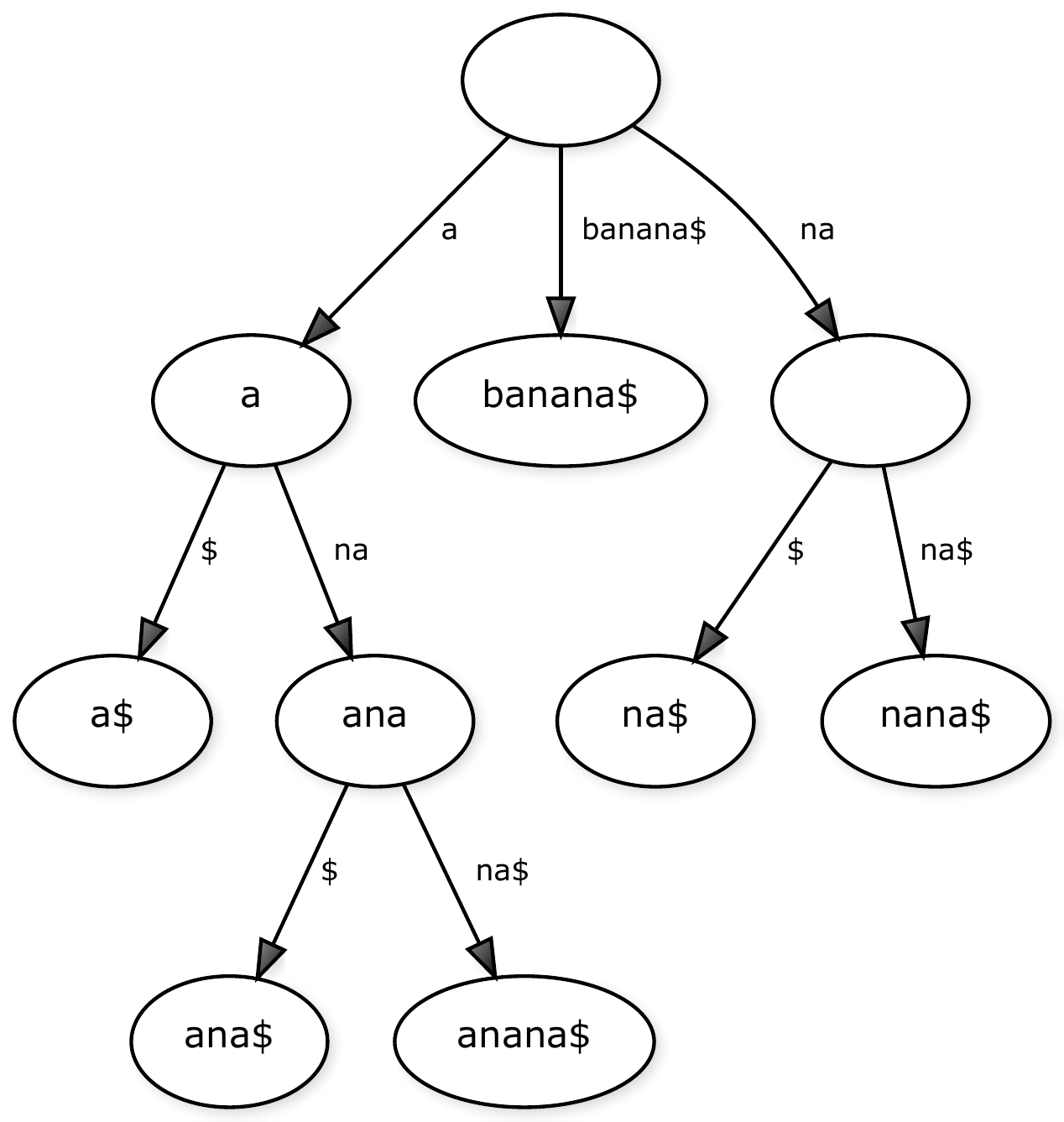}

    \caption[A \acf{st} which stores all suffixes of a given text.]{A \acf{st} which stores all suffixes of the text \texttt{banana} with an appended terminating character \texttt{\$}, which prevents a situation where a suffix could be a prefix of another suffix.}
    \label{Fig:suf_tree}
\end{figure}

The \ac{st} was introduced by Weiner in 1973~\cite{weiner1973linear}.
It is a trie (see Subsection~\ref{Sec:trie}) which stores all suffixes of the input string, that is $n$ suffixes in total for the string of length $n$.
Moreover, the suffix tree is compressed, which in this context means that each node which has only one child is merged with this child, as shown in Figure~\ref{Fig:suf_tree}.
Searching for a pattern takes $O(m)$ time, since we proceed in a way similar to the search in a regular trie.
Suffix trees offer a lot of additional functionality beyond string searching such as calculating the Lempel--Ziv compression~\cite[p.~166]{gusfield1997algorithms} or searching for string repeats~\cite{abouelhoda2002enhanced}.
The \ac{st} takes linear space with respect to the total input size ($O(n^2)$ if uncompressed), however, it occupies significantly more space than the original string --- in a space-efficient implementation around $10.1n$ bytes on average in practice and even up to $20n$ in the worst case~\cite{kurtz1999reducing, abouelhoda2002optimal}, which might be a bottleneck when dealing with massive data.
Moreover, the space complexity given in bits is actually equal to $O(n \log n)$ (which is also the case for the suffix array) rather than $O(n \log \sigma)$ required to store the original text.
When it comes to preprocessing, there exist algorithms which construct the \ac{st} in linear time~\cite{ukkonen1995line, Farach97}.

As regards the implementation, an important consideration is how to represent the children of each node.
A straightforward approach such as storing them in a linked list would degrade the search time, since in order to achieve the overall time of $O(m)$, we have to be able to locate each child in constant time.
This can be accomplished for example with a hash table which offers an $O(1)$ average time for a lookup.

A common variation is called \emph{generalized suffix tree} and it refers to a \ac{st} which stores multiple strings, that is all suffixes for each string $S_1,\ldots,S_x$.
Additional information which identifies the string $S_i$ is stored in the nodes, and the complexities are the same as for a regular \ac{st}~\cite[p.~116]{gusfield1997algorithms}.
Compressed suffix trees which reduce the space requirements were also described (they are usually based on a compressed suffix array)~\cite{gog2011compressed, gagie2015relative}.

\subsection{Suffix array}
\label{Sec:suf_array}
The \ac{sa} comes from Manber and Myers~\cite{manber1993suffix} and it stores indexes of sorted suffixes of the input text, see Figure~\ref{Fig:suf_array} for an example.
According to K{\"a}rkk{\"a}inen~\cite{karkkainen1995suffix}, suffix arrays perform comparably to suffix trees when it comes to string matching, however, they are slower for other kinds of searches such as regular expression matching.
Even though the \ac{sa} takes more space than the original string ($4n$ bytes in its basic form, and the original string has to be stored as well), it is significantly smaller than the suffix tree ($5n < 10.1n$) and it has better locality properties~\cite{abouelhoda2002enhanced}.
The search over a \ac{sa} takes $O(m \log n)$ time, since we perform a binary search over $n$ suffixes and each comparison takes at most $m$ time (although the comparison is constant on average, see Appendix~\ref{App:str_comp}).
The space complexity is equal to $O(n)$ (there are $n$ suffixes and we store one index per suffix), and it is possible to construct a \ac{sa} in linear time.
See Puglisi et al.~\cite{puglisi2007taxonomy} for an extensive survey of multiple construction algorithms with a practical evaluation, which concludes that the algorithm by Maniscalco and Puglisi~\cite{maniscalco2008efficient} is the fastest one.
Parallel construction algorithms which use the GPU were also considered~\cite{deo2013parallel}.
Let us point out a similarity of \ac{sa} to the \ac{st}, since sorted suffixes correspond to the depth-first traversal over the \ac{st}.
Main disadvantage with respect to the \ac{st} is a lack of additional functionality such as this mentioned in the previous subsection.

\begin{figure}[ht]
\vspace{1em}
\centering
\begin{tabular}{c|c}
Suffix & Index \\
\hline
\texttt{\$} & 6 \\
\texttt{a\$} & 5 \\
\texttt{ana\$} & 3 \\
\texttt{anana\$} & 1 \\
\texttt{banana\$} & 0 \\
\texttt{na\$} & 4 \\
\texttt{nana\$} & 2 \\
\end{tabular}
    \caption[A \acf{sa} which stores indexes of sorted suffixes of a given text.]{A \acf{sa} which stores indexes (0-based) of sorted suffixes of the text \texttt{banana\$}. The suffixes are not stored explicitly, although the entire input text has to be stored.}
    \label{Fig:suf_array}
\end{figure}

\subsubsection{Modifications}
\label{Sec:sa_mod}

Multiple modifications of the original \ac{sa} have been proposed over the years.
Their aim is to either speed up the searches by storing additional information or to reduce space requirements by compressing the data or omitting a subset of the data.
Most notable examples are presented below.

The \emph{\aclu{esa}}~(\acs{esa}) is a variant where additional information in the form of a \ac{lcp} table is stored~\cite{abouelhoda2002enhanced}.
For a suffix array $S_A$ over the string of length $n$, the \ac{lcp} table $L$ holds integers from the range $[0,n]$, and it has the following properties: $L[0] = 0$, and $L[i]$ holds the length of the longest common prefix of suffixes from $S_A[i]$ and $S_A[i - 1]$.
\acs{esa} can essentially replace the suffix tree since it offers the same functionality, and it can deal with the same problems in the same time complexity (although constant alphabet size is assumed in the analysis)~\cite{abouelhoda2004replacing}.
For certain applications, it also required to store the \acl{bwt} (see Subsection~\ref{Sec:BWT}) of the input string and an inverse suffix array ($S_A^{-1}[S_A[i]] = i$).

The size of the index can be always reduced using any compression method.
However, such a naive approach would certainly have a negative impact on search performance because of the overhead associated with decompression, and a much better approach is to use a dedicated solution.
M{\"a}kinen~\cite{makinen2000compact} presented a \emph{\aclu{cosa}}~(\acs{cosa}) with average search time $O(((2n - n^\prime) / n^\prime)^2 (m + \log n))$ where $n^\prime$ is the length of the \ac{cosa} and practical space reduction of up to 50\% by replacing repetitive suffixes with links to other suffixes.
Grossi and Vitter~\cite{grossi2005compressed} introduced a \emph{\aclu{csa}}~(\acs{csa}) which uses $O(n \log \sigma)$ bits instead of $O(n \log n)$ bits.
It is based on a transformation of the \ac{sa} into the array which points to the position of the next suffix in the text.
For instance, for the text \texttt{banana\$} and the suffix \texttt{ana\$}, the next suffix is \texttt{na\$}, see Figure~\ref{Fig:cosa}.
These transformed values are compressible because of certain properties such as the fact that number of increasing sequences is in $O(\sigma)$.
The search takes $O(m / \log_{\sigma} n + \log_{\sigma} n)$ time, and the relation between search time and space can be fine-tuned using certain parameters.

For more information on compressed indexes, including the modifications of the \ac{sa}, we refer the reader to the survey by Navarro and M{\"a}kinen~\cite{navarro2007compressed}.
The FM-index which is presented in Subsection~\ref{Sec:FMindex} can be also regarded as a compressed variant of the \ac{sa}~\cite{ferragina2005indexing}.

\begin{figure}[ht]
\vspace{1em}
\centering
\begin{tabular}{c|ccc}
i & Suffix & \ac{sa} index & \ac{csa} index \\
\hline
0 & \texttt{\$} & 6 & 4 \\
1 & \texttt{a\$} & 5 & 0 \\
2 & \texttt{ana\$} & 3 & 5 \\
3 & \texttt{anana\$} & 1 & 6 \\
4 & \texttt{banana\$} & 0 & 3 \\
5 & \texttt{na\$} & 4 & 1 \\
6 & \texttt{nana\$} & 2 & 2 \\
\end{tabular}
    \caption[A \acf{csa} which stores indexes pointing to the next suffix from the text.]{A \acf{csa} for the text \texttt{banana\$} which stores indexes pointing to the next suffix from the text (the \ac{sa} is shown for clarity and it is not stored along with the \ac{csa}).}
    \label{Fig:cosa}
\end{figure}

The \emph{sparse suffix array} stores only suffixes which are located at the positions in the form $i q$ for a fixed $q$ value~\cite{karkkainen1996sparse}.
In order to answer a query, $q$ searches and $q - 1$ explicit verifications are required, and it must hold that $m \geqslant q$.
Another notable example of a modified suffix array which stores only a subset of data is the \emph{sampled suffix array}~\cite{claude2012string}.
The idea is to select a subset of the alphabet (denoted with $\Sigma_S$) and extract corresponding substrings from the text.
The array is constructed only over those suffixes which start with a symbol from the chosen subalphabet (although the sorting is performed on full suffixes).
Only the part of the pattern which contains a character $c \in \Sigma_S$ is searched for (i.e.~there is one search in total) and the matches are verified by comparing the rest of the pattern with the text.
The disadvantage is that the following must hold, $\exists c \in P : c \in \Sigma_S$.
Practical reduction in space in the order of 50\% was reported.
Recently, Grabowski and Raniszewski~\cite{grabowski2014sampling} proposed an alternative sampling technique based on minimizers (see Section~\ref{Sec:minimizers}) which allows for matching all patterns $P$ s.t. $|P| \geqslant q$ where $q$ is the minimizer window length and requires only one search.

\subsection{Other suffix-based structures}
The \emph{suffix tray} combines --- just as the name suggests --- the suffix tree with the suffix array~\cite{cole2006suffix}.
The top-level structure is a \ac{st} whose nodes are divided into heavy and light, depending on whether their subtrees have more or fewer leaves than some predefined threshold.
Light children of heavy nodes store their corresponding \ac{sa} interval.
The query time equals $O(m + \log \sigma)$, and preprocessing and space complexities are equal to $O(n)$.
The authors also described a dynamic variant which is called a \emph{suffix trist} and allows updates.

Yet another modification of the classical \ac{st} is called \emph{\aclu{sc}}~(\acs{sc})~\cite{karkkainen1995suffix}.
Here, K{\"a}rkk{\"a}inen reworks the compaction procedure, which is a part of the construction of the \ac{st}.
Instead of collapsing only the nodes which have only one child, every internal node is combined with one of its children.
Various methods of selecting such a child exist (e.g.,~alphabetical ordering), and thus the \ac{sc} can take multiple forms for the same input string.
The original article reports the best search times for the DNA, whereas the \ac{sc} performed worse than both \ac{st} and \ac{sa} for the English language and random data.
The space complexity is equal to $O(n)$.

\subsection{FM-index}
\label{Sec:FMindex}

The FM-index is a compressed (succinct) full-text index which was introduced by Ferragina and Manzini~\cite{ferragina2000opportunistic} in the year 2000.
It was applied in a variety of situations, for instance for sequence assembly~\cite{langmead2009ultrafast, simpson2010efficient} or for ranked document retrieval~\cite{culpepper2012efficient}.
Multiple modifications of the FM-index were described throughout the years (some are introduced in the following subsections).
The strength of the original FM-index lies in the fact that it occupies less space than the input text while still allowing fast queries.
The search time of its unmodified version is linear with respect to the pattern length (although a constant-size alphabet is assumed), and the space complexity is equal to $O(H_k(T) + \log \log n / \log n)$ bits per input symbol.
Taking the alphabet size into account, Grabowski et al.~\cite{grabowski2006simple} provide a more accurate total size bound of $O(H_k(T)n + (\sigma \log \sigma + \log \log n) \frac{n}{\log n} + n^\gamma \sigma^{\sigma +1})$ bits for $0 < \gamma < 1$.

\subsubsection{Burrows--Wheeler Transform}
\label{Sec:BWT}

FM-index is based on the \ac{bwt}~\cite{burrows1994block}, which is an ingenious method of transforming a string $S$ in order to reduce its entropy.
\ac{bwt} permutes the characters of $S$ in such a way that duplicated characters often appear next to each other, which allows for easier processing using methods such as run-length or move-to-front encoding (as is the case in, e.g.,~the bzip2 compressor~\cite{bzip2, pankratius2009parallelizing}).
Most importantly, this transformation is reversible (as opposed to straightforward sorting), which means that we can extract the original string from the permuted order.
\ac{bwt} could be also used for compression based on the $k$-th order entropy (described in Subsection~\ref{Sec:entropy}) since basic context information can be extracted from BWT, however, the loss of speed renders such an approach impractical~\cite{deorowicz2005context}.

In order to calculate the \ac{bwt}, we first append a special character (we describe it with \texttt{\$}, but in practice it can be any character $c \notin S$) to $S$ in order indicate its end.
The character \texttt{\$} is lexicographically smaller than all $\{ c : c \in S \}$.
The next step is to take all rotations of $S$ ($|S|$ rotations in total) and sort them in a lexicographic order, thus forming the \ac{bwt} matrix, where we denote the first column (sorted characters) with $F$ and the last column (the result of the \ac{bwt}, i.e.~$T^{bwt}$) with $L$.
In order to finish the transform, we take the last character of each rotation, as demonstrated in Figure~\ref{Fig:BWT}.
Let us note the similarities between the \ac{bwt} and the suffix array described in Subsection~\ref{Sec:suf_array}, since the sorted rotations correspond directly to sorted suffixes (see Figure~\ref{Fig:bwt_suf}).
The calculation takes $O(n)$ time, assuming that the prefixes can be sorted in linear time, and the space complexity of the naive approach is equal to $O(n^2)$ (but it is linear if optimized).

\begin{figure}[ht]
\vspace{1em}
\centering
\begin{tabular}{c|ccccccc|c}
$R_1$ & \$ & p & a & t & t & e & r & n \\
$R_2$ & a & t & t & e & r & n & \$ & p \\
$R_3$ & e & r & n & \$ & p & a & t & t \\
$R_4$ & n & \$ & p & a & t & t & e & r \\
$R_5$ & p & a & t & t & e & r & n & \$ \\
$R_6$ & r & n & \$ & p & a & t & t & e \\
$R_7$ & t & e & r & n & \$ & p & a & t \\
$R_8$ & t & t & e & r & n & \$ & p & a \\

\end{tabular}
\caption[Calculating the \acf{bwt}.]{Calculating a \acf{bwt} for the string \texttt{pattern} with an appended terminating character \texttt{\$} (it is required for reversing the transform). The rotations are already sorted and the result is in the last column, i.e. BWT(\texttt{pattern\$}) = \texttt{nptr\$eta}.}
\label{Fig:BWT}
\end{figure}

In order to reverse the \ac{bwt}, we first sort all characters and thus obtain the first column of the matrix.
At this point, we have two columns, namely the first and the last one, which means that we also have all character 2-grams from the original string $S$.
Sorting these 2-grams gives us the first and the second column, and we proceed in this manner (later we sort 3-grams, 4-grams, etc) until we reach $|S|$-grams and thus reconstruct the whole transformation matrix.
At this point, $S$ can be found in the row where the last character is equal to \texttt{\$}.

\subsubsection{Operation}

Important aspects of the FM-index are as follows:

\begin{itemize}
\item
\textbf{Count table} $C$, which describes the number of occurrences of lexicographically smaller characters for all $c \in S$ (see Figure~\ref{Fig:FM_C}).
\item
\textbf{Rank} operation, which counts the number of set bits in a bit vector $v$ before a certain position $i$ (we assume that $v[i]$ is included as well), that is $rank(i, v) = |\{i^\prime : 0 \leqslant i^\prime \leqslant i \land v[i^\prime] = 1\}|$.
\item
\textbf{Select} operation (used only in some variants, e.g.,~the RLFM~\cite{makinen2005succinct}), which reports the position of the $i$-th set bit in the bit vector $v$, that is $select(i, v) = p$ if and only if $|\{i^\prime : 0 \leqslant i^\prime < p \land v[i^\prime] = 1\}| = i - 1$.
\end{itemize}

Note that both $rank$ and $select$ operations can be generalized to any finite alphabet $\Sigma$.
When we perform the search using the FM-index, we iterate the pattern characterwise in a reverse order while maintaining a current range $r = [s, e]$.
Initially, $i = m - 1$ and $r = [0, n - 1]$, that is we start from the last character in the pattern and the range covers the whole input string (here the input string corresponds to $T^{bwt}$, that is a text after the BWT).
At each step we update $s$ and $e$ using the formulae presented in Figure~\ref{Fig:fm_iter}.
The size of the range after the last iteration gives us the number of occurrences of $P$ in $T$, or it turns out that $P \not\subset T$ if $s > e$ at any point.
This mechanism is also known as the LF-mapping.

\begin{figure}[ht]
\vspace{1em}
\centering
\begin{tabular}{c|c|c|c}
i & \ac{bwt} & \ac{sa} & suffix \\
\hline
0 & n & 7  & \texttt{\$} \\
1 & p & 1  & \texttt{attern\$} \\
2 & t & 4  & \texttt{ern\$} \\
3 & r & 6  & \texttt{n\$} \\
4 & \$ & 0 & \texttt{pattern\$} \\
5 & e & 5  & \texttt{rn\$} \\
6 & t & 3  & \texttt{tern\$} \\
7 & a & 2  & \texttt{ttern\$} \\
\end{tabular}
\caption[A relation between the \ac{bwt} and the \ac{sa}.]{A relation between the \ac{bwt} and the \ac{sa} for the string \texttt{pattern} with an appended terminating character \texttt{\$}. Let us note that $BWT[i] = S[SA[i] - 1]$ (where $S[-1]$ corresponds to the last character in $S$), that is a character at the position $i$ in \ac{bwt} is a character preceding a suffix which is located at the same position in the \ac{sa}.}
\label{Fig:bwt_suf}
\end{figure}

\begin{figure}[ht]
\vspace{1em}
\centering
\begin{tabular}{c|c|c|c|c|c}
c & \$ & i & m & p & s \\ 
\hline
C & 0 & 1 & 5 & 6 & 8
\end{tabular}
\caption[Count table $C$ which is a part of the FM-index.]{Count table $C$ which is a part of the FM-index for the text \texttt{mississippi\$}. The entries describe the number of occurrences of lexicographically smaller characters for all $c \in S$. For instance for the letter \texttt{m}, there are 4 occurrences of \texttt{i} and 1 occurrence of \texttt{\$} in $S$, hence $C[\texttt{m}] = 5$. It is worth noting that $C$ is actually a compact representation of the $F$ column.}
\label{Fig:FM_C}
\end{figure}

\begin{figure}[ht]
\centering

\[
s = C[P[i]] + rank(s - 1, P[i])
\]
\[
e = C[P[i]] + rank(e, P[i]) - 1
\]

\caption[Formulae for updating the range during the search procedure in the FM-index.]{Formulae for updating the range during the search procedure in the FM-index, where $P[i]$ is the current character and $C$ is the count table. $Rank$ is invoked on $T^{bwt}$ and it counts occurrences of the current character $P[i]$.}
\label{Fig:fm_iter}
\end{figure}

\subsubsection{Efficiency}
\label{Sec:FMeff}

We can see that the performance of $C$ lookup and $rank$ is crucial to the complexity of the search procedure.
In particular, if these operations are constant, the search takes $O(m)$ time.
For the count table, we can simply precompute the values and store them an array of size $\sigma$ with $O(1)$ lookup.
As regards $rank$, a naive implementation which would iterate the whole array would clearly take $O(n)$ time.
On the other hand, if we were to precompute all values, we would have to store a table of size $O(n \sigma)$.
One of the popular solutions for an efficient $rank$ uses two structures which are introduced in the following paragraphs.

The RRR (from authors' names: Raman, Raman, and Rao) is a data structure which can answer the $rank$ query in $O(1)$ time for bit vectors (i.e.~where $\Sigma = \{0, 1\}$), while providing compression at the same time~\cite{raman2002succinct}.
It divides a bit vector $v$ of size $n$ into $n/b$ blocks each of size $b$, and groups each consecutive $s$ blocks into one superblock (see Figure~\ref{Fig:RRRblocks}).
For each block, we store a weight $w$ which describes the number of set bits and offset $o$ which describes its position in a table $T_R$ (the maximum value of $o$ depends on $w$).
In $T_R$, for each $w$ and each $o$, we store a value of $rank$ for each index $i$, where $0 \leqslant i \leqslant b$ (see Figure~\ref{Fig:RRRtable}).
This means that we have to keep $\binom{b}{w}$ entries each of size $b$ for each of the $b + 1$ consecutive weights.
Such a scheme provides compression with respect to storing all $n$ bits explicitly.
We achieve the $O(1)$ query time by storing a $rank$ value for each superblock, and thus during the search we only iterate at most $s$ blocks ($s$ is constant).
The space complexity is equal to $n H_0(v) + O(n \log \log n / \log n)$ bits. 

\begin{figure}[ht]
\vspace{1em}
\centering
\begin{tabular}{|c|c||c|c|}

001 & 001 & 101 & 000
\end{tabular}
\caption[An example of RRR blocks.]{An example of RRR blocks for $b = 3$ and $s = 2$, where the first superblock is equal to \texttt{001001} and the second superblock is equal to \texttt{101000}.}
\label{Fig:RRRblocks}
\end{figure}

\begin{figure}[ht]
\vspace{1em}
\centering
\begin{tabular}{c|c|c}
offset & block value & rank \\
\hline
0 & 011 & 0~1~2 \\
1 & 101 & 1~1~2 \\
2 & 110 & 1~2~2 \\
\end{tabular}
\caption[An example of an RRR table.]{An example of an RRR table for $w = 2$ and $b = 3$, where $\binom{3}{2} = 3$ (i.e.~the number of all block values of length 3 with weight 2 is equal to 3), with rank presented for successive indexes $i \in [0, 2]$. Block values do not have to be stored explicitly.}
\label{Fig:RRRtable}
\end{figure}

\begin{figure}[ht]
\vspace{1em}
\centering
\includegraphics[scale=0.65]{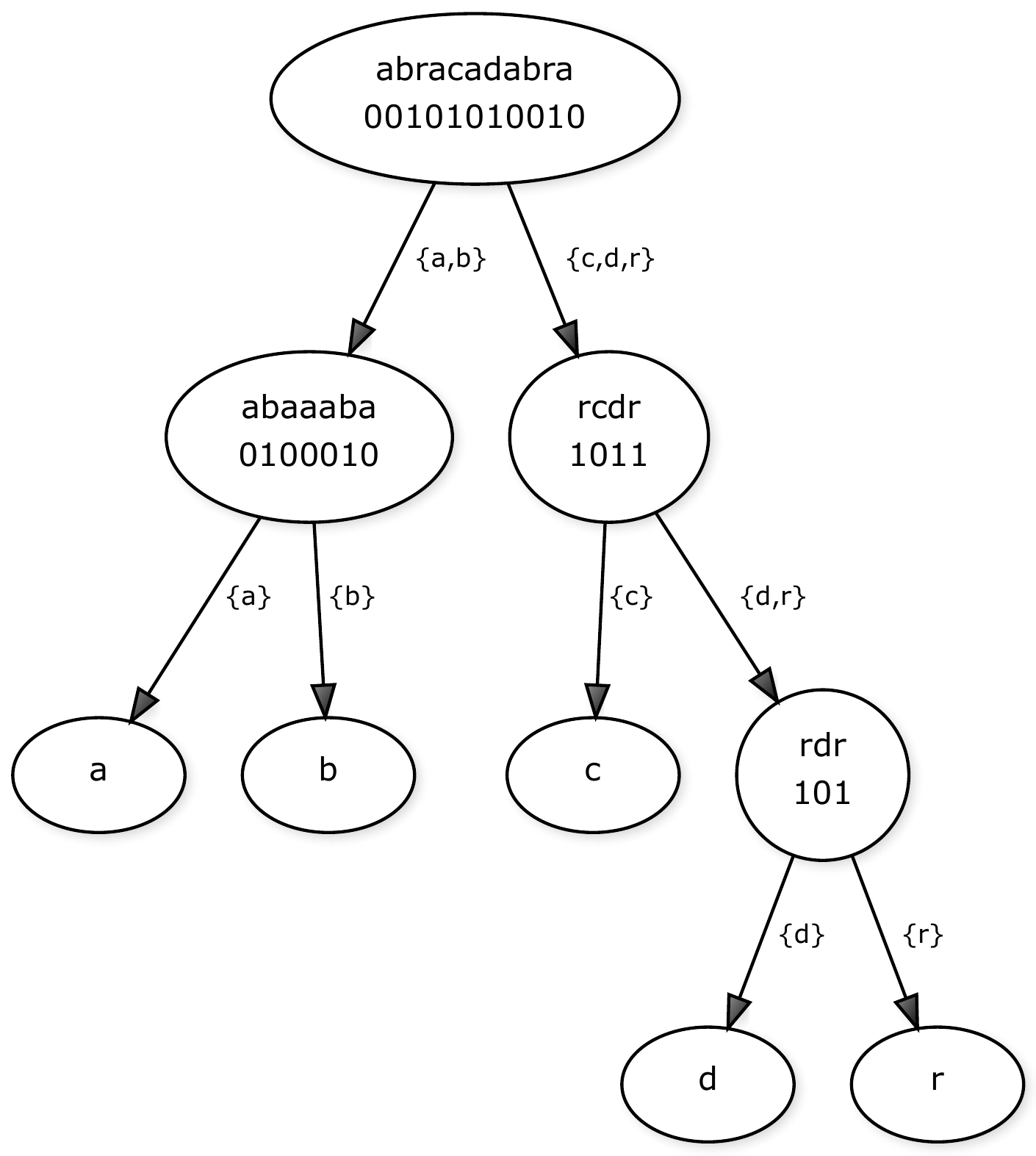}
\caption[A \acf{wt}.]{A \acf{wt} over the string \texttt{abracadabra}. The alphabet is divided into two subsets at each level, with \texttt{0} corresponding to one subset and \texttt{1} to the other.}
\label{Fig:wavelet}
\end{figure}

The \emph{\aclu{wt}}~(\acs{wt}) from Grossi et al.~\cite{grossi2003high} is a balanced tree data structure that stores a hierarchy of bit vectors instead of the original string, which allows the use of RRR (or any other bit vector with an efficient $rank$ operation).
Starting from the root, we recursively partition the alphabet into two subsets of equal length (if the number of distinct characters is even), until we reach single symbols which are stored as leaves.
Characters belonging to the first subset are indicated with 0s, and characters belonging to the second subset are indicated with 1s (consult Figure~\ref{Fig:wavelet} for an example).

Thanks to the \ac{wt} we can implement a rank query for any fixed size alphabet in $O(\log \sigma)$ time (assuming that a binary $rank$ is calculated in constant time), since the height of the tree is equal to $\log \sigma$.
For a given character $c$, we query the \ac{wt} at each node and proceed left or right depending on the subset to which $c$ belongs.
Each subsequent rank is called in the form $rank(c, p)$, where $p$ is the result of the rank at the previous level.
Ferragina et al.~\cite{ferragina2007compressed} described generalized \ac{wt}s, for instance a multiary \ac{wt} with $O( \log \sigma / \log \log n)$ traversal time (consult Bowe's thesis~\cite{bowe2010multiary} for more information and a practical evaluation).

\subsection{FM-index flavors}

Multiple flavors of the FM-index were proposed over the years with the goal of decreasing the query time, e.g.,~having $O(m)$ time without the dependence on $\sigma$, or reducing the occupied space.
The structures which provide asymptotically optimal bounds are often not practical due to the very large constants which are involved~\cite{vigna2008broadword}.
For this reason, many authors focus on practical performance, and these structures are usually based on a fast $rank$ operation and take advantage of compressed representations of bit vectors~\cite{navarro2007compressed}.
The following paragraphs present selected examples, consult Navarro and M{\"a}kinen~\cite{navarro2007compressed} for an extensive survey of compressed full-text indexes which discusses the whole family of FM-indexes.

One of the notable examples where the query time does not depend on $\sigma$ is the alphabet-independent FM-index by Grabowski et al.~\cite{grabowski2004first}.
The idea is to first compress the text using Huffman coding and then apply the \ac{bwt} transform over it, obtaining a bit vector $v$.
This vector is then used for searching in a manner corresponding to the FM-index --- the array $C$ stores the number of zeros up to a certain position, and the relation $C[c] + rank(T^{bwt}, i, c)$ is replaced with $i - rank(v, i)$ if $c = 0$ and $n^\prime - rank(v, n^\prime) + rank(v, i)$ if $c = 1$, where $n^\prime$ is the length of the text compressed with Huffman.
The space complexity is equal to $O(n(H_0(T) + 1))$ bits, and the average search time is equal to $O(m(H_0(T) + 1))$ under ``reasonable assumptions''.

On the practical front, Grabowski et al.~\cite{grabowski2015fm} recently described a cache-aligned rank with 1 cache miss.
Moreover, they proposed the so-called FM-dummy index with several variants, for instance one which stores a separate bit vector for each alphabet symbol ($\sigma$ vectors in total).
Other variants include using certain dense codes as well as using Huffman-shaped multiary wavelet trees with different arity values.
A Huffman-shaped wavelet tree is unbalanced, and the paths for frequent characters are shorter (which translates to a smaller number of $rank$ queries on bit vectors).
Moreover, the operations (which are performed in the same manner as for the regular wavelet tree) are faster on average~\cite{claude2014efficient}.
They reported search times which are 2--3 times faster than those for other state-of-the-art methods at the cost of using 1.5--5 times more space.

Data structures which concentrate on reducing space requirements rather than the query time include the compressed bit vectors from K{\"a}rkk{\"a}inen et al.~\cite{karkkainen2014hybrid}, where different compression methods are used for blocks depending on the type of the block, for instance run-length encoding for blocks with a small number of runs.
Another notable example is a data-aware FM-index by Huo et al.~\cite{DBLP:conf/alenex/HuoCZVNY15}, which encodes the bit vectors (resulting from the \ac{wt}) using Gamma coding (a kind of variable-length coding), and thus obtain one of the best compression ratios in practice.

\subsubsection{Binary \emph{rank}}
As described in the previous subsection, in order to achieve good overall performance, it is sufficient to design a data structure which supports an efficient $rank$ query for bit vectors thanks to the use of a wavelet tree (RRR being a notable example).
Jacobson~\cite{jacobson1989space} originally showed that it is possible to obtain a constant-time $rank$ operation using $o(n)$ extra bits for $|v| = n$ (the same holds for $select$~\cite{clark1996}).
Vigna~\cite{vigna2008broadword} proposed to interleave (i.e. store next to one another) blocks and superblocks (concepts which were introduced for the RRR structure) for uncompressed bit vectors  in order to reduce the number of cache and \ac{tlb} misses from 3 to 2.
This was extended by Gog and Petri~\cite{gog2014optimized}, who showed better practical performance by using a slightly different layout with 64-bit counters.
Gonzalez and Navarro~\cite{gonzalez2009rank} provided a discussion of the dynamic scenario where insertions and deletions to the vector are allowed, and they obtain a space bound of $n H_0(v) + o(n \log \sigma)$ bits and $O(\log n (1 + \log \sigma / \log \log n))$ time for all operations (i.e. queries and updates).

\subsection{FM-bloated}
\label{Sec:fm-bloated}

One of the crucial issues when it comes to the performance of the FM-index is the number of CPU cache misses which occur during the search.
This comes from the fact that in order to calculate the LF-mapping, non-local access to the \ac{bwt} sequence is often required (in the order of $\Omega(m)$ misses during the search for a pattern of length $m$), even for a small alphabet.
The problem of cache misses during the FM-index backward search was identified as the main performance limiter by Chac{\'o}n et al.~\cite{chacon2013n}, who proposed to perform the LF-mapping with several symbols at a time (in practice, at most 4 for the 4-symbol DNA alphabet for which the scheme was described).
This solution allowed, for example, to improve the search speed by a factor of 1.5 for the price of occupying roughly 2 times the size of the 1-step FM-index.
Here, we address the problem of cache misses during the pattern search ($count$ query) in a way related to the Chac{\'o}n et al.~solution --- we also work on $q$-grams, yet the algorithmic details are different.
Two following subsections describe two variants of our approach, and experimental results can be found in Section~\ref{Sec:fm-bloated_res}.

\subsubsection{Superlinear space}

FM-bloated is a variation of the FM-index which aims to speed up the queries at the cost of additional space.
We start by calculating the \ac{bwt} for the input string in the same way as for the regular FM-index, however, the difference is that we operate on $q$-grams rather than on individual characters, and the count table stores results for each $q$-gram sampled from the \ac{bwt} matrix.
This is the case for all $q$, where $q$ is the power of 2, up to some predefined value $q_{max}$ (for instance 128).
Namely, for each suffix $T[i, n - 1]$, we take all $q$-grams in the following form: $T[i - 1]$ (1-gram), $T[i - 2, i - 1]$ (2-gram), $T[i - 4, i - 1]$ (4-gram), etc.
The $q$-grams are extracted until we reach $q_{max}$ or one of the $q$-grams contains the terminating character (such a $q$-gram with the terminating character is discarded), consult Figure~\ref{Fig:bloated_qgrams} for an example.
Let $\mathcal{Q}$ denote a collection of all $q$-grams for all $i$.
For each distinct item $s$ from $\mathcal{Q}$, we create a list $L_s$ of its occurrences in the sorted suffix order (simply called \ac{sa} order).
This resembles an inverted index on $q$-grams, yet the main difference is that the elements in the lists are arranged in \ac{sa} rather than the text order, e.g.,~for the $q$-gram \texttt{t} in Figure~\ref{Fig:bloated_qgrams}, the 1-based list of occurrences corresponding to rows would be as follows: $\{3, 7\}$.

\begin{figure}[ht]
\vspace{1em}
\centering
\begin{tabular}{c|cc}
row & BWT & $q$-grams \\
\hline
$R_1$ & \texttt{\$pattern} & \texttt{n}, \texttt{rn}, \texttt{tern} \\
$R_2$ & \texttt{attern\$p} & \texttt{p} \\
$R_3$ & \texttt{ern\$patt} & \texttt{t}, \texttt{tt}, \texttt{patt}\\
$R_4$ & \texttt{n\$patter} & \texttt{r}, \texttt{er}, \texttt{tter} \\
$R_5$ & \texttt{pattern\$} & \\
$R_6$ & \texttt{rn\$patte} & \texttt{e}, \texttt{te}, \texttt{atte} \\
$R_7$ & \texttt{tern\$pat} & \texttt{t}, \texttt{at} \\
$R_8$ & \texttt{ttern\$pa} & \texttt{a}, \texttt{pa} \\
\end{tabular}
\caption[$Q$-gram extraction in the FM-bloated structure with superlinear space.]{$Q$-gram extraction in the FM-bloated structure with superlinear space for the text $T = \texttt{pattern\$}$. All $q$-grams are extracted, i.e.~$q_{max} \geqslant \lceil n / 2 \rceil$.}
\label{Fig:bloated_qgrams}
\end{figure}

For a given pattern $P$, we start the LF-mapping with its longest suffix $P_s$ s.t. $|P_s| \leqslant q_{max} \land |P_s| = 2^c$ for some $c \in \mathbb{Z}$.
The following backward steps deal with the remaining prefix of $P$ in a similar way.
Note that the number of LF-mapping steps is equal to the number of 1s in the binary representation of $m$, i.e.~it is in the order of $O(\log m)$, and if $m$ is a power of two, then the result for $match$ and $count$ queries can be reported in constant time (we simply return $|L_s|$).
When $q_{max}$ is bigger, the overall index size is bigger, but the search is faster (for patterns of sufficient length) because it allows for farther jumps towards the beginning of the pattern.
In our representation, each LF-mapping step translates to performing two predecessor queries on a list $L_s$.
A naive solution is a binary search with $O(\log n)$ worst-case time (or even a linear search, which may be faster if the list is short), yet the predecessor query can be also handled in $O(\log \log n)$ time using a y-fast trie~\cite{willard1983log}.
Hence, the overall average search complexity is equal to $O(m + \log m \log \log n)$, with $O(m / C_L + \log m \log \log n)$ cache misses where $C_L$ is the cache line size in bytes (provided that each symbol from the pattern occupies one byte).
As regards the space complexity, there is a total of $n \log n$ $q$-gram occurrences ($\log n$ $q$-gram positions for each of $n$ rows of the \ac{bwt} matrix).
Hence, the total length of all occurrence lists is equal to $n \log n$, and the total complexity is equal to $O(n \log^2 n)$ bits (since we need $\log n$ bits to store one position from the BWT matrix of $n$ rows).

As regards the implementation (in the C++ language), our focus is on data compaction.
Each $q$-gram acts as a key in a hash table where collisions are resolved with chaining, and the $q$-grams are stored implicitly, i.e.~as a (pointer, length) pair, where the pointer refers to the original string.
The values in the hash table include the count and the list of occurrences which are stored in one, contiguous array.
We use a binary search for calculating $rank$ on lists whose length is greater than or equal to 16 (an empirically determined value) and a linear search otherwise.

\subsubsection{Linear space}
\label{Sec:bloated_lin_space}

In this variant, instead of extracting all 1-, 2-, etc, $q$-grams for each row of the \ac{bwt} matrix, we extract only selected $q$-grams with the help of minimizers (consult Subsection~\ref{Sec:minimizers} for the description of minimizers).
The first step is to calculate all $(\alpha, q)$-minimizers for the input text $T$, with some fixed $\alpha$ and $q$ parameters and lexicographic ordering, where ties are resolved in favor of the leftmost of the smallest substrings.
Next, we store both the count table and the occurrence lists for all single characters in the same way as for the regular FM-index (using, e.g.,~a wavelet tree).
Moreover, we store information about the counts and occurrences of all $q$-grams which are located in between the minimizers from the set $M(T)$ --- these $q$-grams are referred to as phrases.
For the set of minimizer indexes $\mathcal{I}(T)$, consecutive phrases $p_i$ are constructed in the following manner: $\forall^{|\mathcal{I}|-1}_{i=1} p_i = T[\mathcal{I}[i], \mathcal{I}[i + 1] - 1]$, consult Figure~\ref{Fig:mini_phrases}.
It is worth noting that this approach resembles the recently proposed SamSAMi index, a sampled suffix array on minimizers~\cite{grabowski2014sampling}.

\begin{figure}[ht]
\vspace{1em}
\centering
\begin{tabular}{c|c}
T             & \texttt{appearance} \\
\hline
$M$           & \texttt{ap}, \texttt{ar}, \texttt{an} \\
$\mathcal{I}$ & $0, 4, 6$ \\
Phrases       & \texttt{appe}, \texttt{ar}  \\
Phrase ranges & $[0, 3]$, $[4, 5]$ \\
\end{tabular}
\caption[Constructing FM-bloated phrases with the use of minimizers.]{Constructing FM-bloated phrases for the text \texttt{appearance} with the use of $(4, 2)$-minimizers.}
\label{Fig:mini_phrases}
\end{figure}

The search proceeds as follows:

\begin{enumerate}

\item
We calculate all minimizers for the pattern.

\item
We search for the pattern suffix $P_S = P[s_r, -1]$, where $s_r$ is the starting position of the rightmost minimizer using the regular FM-index mechanism, i.e.~processing 1 character at a time.

\item
We operate on the phrases between the minimizers rather than individual characters and the search for these $q$-grams is performed in the same way as for the superlinear variant.
If it turns out that the phrase is a 1-gram, a faster FM-index mechanism for single characters can be used.

\item
We search for the pattern prefix $P_P = P[0, s_l - 1]$, where $s_l$ is the starting position of the leftmost minimizer using the regular FM-index mechanism, i.e.~processing 1 character at a time.

\end{enumerate}

The use of minimizers ensures that the phrases are selected from $P$ in the same way as they are selected from $T$ during the index construction.
The overall average search complexity is equal to $O(m \log \log n)$ (again, assuming that a y-fast trie~\cite{willard1983log} is used), and the space complexity is linear.

\section{Approximate}
\label{Sec:full-approx}

Navarro et al.~\cite{navarro2001indexing} provided an extensive survey of full-text indexes for approximate string matching.
They categorized the algorithms into three categories based on the search procedure:

\begin{itemize}

\item
\textbf{Neighborhood generation} --- all strings in $\{ S : S \in \Sigma^* \land D(S, P) \leqslant k \}$ for a given pattern $P$ are searched for directly.

\item
\textbf{Partitioning into exact searching (\acs{pies})} --- substrings of the pattern are searched for in an exact manner and these matches are extended into approximate matches.

\item
\textbf{Intermediate partitioning} --- substrings of the pattern are searched for approximately but with a fewer number of errors. This method lies in between the two other ones.

\end{itemize}

In the neighborhood generation approach, we generate the $k$-neighborhood $K$ of the pattern, which contains all strings which could be possible matches over a specified alphabet $\Sigma$ (if the alphabet is finite, the amount of such strings is finite as well).
These strings can be searched for using any exact index such as a suffix tree or a suffix array.
The main issue is the fact that the size of $K$ grows exponentially, $|K| = O(m^k \sigma^k)$~\cite{ukkonen1985finding}, which means that basically all factors (and especially $k$) should be small.
When the suffix tree is used as an index for the input text, Cobbs~\cite{cobbs1995fast} proposed a solution which reduces the amount of nodes that have to be processed.
It runs in $O(mq + |o|)$ time and occupies $O(q)$ space, where $q \leqslant n$ ($q$ depends on the problem instance) and $|o|$ is the size of the output.

When the pattern is partitioned and searched for exactly (\acs{pies}), we have to again store the index which can answer these exact queries.
Let us note that this approach is based on the pigeonhole principle.
In the context of approximate string searching this means that for a given $k$, at least one of $k + 1$ parts of average length $|P|/(k + 1)$ must match the text exactly (more generally, $s$ parts match if $k + s$ parts are created).
The value of $k$ should not be too large, otherwise it could be the case that a substantial part of the input text has to be verified (especially if the pattern is small).
Alternatively, the pattern can be divided into $m - q + 1$ overlapping $q$-grams and these $q$-grams are searched for (using the locate query) against the index of $q$-grams extracted from the text (see Figure~\ref{Fig:qgrams} for an example of $q$-gram extraction).
These $q$-grams which are stored by the index are situated at fixed positions with an interval $h$ and it must hold that $h \leqslant \lfloor (m-k-q+1)/(k+s) \rfloor$ for occurrences of $P$ in $T$ to contain $s$ samples.
Sutinen and Tarhio~\cite{SutinenT95} suggested that the optimal value for $q$ is in the order of $\Theta(\log_{\sigma} m)$.
If it turns out that the positions of subsequent $q$-grams may correspond to a match, explicit verification is performed.
Similarly to the $k$-neighborhood scenario, any index can be used in order to answer the exact queries.
Let us note that this approach with pattern substring lookup and verification can be also used for exact searching.

In the case of intermediate partitioning, we split the pattern into $s$ pieces, and we search for these pieces in an approximate manner using neighborhood generation.
The case of $s = 1$ corresponds to pure neighborhood generation, whereas the case of $s = k + 1$ is almost like \acs{pies}.
In general, this method requires more searching but less verification when compared to \acs{pies}, and thus lies in between the two approaches which were previously described.

Consult Maa{\ss} and Nowak~\cite{maass2005text} in order to see a detailed comparison of the complexities of modern text indexing methods for approximate matching.
Notable structures from the theoretical point of view include the \emph{k-errata trie} by Cole et al.~\cite{cole2004dictionary} which is based on the suffix tree and the LCP structure (see Subsection~\ref{Sec:sa_mod} for a description of the LCP).
It can be used in various contexts, including full-text and keyword indexing as well as wildcard matching.
For full-text indexing and the $k$-mismatches problem, it uses $O(n \log^k n / k!)$ space and offers $O(m + \log^k n / k! + occ)$ query time.
This was extended by Tsur~\cite{tsur2010fast} who described a structure similar to the one from Cole et al.~with time complexity $O(m + \log \log n + occ)$ (for constant $k$) and $O(n^{1+\epsilon})$ space for a constant $\epsilon > 0$.
As regards a solution which is dedicated for the Hamming distance, Gabriele et al.~\cite{gabriele2003indexing} provided an index with average search time $O(m + occ)$ and $O(n \log^l n)$ space (for some $l$).

Let us note that these full-text indexes can be usually easily adapted to the keyword matching scenario which is described in the following chapter.

\section{Word-based}
\label{Sec:full-word_based}

An interesting category of data structures are word-based indexes, which can be used for approximate matching and especially sequence alignment.
They employ heuristic approaches in order to speed up the searching, and for this reason they are not guaranteed to find the optimal match.
This means that they are also approximate in the mathematical sense, i.e.~they do not return the true answer to the problem.
Their popularity is especially widespread in the context of bioinformatics, where the massive sizes of the databases often force the programmers to use efficient filtering techniques.
Notable examples include BLAST~\cite{altschul1990basic} and FASTA~\cite{lipman1985rapid} tools.

\subsection{BLAST}
\label{Sec:BLAST}

BLAST stands for \emph{Basic Local Alignment Search Tool}, and it was published by Altschul et al.~\cite{altschul1990basic} in 1990 with the purpose of comparing biological sequences (see Subsection~\ref{Sec:bioinformatics} for more information about biological data).
The name may refer to the algorithm or to the whole suite of string searching tools for bioinformatics which are based on the said algorithm.
BLAST relies heavily on various heuristics and for this reason it is highly domain specific.
In fact there exist various flavors of BLAST for different data sets, for instance one for protein data (blastp) and one for the DNA (blastn).
Another notable modification is the PSI-BLAST which is combined with dynamic programming in order to identify distant protein relationships.

The basic algorithm proceeds as follows:

\begin{enumerate}
\item
Certain regions are removed from the pattern. These include repeated substrings and regions of low complexity (measured statistically using, e.g.,~DUST for DNA \cite{morgulis2006fast}).
\item
We create a set $\mathcal{Q}$ containing $q$-grams with overlaps (that is all available $q$-grams, see Figure~\ref{Fig:qgrams}) which are extracted from the pattern.
Each $s \in \mathcal{Q}$ is scored against all possible $q$-grams (these can be precomputed) and ones with the highest scores are retained creating a candidate set $\mathcal{Q}_C$.
\item
Each word from $\mathcal{Q}_C$ is searched for in an exact manner against the database (using for instance an inverted index, see Subsection~\ref{Sec:inv_index}).
These exact matches create the \emph{seeds} which are later used for extending the matches.
\item
The seeds are extended to the left and to the right as long as the alignment score is increasing.
\item
Alignment significance is assessed using domain-specific statistical tools.
\end{enumerate}

\begin{figure}[ht]
\vspace{1em}
\centering
\begin{tabular}{c|c}
size & $q$-grams \\
\hline
2 & \texttt{te}, \texttt{ex}, \texttt{xt}, \texttt{ti}, \texttt{in}, \texttt{ng}\\
3 & \texttt{tex}, \texttt{ext}, \texttt{xti}, \texttt{tin}, \texttt{ing}\\
4 & \texttt{text}, \texttt{exti}, \texttt{xtin}, \texttt{ting} \\
5 & \texttt{texti}, \texttt{extin}, \texttt{xting} \\

\end{tabular}
\caption[Selecting all overlapping $q$-grams from a given text.]{Selecting all overlapping $q$-grams (with the shift of 1) from the text $T = \texttt{texting}$. It must always hold that $q \leqslant |T|$.}
\label{Fig:qgrams}
\end{figure}

In general, BLAST is faster than other alignment algorithms such as the \ac{sw} algorithm (see Subsection~\ref{Sec:online_approx}) due to its heuristic approach.
However, this comes at a price of reduced accuracy, and Shpaer et al.~\cite{shpaer1996sensitivity} state that there is a substantial chance that BLAST will miss a distant sequence similarity.
Moreover, hardware-oriented implementations of the SW have been created, and in certain cases they can match the performance of BLAST~\cite{manavski2008cuda}.
Still, BLAST is currently the most common tool for sequence alignment using massive biological data, and it is openly available via its website~\cite{BLAST}, which means that it can be conveniently run without consuming local resources.

%
%

\chapter{Keyword Indexes}
\label{Chap:keyword}
\lhead{\emph{Keyword Indexes}}

Keywords indexes operate on individual words rather than the whole input string.
Formally, for a collection $\mathcal{D} = \{d_1, \ldots, d_{x}\}$ of $x$ strings (words, $q$-grams) of total length $n$ over a given alphabet $\Sigma$, $I(\mathcal{D})$ is a keyword index supporting matching with a specified distance $D$.
For any query pattern $P$, it returns all words $d$ from $\mathcal{D}$ s.t. $D(P, d) \leqslant k$ (with $k=0$ for exact matching).
Approximate dictionary matching was introduced by Minsky and Papert in 1969~\cite{minsky1969, cole2004dictionary}.

In the following sections, we describe algorithms from this category, divided into exact (Section~\ref{Sec:keyword-exact}) and approximate ones (Section~\ref{Sec:keyword-approx}).
Our contribution in this field is presented in Subsection~\ref{Sec:split_index}, which describes an index for approximate matching with few mismatches (especially 1 mismatch).

\section{Exact}
\label{Sec:keyword-exact}

If the goal were to support only the $match$ query for a finite number of keywords, we could use any efficient set data structure such as a hash table or a trie (see Subsections~\ref{Sec:hashing}~and~\ref{Sec:trie}) in order to store all those keywords.
Boytsov~\cite{boytsov2011indexing} reported that depending on the data set, either one of these two may be faster.
In order to reduce space requirements we could use minimal perfect hashing (see Subsection~\ref{Sec:hashing}), and we could also compress the entries in the buckets.

\subsection{Bloom filter}
\label{Sec:bloom}

\begin{figure}[ht]
    \centering
    \includegraphics[scale=0.5]{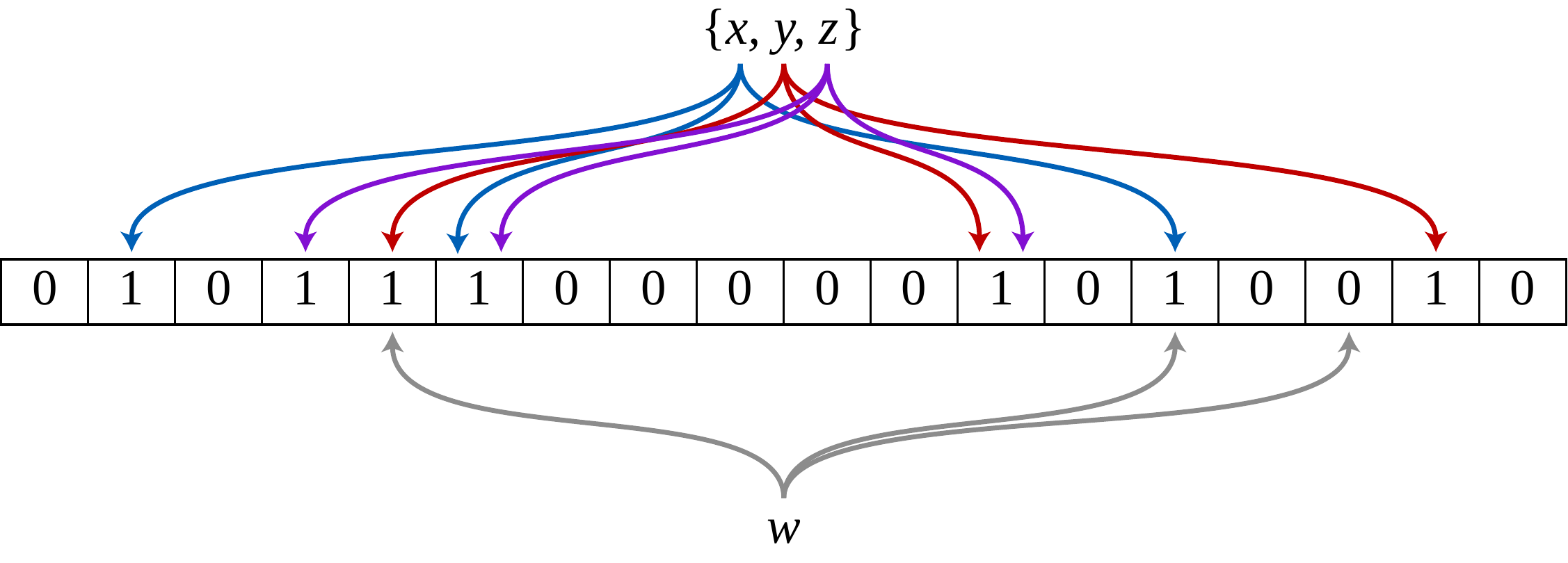}

    \caption[A \acf{bf} for approximate membership queries.]{A \acf{bf} for approximate membership queries with $n = 18$ and $k = 3$, holding the elements from the set $\{x, y, z\}$. The element $w$ is not in the set since $H_2(w) = 15$ and $A[15] = 0$; reproduced from Wikimedia Commons\protect\footnotemark.}
    \label{Fig:bloom}
\end{figure}

Alternatively, we could provide only approximate answers (in a mathematical sense) in order to occupy even less space.
A relevant data structure is the \ac{bf}~\cite{bloom1970space}, which is a space-efficient, probabilistic data structure with possible false positive matches, but no false negatives and an adjustable error rate.
The \ac{bf} uses a bit vector $A$ of size $n$, where no bits are initially set.
Each element $e$ is hashed with $k$ different hash functions $H_i$ in the form $H(e) = i$, where $i \in \mathbb{Z} \land 0 \leqslant i < n$, and $\forall_{i=1}^k A[H_i(e)] = 1$.
When the lookup is performed, the queried element is hashed with the same functions and it is checked whether $A[i] = 1$ for all $i$, and if that is the case a possible match is reported (consult Figure~\ref{Fig:bloom}).
Broder and Mitzenmacher~\cite{broder2004network} provided the following formula for the expected false positive rate: $F_P = 0.5^k \geqslant 0.5^{m \ln 2/n}$, where $m$ is the size of the filter in bits and $n$ is the number of elements.
They note that for example when $m = 8n$, the false positive probability is slightly above 0.02.
Recently, Fan et al.~\cite{fan2014cuckoo} described a structure based on cuckoo hashing which takes even less space than the \ac{bf} and supports deletions (unlike the \ac{bf}).

\setcounter{footnote}{0}

\subsection{Inverted index}
\label{Sec:inv_index}

An inverted index is a keyword index which contains a mapping from words $d \in \mathcal{D}$ to the lists which store all positions $p_i$ of their occurrences in the text ($d \to p_1, \ldots, p_n$).
These positions can be for instance indexes in a string of characters, or if a more coarse-grained approach were sufficient, they could identify individual documents or databases.
See Figure~\ref{Fig:inv_index} for an example with a single input string.
The positions allow a search on the whole phrase (i.e.~multiple words) by searching for each word separately and checking whether the positions describe consecutive words in the text (that is by looking for list intersections with a shift).
It could be also used for searching for a query which may cross the boundaries of the words by searching for substrings of a pattern and comparing the respective positions (consult Section~\ref{Sec:keyword_sel} for more information).
This means that the goal of an inverted index is to support various kinds of queries (e.g.,~locate, see Section~\ref{Sec:problem_class}) efficiently.

\footnotetext{David Eppstein, available at \url{http://en.wikipedia.org/wiki/Bloom_filter\#/media/File:Bloom_filter.svg}, in public domain.}

\begin{figure}[ht]
\vspace{1em}
\centering
\begin{tabular}{c|c}
Word & Occurrence list \\
\hline
\texttt{This} & 0 \\
\texttt{is} & 12 \\
\texttt{a} & 15 \\
\texttt{banana} & 5, 17 \\
\end{tabular}
    \caption[An inverted index which stores a mapping from words to their positions.]{An inverted index which stores a mapping from words to their positions (0-based) in the text ``This banana is a banana''.}
    \label{Fig:inv_index}
\end{figure}

Main advantage of the inverted index are fast single-word queries, which can be answered in constant average time using for example a hash table (an inverted index is a rather generic idea, which means that it could be also implemented with other data structures such as binary trees). 
On the other hand, there is a substantial space overhead (in the order of $O(n)$) and the original string has to be stored as well.
For this reason, one of the key challenges for inverted indexes is how to succinctly represent the lists of positions while still allowing fast access.
Multiple methods were proposed, and they are often combined with one other~\cite{anh2005inverted}.

The most popular one is to store gaps, that is differences between subsequent positions.
For the index in Figure~\ref{Fig:inv_index}, the list for banana would be equal to $\{5, 12\}$ instead of $\{5, 17\}$ ($17-5=12$).
The values of the gaps are usually smaller than the original positions and for this reason they can stored using a fewer amount of bits.
Another popular approach is to use byte-aligned coding.
Here, each byte contains a one-bit flag which is set if the number is bigger or equal to $2^7$ (that is when it does not fit into 7 bits), and the
other seven bits are used for the data.
If the number does not fit, 7 least-significant bits are stored in the original byte, and the algorithm tries to store the remaining bits in the next byte, proceeding until the whole number has been exhausted.
In order to reduce the average length (in bits) of the occurrence list, one could also divide the original text into multiple blocks of fixed size.
Instead of storing exact positions only block indexes are stored, and after the index is retrieved the word is searched for explicitly within the block~\cite{manber1994glimpse}.
If the size of the data is so massive that it is infeasible to construct a single index (as is often the case for web search engines), sometimes only the most relevant data is selected for being stored in the index (thus forming a pruned index)~\cite{ntoulas2007pruning}.

\section{Approximate}
\label{Sec:keyword-approx}

Boytsov~\cite{boytsov2011indexing} presented an extensive survey of keyword indexes for approximate searching (including a practical evaluation).
He divided the algorithms into two following categories:

\begin{itemize}
\item
\textbf{Direct} methods --- like neighborhood generation (see Section~\ref{Sec:full-approx}), where certain candidates are searched for exactly.

\item
\textbf{Sequence-based filtering} methods --- the dictionary is divided into many (disjoint or overlapping) clusters.
During the search, a query is assigned to one or several clusters containing candidate strings and thus an explicit verification is performed only on a fraction of the original dictionary.

\end{itemize}

Notable results from the theoretical point of view include the \emph{k-errata trie} by Cole et al.~\cite{cole2004dictionary} which was already mentioned in the previous chapter.
For the Hamming distance and dictionary matching, it uses $O(n + d \frac{(\log d)^k}{k!})$ space and offers $O(m + \frac{(\log d)^k}{k!} \log \log n + occ)$ query time, where $d = |\mathcal{D}|$ (this also holds for the edit distance but with larger constants).
Another theoretical work describing the algorithm which is similar to our split index (which we describe in Subsection~\ref{Sec:split_index}) was given by Shi and Widmayer~\cite{shi1996approximate}, who obtained $O(n)$ preprocessing time and space complexity and $O(n)$ expected time if $k$ is bounded by $O(m / \log m)$.
They introduced the notion of home strings for a given $q$-gram, which is the set of strings in $\mathcal{D}$ that contain the $q$-gram in the exact form (the value of $q$ is set to $|P|/(k + 1)$).
In the search phase, they partition $P$ into $k + 1$ disjoint $q$-grams and use a candidate inspection order to speed up finding the matches with up to $k$ edit distance errors.

On the practical front, Bocek et al.~\cite{bocek2007fast} provided a generalization of the \ac{mf}~\cite{mor1982hash} algorithm for $k \geqslant 1$ which is called \emph{FastSS}.
To check if two strings $S_1$ and $S_2$ match with up to $k$ errors, we first delete all possible ordered subsets of $k^\prime$ symbols for all $0 \leqslant k^\prime \leqslant k$ from $S_1$ and $S_2$.
Then we conclude that $S_1$ and $S_2$ \emph{may} be in edit distance at most $k$ if and only if the intersection of the resulting lists of strings
is non-empty (explicit verification is still required).
For instance, if $S_1$ = \texttt{abbac} and $k = 2$, then its neighborhood is as follows: \texttt{abbac}, \texttt{bbac}, \texttt{abac}, \texttt{abac}, \texttt{abbc}, \texttt{abba}, \texttt{abb}, \texttt{aba}, \texttt{abc}, \texttt{aba}, \texttt{abc}, \texttt{aac}, \texttt{bba}, \texttt{bbc}, \texttt{bac} and \texttt{bac} (of course, some of the resulting strings are repeated and they may be removed).
If $S_2$ = \texttt{baxcy}, then its respective neighborhood for $k = 2$ will contain, e.g., the string \texttt{bac}, but the following verification will show that
$S_1$ and $S_2$ are in edit distance greater than 2.
If, however, $Lev(S_1 , S_2) \leqslant 2$, then it is impossible not to have in the neighborhood of $S_2$ at least one string from the neighborhood of $S_1$, hence we will never miss a match.
The lookup requires $O(k m^k \log n^k)$ time (where $m$ is the average word length from the dictionary) and the index occupies $O(n^k)$ space.
Another practical filter was presented by Karch et al.~\cite{karch2010improved} and it improved on the FastSS method.
They reduced space requirements and query time by splitting long words (similarly to FastBlockSS which is a variant of the original method) and storing the neighborhood implicitly with indexes and pointers to original dictionary entries.
They claimed to be faster than other approaches such as the aforementioned FastSS and the BK-tree~\cite{burkhard1973some}.
Recently, Chegrane and Belazzougui~\cite{chegrane2014simple} described another practical index and they reported better results when compared to Karch et al.
Their structure is based on the dictionary by Belazzougui~\cite{belazzougui2009faster} for the edit distance of 1 (see the following subsection).
An approximate (in the mathematical sense) data structure for approximate matching which is based on the Bloom filter (see Subsection~\ref{Sec:bloom}) was also described~\cite{manber1994algorithm}.

\subsection{The 1-error problem}

It is important to consider methods for detecting a single error, since over 80\% of errors (even up to roughly 95\%) are within $k=1$ for the edit distance with transpositions~\cite{damerau1964technique, pollock1984automatic}.
Belazzougui and Venturini~\cite{belazzougui2012compressed} presented a compressed index whose space is bounded in terms of the $k$-th order empirical entropy of the indexed dictionary.
It can be based either on perfect hashing, having $O(m + occ)$ query time or on a compressed permuterm index with $O(m \min(m, \log_{\sigma} n \log \log n) + occ)$ time (when $\sigma = \log^c n$ for some constant $c$) but improved space requirements.
The former is a compressed variant of a dictionary presented by Belazzougui~\cite{belazzougui2009faster} which is based on neighborhood generation and occupies $O(n \log \sigma)$ space and can answer queries in $O(m)$ time.
Chung et al.~\cite{chung2014efficient} showed a theoretical work where external memory is used, and their focus is on I/O operations.
They limited the number of these operations to $O(1 + m / wB + occ / B)$, where $w$ is the size of the machine word and $B$ is the number of words within a block (a basic unit of I/O), and their structure occupies $O(n/B)$ blocks.
In the category of filters, Mor and Fraenkel~\cite{mor1982hash} described a method which is based on the deletion-only 1-neighborhood.

For the 1-mismatch problem, Yao and Yao~\cite{YaoY95} described a data structure for binary strings of fixed length $m$ with $O(m \log\log |\mathcal{D}|)$ query time and $O(|\mathcal{D}| m \log m)$ space requirements.
This was later improved by Brodal and G\k{a}sieniec~\cite{brodal1996approximate} with a data structure with $O(m)$ query time which occupies $O(n)$ space.
This was improved with a structure with $O(1)$ query time and $O(|\mathcal{D}| \log m)$ space in the cell probe model (where only memory accesses are counted)~\cite{brodal2000improved}.
Another notable example is a recent theoretical work of Chan and Lewenstein~\cite{chan2015fast}, who introduced the index with the optimal query time, i.e.~$O(m/w + occ)$, which uses additional $O(w d \log^{1+\epsilon} d)$ bits of space (beyond the dictionary itself), assuming a constant-size alphabet.

\subsection{Permuterm index}
\label{Sec:permuterm}

A permuterm index is a keyword index which supports queries with one wildcard symbol~\cite{garfield1976permuterm}.
The idea is store all rotations of a given word appended with the terminating character, for instance for the word \texttt{text}, the index would consist of the following permuterm vocabulary: \texttt{text\$, ext\$t, xt\$te, t\$tex, \$text}.
When it comes to searching, the query is first rotated so that the wildcard appears at the end, and subsequently its prefix is searched for using the index.
This could be for example a trie, or any other data structure which supports a prefix lookup.

The main problem with the standard permuterm index is its space usage, as the number of strings inserted into the data structure is the number of words multiplied by the average string length.
Ferragina and Venturini~\cite{ferragina2010compressed} proposed a \emph{compressed} permuterm index in order to overcome the limitations of the original structure with respect to space.
They explored the relation between the permuterm index and the BWT (see Subsection~\ref{Sec:BWT}), which is applied to the concatenation of all strings from the input dictionary, and they provided a modification of the LF-mapping known from FM-indexes in order to support the functionality of the permuterm index.

\subsection{Split index}
\label{Sec:split_index}

One of the practical approximate indexes was described by Cis{\l}ak (thesis author) and Grabowski~\cite{cislak2015practical}.
Experimental results for this structure can be found in Section~\ref{Sec:split_index_res}.

As indexes supporting approximate matching tend to grow exponentially in $k$, the maximum number of allowed errors, it is also a worthwhile goal to design efficient indexes supporting only a small $k$.
For this reason, we focus on the problem of dictionary matching with few mismatches, especially one mismatch, where $Ham(d_i, P) \leqslant k$ for a collection of words $d \in \mathcal{D}$, a pattern $P$, and the Hamming distance $Ham$.
The algorithm that we are going to present is uncomplicated and based on the Dirichlet principle, ubiquitous in approximate string matching techniques.
We partition each word $d$ into $k+1$ disjoint pieces $p_1, \ldots, p_{k+1}$, of average length $|d|/(k+1)$ (hence the name ``split index''), and each such piece acts as a key in a hash table $H_T$.
The size of each piece $p_i$ of word $d$ is determined using the following formula: $|p_i| = \lfloor |d| / (k+1) \rceil$ (or $|p_i| = \lfloor |d| / (k+1) \rfloor$, depending on the practical evaluation) and $|p_{k+1}| = |d| - \sum_{i = 1}^k |p_i|$, i.e.~the piece size is rounded to the nearest integer and the last piece covers the characters which are not in other pieces.
This means that the pieces might be in fact unequal in length, e.g., 3 and 2 for $|d| = 5 \land k = 1$.
The values in $H_T$ are the lists of words which have one of their pieces as the corresponding key.
In this way, every word occurs on exactly $k+1$ lists.
This seemingly bloats the space usage, still, in the case of small $k$ the occupied space is acceptable.
Moreover, instead of storing full words on the respective lists, we only store their ``missing'' prefix or suffix.
For instance for the word \texttt{table} and $k=1$, we would have a relation \texttt{tab} $\to$ \texttt{le} on one list (i.e.~\texttt{tab} would be the key and \texttt{le} would be the value) and \texttt{le} $\to$ \texttt{tab} on the other.

In the case of $k=1$, we first populate each list with the pieces without the prefix and then with the pieces without the suffix; 
additionally we store the position on the list (as a 16-bit index) where the latter part begins.
In this way, we traverse only a half of a list on average during the search.
We can also support $k$ larger than 1 --- in this case, we ignore the piece order on a list, and we store $\lceil \log_2(k+1) \rceil$ bits with each piece that indicate which part of the word is the list key.
Let us note that this approach would also work for $k=1$, however, it turned out to be less efficient.

\begin{figure}[ht]
    \centering
    \includegraphics[scale=0.3]{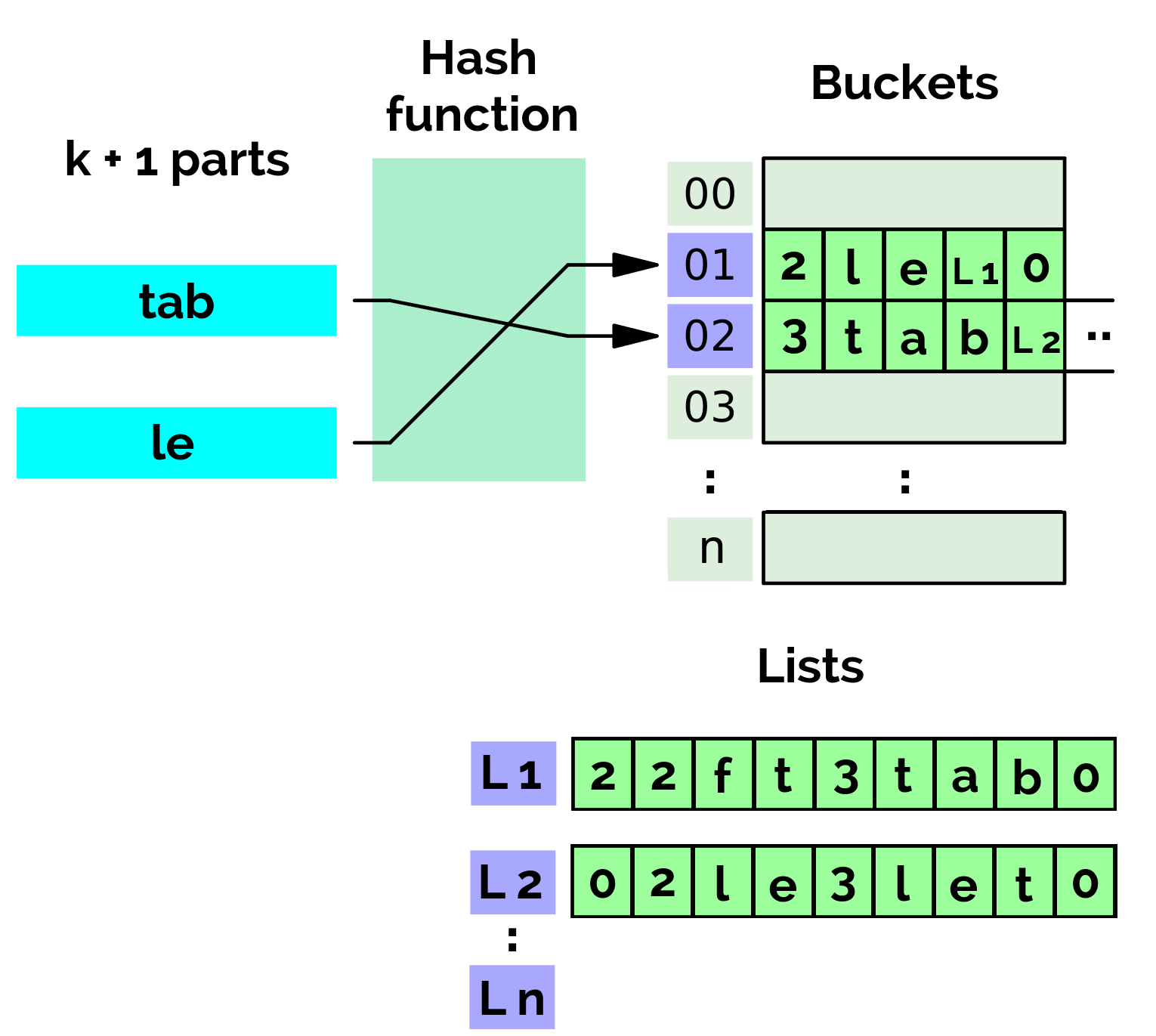}

    \caption[Split index for keyword indexing.]{Split index for keyword indexing which shows the insertion of the word \texttt{table} for $k = 1$. The index also stores the words \texttt{left} and \texttt{tablet} (only selected lists containing pieces of these two words are shown), and \texttt{L1} and \texttt{L2} indicate pointers to the respective lists. The first cell of each list indicates a 1-based word position (i.e.~the word count from the left) where the missing prefixes begin ($k = 1$, hence we deal with two parts, namely prefixes and suffixes), and 0 means that the list has only missing suffixes. Adapted from Wikimedia Commons\protect\footnotemark.}
    \label{Fig:split_index}
\end{figure}

As regards the implementation (in the C++ language), our focus is on data compactness.
In the hash table, we store the buckets which contain word pieces as keys (e.g., \texttt{le}) and pointers to the lists which store the missing pieces of the word  (e.g., \texttt{tab}, \texttt{ft}).
These pointers are always located right next to the keys, which means that unless we are very unlucky, a specific pointer should already be present in the CPU cache during the traversal.
The memory layouts of these substructures are fully contiguous. 
Successive strings are represented by multiple characters with a prepended 8-bit counter which specifies the length, and the counter with the value 0 indicates the end of the list.
During the traversal, each length can be compared with the length of the piece of the pattern.
As mentioned before, the words are partitioned into pieces of fixed length. 
This means that on average we calculate the Hamming distance for only half of the pieces on the list, since the rest can be ignored based on their length.
Any hash function for strings can be used, and two important considerations are the speed and the number of collisions, since a high number of collisions results in longer buckets, which may in turn have a negative effect on the query time (this subject is explored in more detail along with the results in Chapter~\ref{Chap:experimental}).
Figure~\ref{Fig:split_index} illustrates the layout of the split index.

\setcounter{footnote}{0}

The preprocessing stage proceeds as follows:

\begin{enumerate}
\item
Duplicate words are removed from the dictionary $\mathcal{D}$.
\end{enumerate}

The following steps refer to each word $d$ from $\mathcal{D}$:

\begin{enumerate}[resume]
\item
The word $d$ is split into $k + 1$ pieces.
\item
For each piece $p_i$: if $p_i \notin H_T$, we create a new list $l_n$ containing the missing pieces (later simply referred to as a missing piece; in the case of $k = 1$, this is always one contiguous piece) $\mathcal{P} = \{p_j : j \in [1, k + 1] \land j \neq i \}$ and add it to the hash table (we append $p_i$ and the pointer to $l_n$ to the bucket).
Otherwise, if $p_i \in H_T$, we append the missing pieces $\mathcal{P}$ to the already existing list $l_i$.
\end{enumerate}

As regards the search:

\begin{enumerate}
\item
The pattern $P$ is split into $k + 1$ pieces.
\item
We search for each piece $p_i$ (the prefix and the suffix if $k = 1$): the list $l_i$ is retrieved from the hash table or we continue if $p_i \notin H_T$.
We traverse each missing piece $p_j$ from $l_i$.
If $|p_j| = |P| - |p_i|$, the verification is performed and the result is returned if $Ham(p_j, P - p_i) \leqslant k$.
\item
The pieces are combined into one word in order to form the answer.
\end{enumerate}

\footnotetext{Jorge Stolfi, available at \url{http://en.wikipedia.org/wiki/File:Hash_table_3_1_1_0_1_0_0_SP.svg}, CC A-SA 3.0.}

\subsubsection{Complexity}

Let us consider the average word length $|d|$, where $|d| = (\sum^{|\mathcal{D}|}_{i=1} |d_i|) / |\mathcal{D}|$.
The average time complexity of the preprocessing stage is equal to $O(k n)$, where $k$ is the allowed number of errors, and $n$ is the total input dictionary size (i.e.~the length of the concatenation of all words from $\mathcal{D}$, $n = \sum_{i=1}^{|\mathcal{D}|}|d_i|$).
This is because for each word and for each piece $p_i$ we either add the missing pieces to a new list or append them to the already existing one in $O(|d|)$ time (let us note that $|\mathcal{D}| \cdot |d| = n$).
We assume that adding a new element to the bucket takes constant time on average, and the calculation of all hashes takes $O(n)$ time in total.
This is true irrespective of which list layout is used (there are two layouts for $k = 1$ and $k > 1$, see the preceding paragraphs).
The occupied space is equal to $O(kn)$, because each piece appears on exactly $k$ lists and in exactly 1 bucket.

The average search complexity is equal to $O(k t)$, where $t$ is the average length of the list.
We search for each of $k + 1$ pieces of the pattern of length $m$, and when the list corresponding to the piece $p$ is found, it is traversed and at most $t$ verifications are performed.
Each verification takes at most $O(\min(m, |d_{max}|))$ time where $d_{max}$ is the longest word in the dictionary (or $O(k)$ time, in theory, 
using the old technique from Landau and Vishkin~\cite{landau1989fast}, after $O(n\log\sigma)$-time preprocessing), but $O(1)$ time on average.
Again, we assume that determining a location of the specific list, that is iterating a bucket, takes $O(1)$ time on average.
As regards the list, its average length $t$ is higher when there is a higher probability that two words $d_1$ and $d_2$ from $\mathcal{D}$ have two parts of the same length $l$
which match exactly, i.e.~$Pr(d_1[s_1, s_1 + l - 1] = d_2[s_2, s_2 + l - 1])$.
Since all words are sampled from the same alphabet $\Sigma$, $t$ depends on the alphabet size, that is $t = f(\sigma)$.
Still, the dependence is rather indirect; in real-world dictionaries which store words from a given language, $t$ will be rather dependent on the $k$-th order entropy of the language.

\subsubsection{Compression}

In order to reduce storage requirements, we apply a basic compression technique. 
We find the most frequent $q$-grams in the word collection and replace their occurrences on the lists with unused symbols, e.g., byte values $128, \ldots, 255$.
The values of $q$ can be specified at the preprocessing stage, for instance $q = 2$ and $q = 4$ are reasonable for the English alphabet and DNA, respectively.
Different $q$ values can be also combined depending on the distribution of $q$-grams in the input text, i.e.~we may try all possible combinations of $q$-grams up to a certain $q$ value and select ones which provide the best compression.
In such a case, longer $q$-grams should be encoded before shorter ones.
For example, a word \texttt{compression} could be encoded as \texttt{\#p*s\textbackslash} using the following substitution list: $\texttt{com} \to \texttt{\#}, \texttt{re} \to \texttt{*}, \texttt{co} \to \texttt{\$}, \texttt{om} \to \texttt{\&}, \texttt{sion} \to \texttt{\textbackslash}$ (note that not all $q$-grams from the substitution list are used).
Possibly even a recursive approach could be applied, although this would certainly have a substantial impact on the query time.
See Section~\ref{Sec:split_index_res} for the experimental results and a further discussion.

The space usage could be further reduced by the use of a different character encoding.
For the DNA (assuming 4 symbols only) it would be sufficient to use 2 bits per character, and for the basic English alphabet 5 bits.
In the latter case there are 26 letters, which in a simplified text can be augmented only with a space character, a few punctuation marks, and a capital letter flag.
Such an approach would be also beneficial for space compaction, and it could have a further positive impact on cache usage.
The compression naturally reduces the space while increasing the search time, and a sort of a middle ground can be achieved by deciding which additional information to store in the index.
This can be for instance the length of an encoded (compressed) piece after decoding, which could eliminate some pieces based on their size without performing the decompression and explicit verification.

\subsubsection{Parallelization}
\label{Sec:split_par}

The algorithm could be sped up by means of parallelization, since index access during the search procedure is read-only.
In the most straightforward approach, we could simply distribute individual queries between multiple threads.
A more fine-grained variation would be to concurrently operate on word pieces after the word has been split up (with the number of pieces being dependent on the $k$ parameter).
We could even access in parallel the lists which contain missing pieces (prefixes and suffixes for $k = 1$), although the gain would be probably limited since these lists usually store at most a few words.
If we had a sufficient amount of threads at our disposal, these approaches could be combined.
Still, it is to be noted that the use of multiple threads has a negative effect on cache utilization.

\subsubsection{Inverted split index}
\label{Sec:split_inv}

The split index could be extended in order to include the functionality of an inverted index for approximate matching.
As mentioned in Subsection~\ref{Sec:inv_index}, the inverted index could be in practice any data structure which supports an efficient word lookup.
Let us consider the compact list layout of the split index presented in Figure~\ref{Fig:split_index}, where each piece is located right next to other pieces.
Instead of storing only the 8-bit counter which specifies the length of the piece, we could also store (right next to this piece) its position in the text.
Such an approach would increase the average length of the list only by constant factor and it would not break the contiguity of the lists, while also keeping the $O(kn)$ space complexity.
Moreover, the position should be already present in the CPU cache during the list traversal.

\section{Keyword selection}
\label{Sec:keyword_sel}

Keyword indexes can be also used in the scenario where there are no explicit boundaries between the words.
In such a case, we would like to select the keywords according to a well-defined set of rules and form a dictionary $\mathcal{D}$ from the input text $T$.
Such an index which stores $q$-grams sampled from the input text may be referred to as a $q$-gram index.
It is useful for answering keyword rather than full-text queries, which might be required for example due to time requirements (i.e.~when we would like to trade space for speed). 
Examples of the input which cannot be easily divided into words include some natural languages (e.g.,~Chinese) where it is not possible to clearly distinguish the words (their boundaries depend on the context) or other kinds of data such as a complete genome~\cite{kim2005n}.

Let us consider the input text $T$ which is divided into $n - q + 1$ $q$-grams.
The issue lies in the amount of space which is occupied by all $q$-gram tuples $(s, l_i)$, where $s$ is the $q$-gram and $l_i$ identifies its positions, which is in the order of $O(n)$ or $O(n q_{max})$ for all possible $q$-grams up to some $q_{max}$ value.
General compression techniques are usually not sufficient and thus a dedicated solution is required.
This is especially the case in the context of bioinformatics where data sets are substantial; the applications could be for instance retrieving the seeds in the seed-and-extend algorithm described in Section~\ref{Sec:full-word_based}.
One of the approaches was proposed by Kim et al.~\cite{kim2005n}, and it aims to eliminate the redundancy in position information.
Consecutive $q$-grams are grouped into subsequences, and each $q$-gram is identified by the position of the subsequence within the documents and the
position of a $q$-gram within the subsequence, which forms a two-level index structure.
This concept was also extended by the original authors to include the functionality of approximate matching~\cite{kim2007n}.

\subsection{Minimizers}
\label{Sec:minimizers}


The idea of minimizers was introduced by Roberts et al.~\cite{roberts2004reducing} (with applications in, e.g.,~genome sequencing with de Bruijn graphs~\cite{chikhi2014representation} and $k$-mer counting~\cite{deorowicz2015kmc}), and it consists in storing only selected rather than all $q$-grams from the input text.
Here, the goal is to choose such $q$-grams from a given string $S$ (a set $M(S)$), so that for two strings $S_1$ and $S_2$, if it holds for a pattern $P$ that $P \subset S_1 \land P \subset S_2$ and $|P|$ is above some threshold, then it should also hold that $|M(S_1) \cap M(S_2)| \geqslant 1$.
In order to find a $(\alpha, q)$-minimizer, we slide a window of length $q + \alpha - 1$ ($\alpha$ consecutive $q$-grams) over $T$, shifting it by 1 character at a time, and at each window position we select a $q$-gram which is the smallest one lexicographically (ties may be resolved for instance in favor of the leftmost of the smallest substrings).
Figure~\ref{Fig:minimizers} demonstrates this process.

\begin{figure}[ht]
\vspace{1em}
\centering
\begin{tabular}{c|ccccccc}
$T$     & t & e & x & t & i & n & g \\
\hline
$W_1$ & t & e & x & t &   &   &  \\
\cline{3-4}
$W_2$ &   & e & x & t & i &   &  \\
\cline{3-4}
$W_3$ &   &   & x & t & i & n &  \\
\cline{6-7}
$W_4$ &   &   &   & t & i & n & g \\
\cline{6-7}
\end{tabular}
\caption[Selecting minimizers from a given text.]{Selecting $(3,2)$-minimizers (underlined), that is choosing 2-grams while sliding a window of length 4 ($3+2-1=4$) over the text \texttt{texting}. The results belong to the following set: $\{\texttt{ex}, \texttt{in}\}$.}
\label{Fig:minimizers}
\end{figure}

Let us repeat an important property of the minimizers which makes them useful in practice.
If for two strings $S_1$ and $S_2$ it holds that $P \subset S_1 \land P \subset S_2 \land |P| \geqslant q + \alpha - 1$, then it is guaranteed that $S_1$ and $S_2$ share an $(\alpha, q)$-minimizer (because they share one full window).
This means that for certain applications we can still ensure that no exact matches are overlooked by storing the minimizers rather than all $q$-grams.

\section{String sketches}
\label{Sec:str_sketches}

We introduce the concept of string sketches, whose goal is to speed up string comparisons at the cost of additional space.
For a given string $S$, a sketch $S^\prime$ is constructed as $S^\prime = f(S)$ using some function $f$ which returns a fixed-sized block of data.
In particular, for two strings $S_1$ and $S_2$, we would like to determine with certainty that $S_1 \neq S_2$ or $Ham(S_1, S_2) \geqslant k$ when comparing only sketches $S^\prime_1$ and $S^\prime_2$.
There exists a similarity between sketches and hash functions, however, hash comparison would work only in the context of exact matching.
When the sketch comparison is not decisive, we still have to perform an explicit verification on $S_1$ and $S_2$, but the sketches allow for reducing the number of such verifications.
Since the sketches refer to individual words, they are relevant in the context of keyword indexes.
Assuming that each word $d \in \mathcal{D}$ is stored along with $d^\prime$, sketches could be especially useful if the queries are known in advance or $|\mathcal{D}|$ is relatively high, since sketch calculation might be time-consuming.

Sketches use individual bits in order to store information about $q$-gram frequencies in the string.
Various approaches exist, and main properties of the said $q$-grams include:

\begin{itemize}
\item
\textbf{Size} --- for instance individual letters (1-grams) are sensible for the English alphabet but pairs (2-grams) might be better for the DNA.
\item
\textbf{Frequency} --- we can store binary information in each bit that indicates whether a certain $q$-gram appears in the string (8 $q$-grams in total for a 1-byte sketch, we call this approach an occurrence sketch), or we can store their count (at most 3) using 2-bits per $q$-gram (4 $q$-grams in total for a 1-byte sketch, we call this approach a count sketch).
\item
\textbf{Selection} --- which $q$-grams should be included in the sketches. These could be for instance $q$-grams which occur most commonly in the sample text.
\end{itemize}

For instance, let us consider an occurrence sketch which is built over 8 most common letters of the English alphabet, namely $\{ \texttt{e}, \texttt{t}, \texttt{a}, \texttt{o}, \texttt{i}, \texttt{n}, \texttt{s}, \texttt{h} \}$ (consult Appendix~\ref{App:letter_freq} to see the frequencies).
For the word \texttt{instance}, the 1-byte sketch where each bit corresponds to one of the letters from the aforementioned set would be as follows: \texttt{11101110}.

We can quickly compare two sketches by taking a binary \texttt{xor} operation and counting the number of bits which are set in the result (calculating the Hamming weight, $H_W$).
Note that $H_W$ can be determined in constant time using a lookup table of size $2^{8n}$ bytes, where $n$ is the sketch size in bytes.
We denote the sketch difference with $H_S$, and $H_S = H_W(S^\prime_1 \oplus S^\prime_2)$.
Let us note that $H_S$ does not determine the number of mismatches, for instance for $S_1 = \texttt{run}$ and $S_2 = \texttt{ran}$, $H_S(S_1, S_2)$ might be equal to 2 (occurrence differences in \texttt{a} and \texttt{u}) but there is still only one mismatch.
On the other extreme, for two strings of length $n$ where each string consists of a repeated occurrence of one different letter, $H_S$ might be equal to 1, but the number of mismatches is $n$.
In general, $H_S$ can be used to provide a lower bound on the true number of errors.
For sketches which record information about single characters (1-grams), the following holds: $Ham(S_1, S_2) \geqslant \lceil H_S(S^\prime_1, S^\prime_2) / 2 \rceil$ (the right-hand side can be calculated quickly using a lookup table, since $0 \leqslant H_S \leqslant 8$).
The true number of mismatches is underestimated especially by count sketches, since we calculate the Hamming weight instead of comparing the counts.
For instance, for the count of 3 (bits \texttt{11}) and the count of 1 (bits \texttt{01}), the difference is 1 instead of 2.
Still, even though the true error is higher than $H_S$, sketches can be used in order to speed up the comparisons because certain strings will be compared (and rejected) in constant time using fast bitwise operations and array lookups.
As regards the space overhead incurred by the sketches, it is equal to $O(|\mathcal{D}| + \sigma)$, since we have to store one constant-size sketch per word together with the lookup tables which are used to speed up the processing.
Consult Section~\ref{Sec:string_sketches_exp} in order to see the experimental results.

%
%


\chapter{Experimental Results}
\label{Chap:experimental}
\lhead{\emph{Experimental Results}}

The results were obtained on the machine with the following specifications:

\begin{itemize}
\item
\textbf{Processor} --- Intel i5-3230M running at 2.6\,GHz
\item
\textbf{RAM} --- 8\,GB DDR3 memory
\item
\textbf{Operating system} --- Ubuntu 14.04 64-bit (kernel version 3.16.0-41)
\end{itemize}

Programs were written in the C++ programming language (with certain prototypes in the Python language) using features from the C++11 standard~\cite{cpp11}.
They use the C++ Standard Library, Boost libraries (version 1.57)~\cite{boost}, and Linux system libraries.
Correctness was analyzed using Valgrind~\cite{valgrind}, a tool for error checking and profiling (no errors or memory leaks were reported).
The source code was compiled (as a 32-bit version) with clang compiler v.~3.4-1, which turned out to be produce a slightly faster executable than the gcc when checked under the \texttt{-O3} optimization flag.

\section{FM-bloated}
\label{Sec:fm-bloated_res}

For the description of the FM-bloated structure consult Subsection~\ref{Sec:fm-bloated}.
Here, we present experimental results for the superlinear index version.
As regards the hash function, xxhash was used (available on the Internet, consult Appendix~\ref{App:hashes}), and the load factor was equal to 2.81.

\begin{figure}[ht]
    \centering
    \includegraphics[scale=0.55]{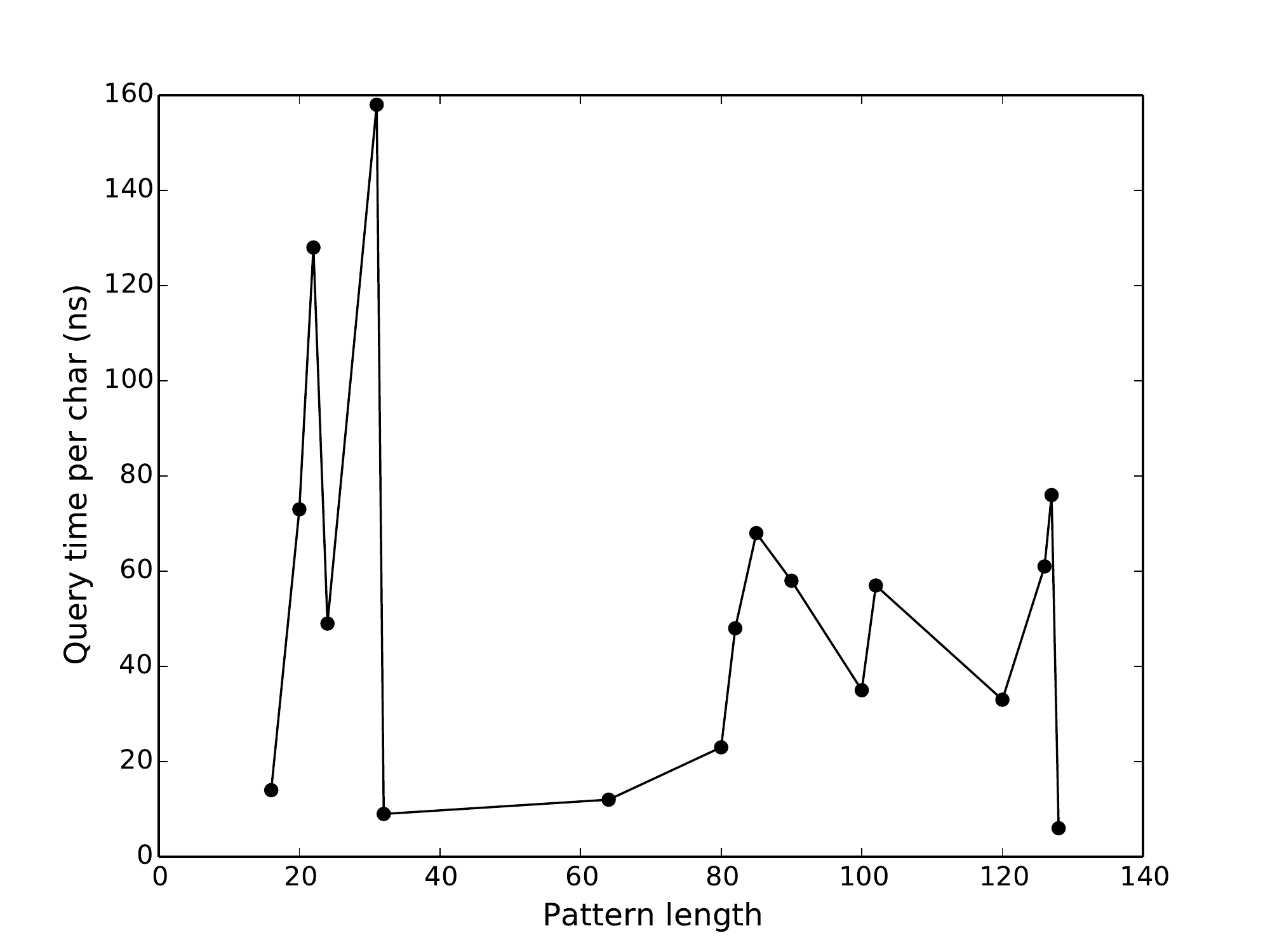}

    \caption[Query time per character vs pattern length for the English text of size 30\,MB (FM-bloated).]{Query time per character vs pattern length (16, 20, 22, 24, 31, 32, 64, 80, 82, 85, 90, 100, 102, 126, 127, and 128) for the English text of size 30\,MB. Let us point out notable differences between pattern lengths 31 and 32, and 127 and 128.}
	\label{Fig:bloated_m}
\end{figure}

The length of the pattern has a crucial impact on the search time, since the number LF-mapping steps is equal the number of 1s in the binary representation of $m$.
This means that the search will be the fastest for $m$ in the form $2^c$ (constant time for $m$ up to a certain maximum value) and the slowest for $m$ in the form $2^c - 1$, where $c \in \mathbb{Z}$.
We can see in Figure~\ref{Fig:bloated_m} that the query time also generally decreases as the pattern length increases, mostly due to the fact that the times are given per character.
The results are the average times calculated for one million queries which were extracted from the input text.

\begin{figure}[ht]
    \centering
    \includegraphics[scale=0.55]{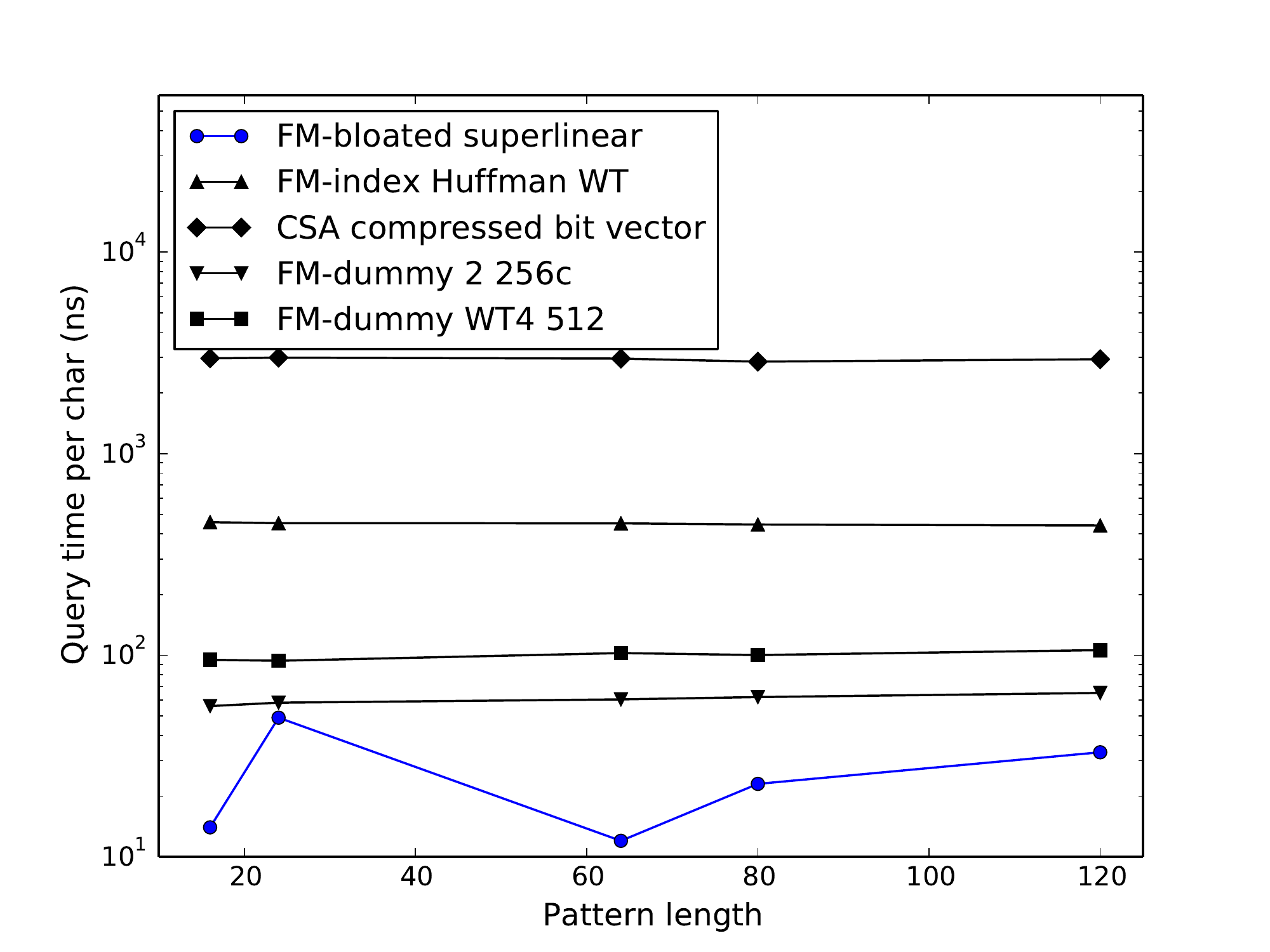}

    \caption[Query time per character vs pattern length for different methods for the English text of size 30\,MB (FM-bloated).]{Query time per character vs pattern length (16, 24, 64, 80, and 120) for different methods for the English text of size 30\,MB. Note the logarithmic y-scale.}
    \label{Fig:bloated_comp}
\end{figure}

We also compare our approach with other FM-index-based structures (consult Figure~\ref{Fig:bloated_comp}).
We used the implementations from the sdsl library~\cite{gbmp2014sea} (available on the Internet~\cite{sdsl-lite}) and the implementations of FM-dummy structures by Grabowski et al.~\cite{grabowski2015fm} (available on the Internet~\cite{ranisz}).
As regards the space, the FM-bloated structure (just as the name suggests) is roughly two order of magnitude bigger than other indexes.
The index size for other methods ranged from approximately $0.6n$ to $4.25n$, where $n$ is the input text size.
FM-bloated, on the other hand, occupied the amount of space equal to almost $85n$ (for $q_{max} = 128$).


\section{Split index}
\label{Sec:split_index_res}

In this section we present the results which appeared in a preprint by Cis{\l}ak (thesis author) and Grabowski~\cite{cislak2015practical}.
For the description of the split index consult Subsection~\ref{Sec:split_index}.

One of the crucial components of the split index is a hash function.
Ideally, we would like to minimize the average length of the bucket (let us recall that we use chaining for collision resolution), however, the hash function should be also relatively fast because it has to be calculated for each of $k + 1$ parts of the pattern (of total length $m$).
We investigated various hash functions, and it turned out that the differences in query times are not negligible, although the average length of the bucket was almost the same in all cases (relative differences were smaller than 1\%).
We can see in Table~\ref{Tab:split_hash} that the fastest function was the xxhash (available on the Internet, consult Appendix~\ref{App:hashes}), and for this reason it was used for the calculation of other results.

\begin{table}[ht]
\vspace{1em}
\centering
\begin{tabular}{c|c}
Hash function & Query time (\SI{}{\micro\second}) \\
\hline
xxhash & 0.93\\
sdbm & 0.95\\
FNV1 & 0.95\\
FNV1a & 0.95\\
SuperFast & 0.96\\
Murmur3 & 0.97\\
City & 0.99\\
FARSH & 1.00\\
SpookyV2 & 1.04\\
Farm & 1.04\\
\end{tabular}
\vspace{4mm}
\caption[Evaluated hash functions and search times per query (split index).]{Evaluated hash functions and search times per query for the English dictionary of size 2.67\,MB and $k = 1$. A list of common English misspellings was used as queries, max \ac{lf} = 2.0.}
\label{Tab:split_hash}
\end{table}

\begin{figure}[ht]
    \centering
    \includegraphics[scale=0.6]{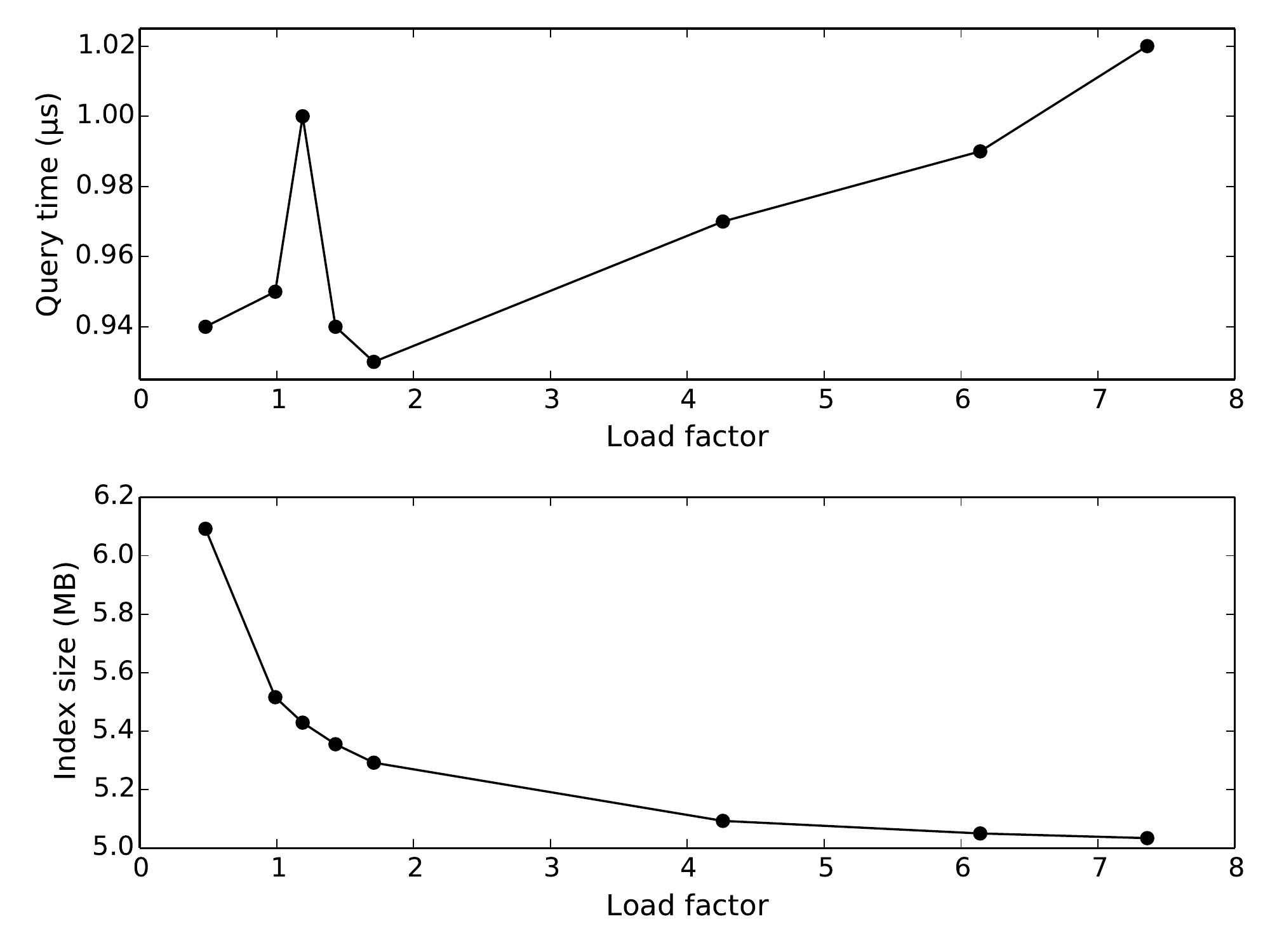}

    \caption[Query time and index size vs the load factor (split index).]{Query time and index size vs the load factor for the English dictionary of size 2.67\,MB and $k = 1$. A list of common English misspellings was used as queries. The value of \ac{lf} can be higher than 1.0 because we use chaining for collision resolution.}
    \label{Fig:split_lf}
\end{figure}

Decreasing the value of the load factor did not strictly provide a speedup in terms of the query time, as demonstrated in Figure~\ref{Fig:split_lf}.
This can be explained by the fact that even though the relative reduction in the number of collisions was substantial, the absolute difference was equal to at most a few collisions per list.
Moreover, when the \ac{lf} was higher, pointers to the lists could be possibly closer to each other, which might have had a positive effect on cache utilization.
The best query time was reported for the maximum \ac{lf} value of 2.0, hence this value was used for the calculation of other results.

In Table~\ref{Tab:split_k} we can see a linear increase in the index size and an exponential increase in query time with growing $k$.
Even though we concentrate on $k = 1$ and the most promising results are reported for this case, our index might remain competitive also for higher $k$ values.

\begin{table}[ht]
\vspace{1em}
\centering

\begin{tabular}{c|cc}
k & Query time (\SI{}{\micro\second}) & Index size (KB) \\
\hline
1 & \hskip 0.5em 0.51 & 1,715 \\ 
2 & 11.49 & 2,248 \\
3 & 62.85 & 3,078 \\
\end{tabular}
\vspace{4mm}
\caption[Query time and index size vs the error value $k$ (split index).]{Query time and index size vs the error value $k$ for the English language dictionary of size 0.79\,MB. A list of common English misspellings was used as queries.}
\label{Tab:split_k}
\end{table}

$Q$-gram substitution coding provided a reduction in the index size, at the cost of increased query time.
$Q$-grams were generated separately for each dictionary $\mathcal{D}$ as a list of 100 $q$-grams which provided the best compression for $\mathcal{D}$, i.e.~they minimized the size of all encoded words, $S_E = \sum_{i=1}^{|\mathcal{D}|} |Enc(d_i)|$.
For the English language dictionaries, we also considered using only 2-grams or only 3-grams, and for the DNA only 2-grams (a maximum of 25 2-grams) and 4-grams, since mixing the $q$-grams of various sizes has a further negative impact on the query time.
For the DNA, 5,000 queries were generated randomly by introducing noise into words sampled from dictionary, and their length was equal to the length of the particular word.
Up to 3 errors were inserted, each with a 50\% probability.
For the English dictionaries we opted for the list of common misspellings, and the results were similar to the case of randomly generated queries.
The evaluation was run 100 times and the results were averaged.

\begin{figure}[ht]
    \centering
    \includegraphics[scale=0.6]{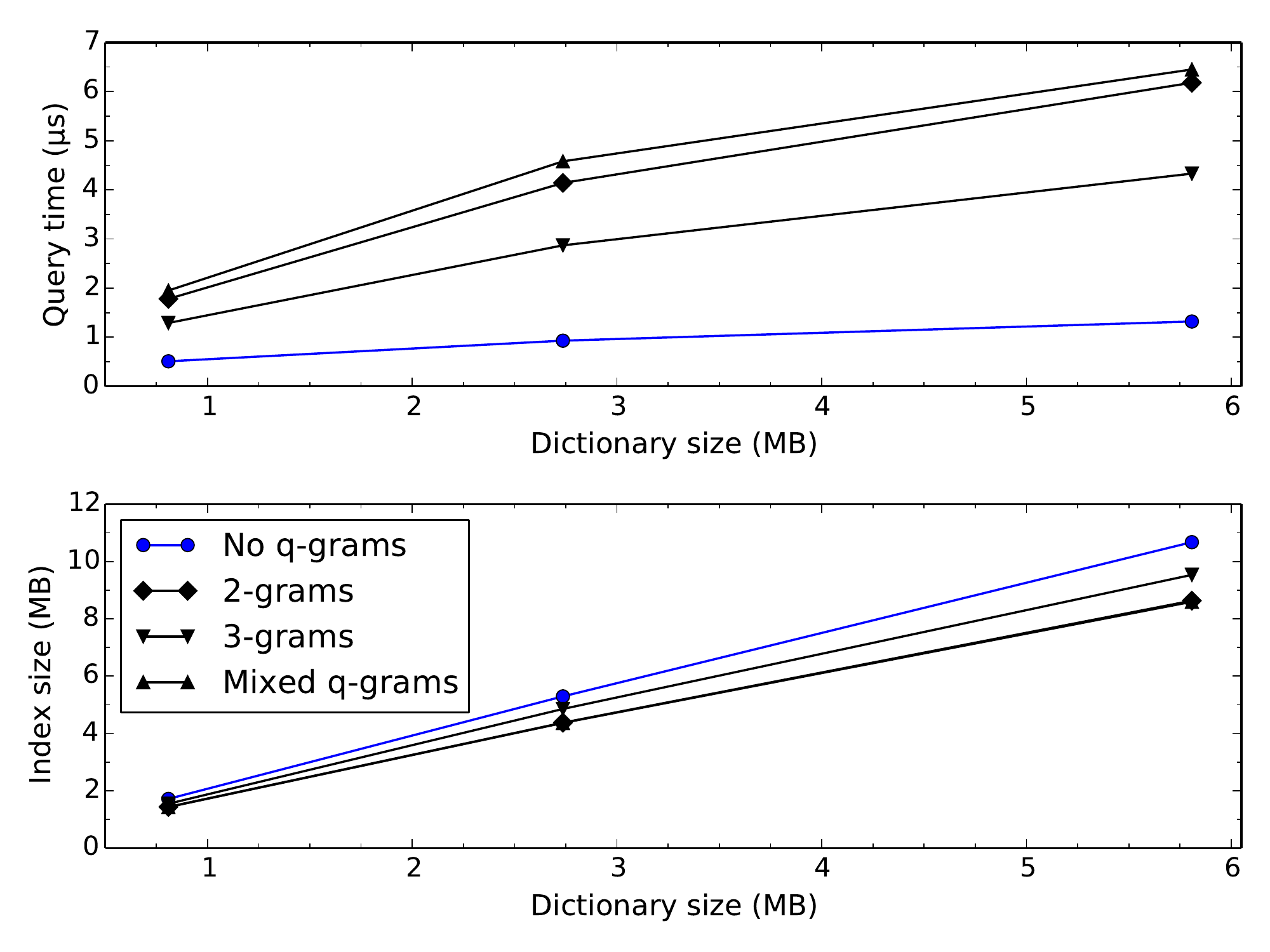}

    \caption[Query time and index size vs dictionary size, with and without $q$-gram coding for English dictionaries (split index).]{Query time and index size vs dictionary size for $k = 1$, with and without $q$-gram coding. Mixed $q$-grams refer to the combination of $q$-grams which provided the best compression, and for the three dictionaries these were equal to ([2-, 3-, 4-] grams): [88, 8, 4], [96, 2, 2], and [94, 4, 2], respectively. English language dictionaries and the list of common English misspellings were used.}
    \label{Fig:split_comp_eng}
\end{figure}

\begin{figure}[ht]
    \centering
    \includegraphics[scale=0.6]{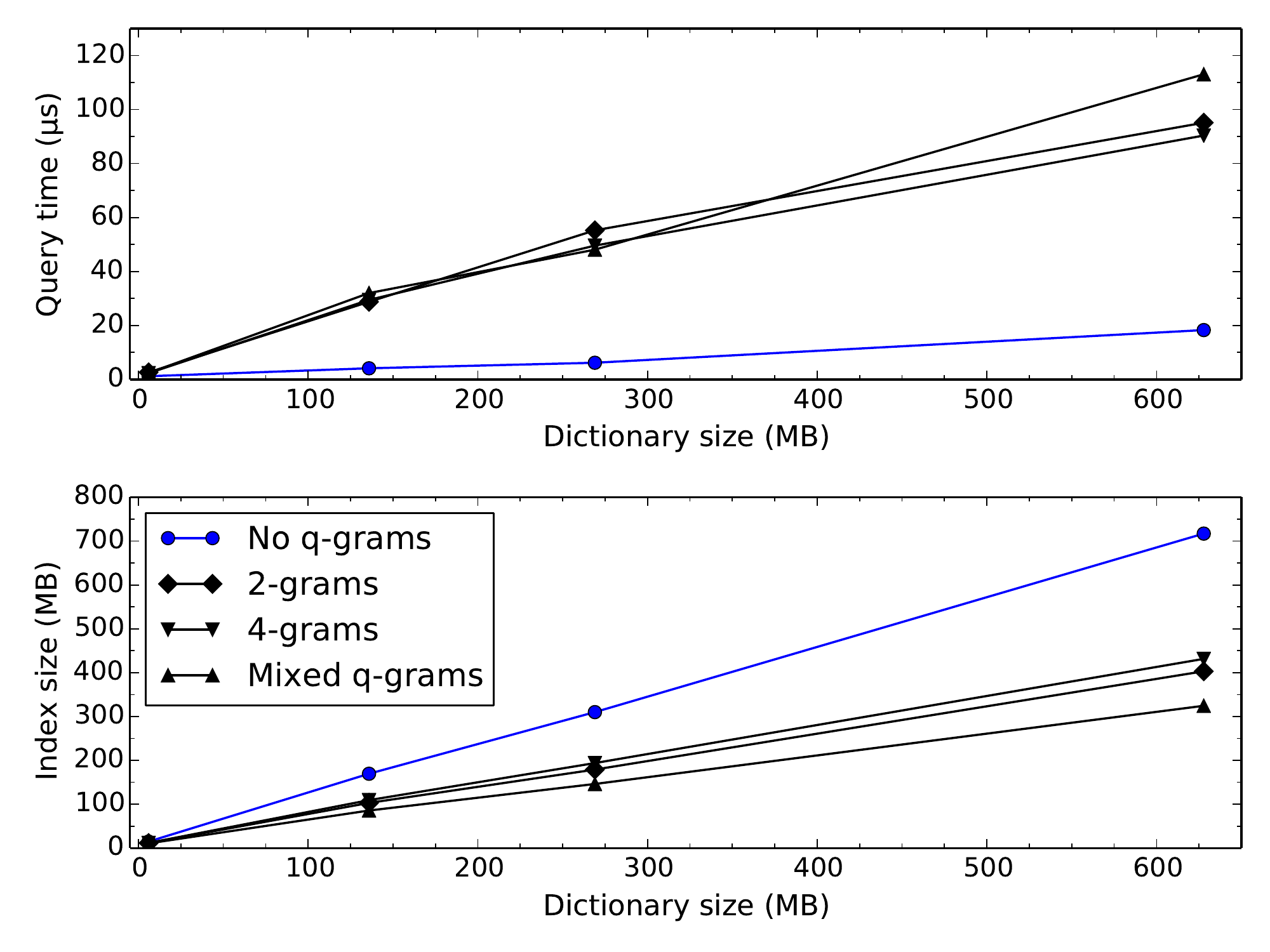}

    \caption[Query time and index size vs dictionary size, with and without $q$-gram coding for DNA dictionaries (split index).]{Query time and index size vs dictionary size for $k = 1$, with and without $q$-gram coding. Mixed $q$-grams refer to the combination of $q$-grams which provided the best compression, and these were equal to ([2-, 3-, 4-] grams): [16, 66, 18] (due to computational constraints, they were calculated only for the first dictionary, but used for all four dictionaries). DNA dictionaries and the randomly generated queries were used.}
    \label{Fig:split_comp_dna}
\end{figure}

We can see the speed-to-space relation for the English dictionaries in Figure~\ref{Fig:split_comp_eng} and for the DNA in Figure~\ref{Fig:split_comp_dna}.
In the case of English, using the optimal (from the compression point of view, i.e.~minimizing the index size) combination of mixed $q$-grams provided almost the same index size as using only 2-grams.
Substitution coding methods performed better for the DNA (where $\sigma = 5$) because the sequences are more repetitive.
Let us note that the compression provided a higher relative decrease in index size with respect to the original text as the size of the dictionary increased.
For instance, for the dictionary of size 627.8\,MB the compression ratio was equal to 1.93 and the query time was still around \SI{100}{\micro\second}.
Consult Appendix~\ref{App:split_comp} for more information about the compression.

\begin{figure}[ht]
    \centering
    \includegraphics[scale=0.55]{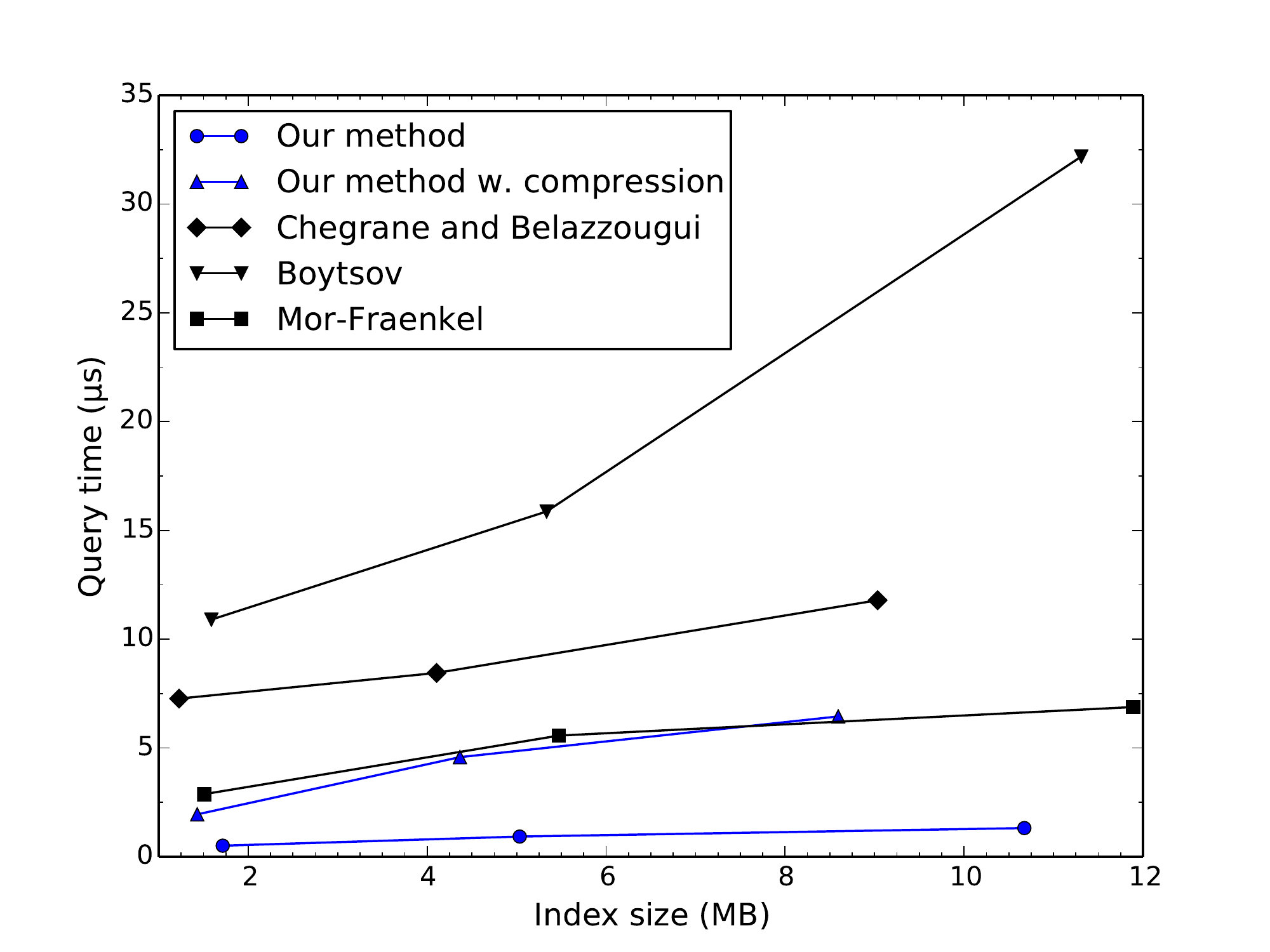}

    \caption[Query time vs index size for different methods (split index).]{Query time vs index size for different methods. The method with compression encoded mixed $q$-grams. We used the Hamming distance, and the other authors used the Levenshtein distance for $k = 1$. English language dictionaries of size 0.79\,MB, 2.67\,MB, and 5.8\,MB were used as input, and the list of common misspellings was used for queries.}
    \label{Fig:split_comp}
\end{figure}

Tested on the English language dictionaries, promising results were reported when compared to methods proposed by other authors.
Others consider the Levenshtein distance as the edit distance, whereas we use the Hamming distance, which puts us at the advantageous position.
Still, the provided speedup is significant, and we believe that the more restrictive Hamming distance is also an important measure of practical use (see Subsection~\ref{Sec:error_metrics} for more information).
The implementations of other authors are available on the Internet~\cite{boy-impl, cheg-bella}.
As regards the results reported for the \ac{mf} and Boytsov's Reduced alphabet neighborhood generation, it was not possible to accurately calculate the size of the index (both implementations by Boytsov), and for this reason we used rough ratios based on index sizes reported by Boytsov for similar dictionary sizes.
Let us note that we compare our algorithm with Chegrane and Belazzougui~\cite{chegrane2014simple}, who published better results when compared to Karch et al.~\cite{karch2010improved}, who in turned claimed to be faster than other state-of-the-art methods.
We have not managed to identify any practice-oriented indexes for matching in dictionaries over any fixed alphabet $\Sigma$ dedicated for the Hamming distance, which could be directly compared to our split index.
The times for the brute-force algorithm are not listed, since they were roughly 3 orders of magnitude higher than the ones presented.
Consult Figure~\ref{Fig:split_comp} for details.

We also evaluated different word splitting schemes.
For instance for $k = 1$, one could split the word into two parts of different sizes, e.g., $6 \to (2, 4)$ instead of $6 \to (3, 3)$, however, unequal splitting methods caused slower queries when compared to the regular one.
As regards Hamming distance calculation, it turned out that the naive implementation (i.e.~simply iterating and comparing each character) was the fastest one.
The compiler with automatic optimization was simply more efficient than other implementations (e.g.,~ones based directly on SSE instructions) that we have investigated.

\section{String sketches}
\label{Sec:string_sketches_exp}

\begin{figure}[ht]
    \centering
    \includegraphics[scale=0.55]{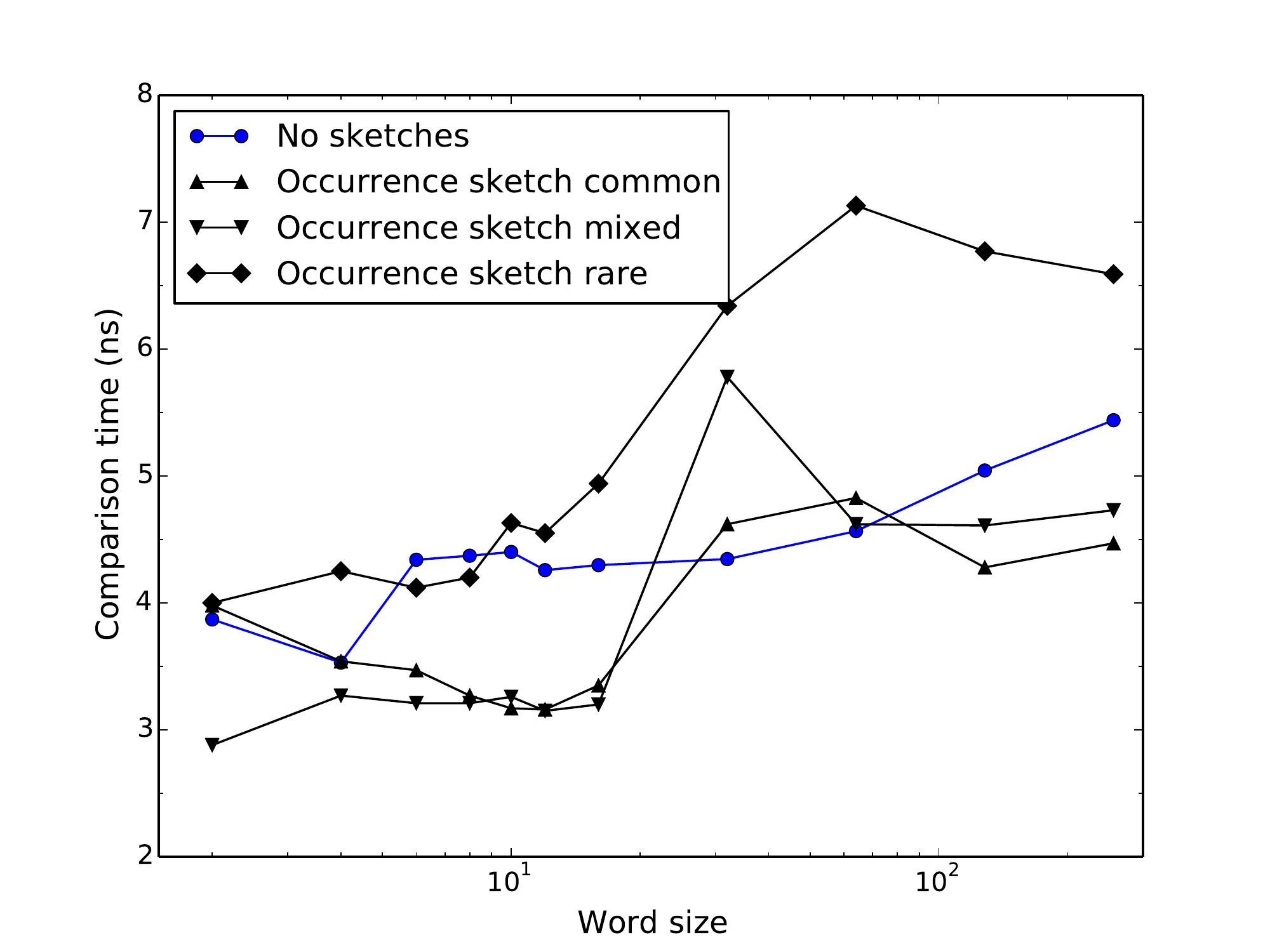}

    \caption[Comparison time vs word size for 1 mismatch using occurrence sketches for words generated over the English alphabet.]{Comparison time vs word size for 1 mismatch using occurrence sketches for words generated over the English alphabet. Each sketch occupies 2 bytes, and time refers to average comparison time between a pair of words. Common sketches use 16 most common letters, rare sketches use 16 least common letters, and mixed sketches use 8 most common and 8 least common letters. Note the logarithmic x-scale.}
    \label{Fig:sketch_word_size_occ}
\end{figure}

\begin{figure}[ht]
    \centering
    \includegraphics[scale=0.55]{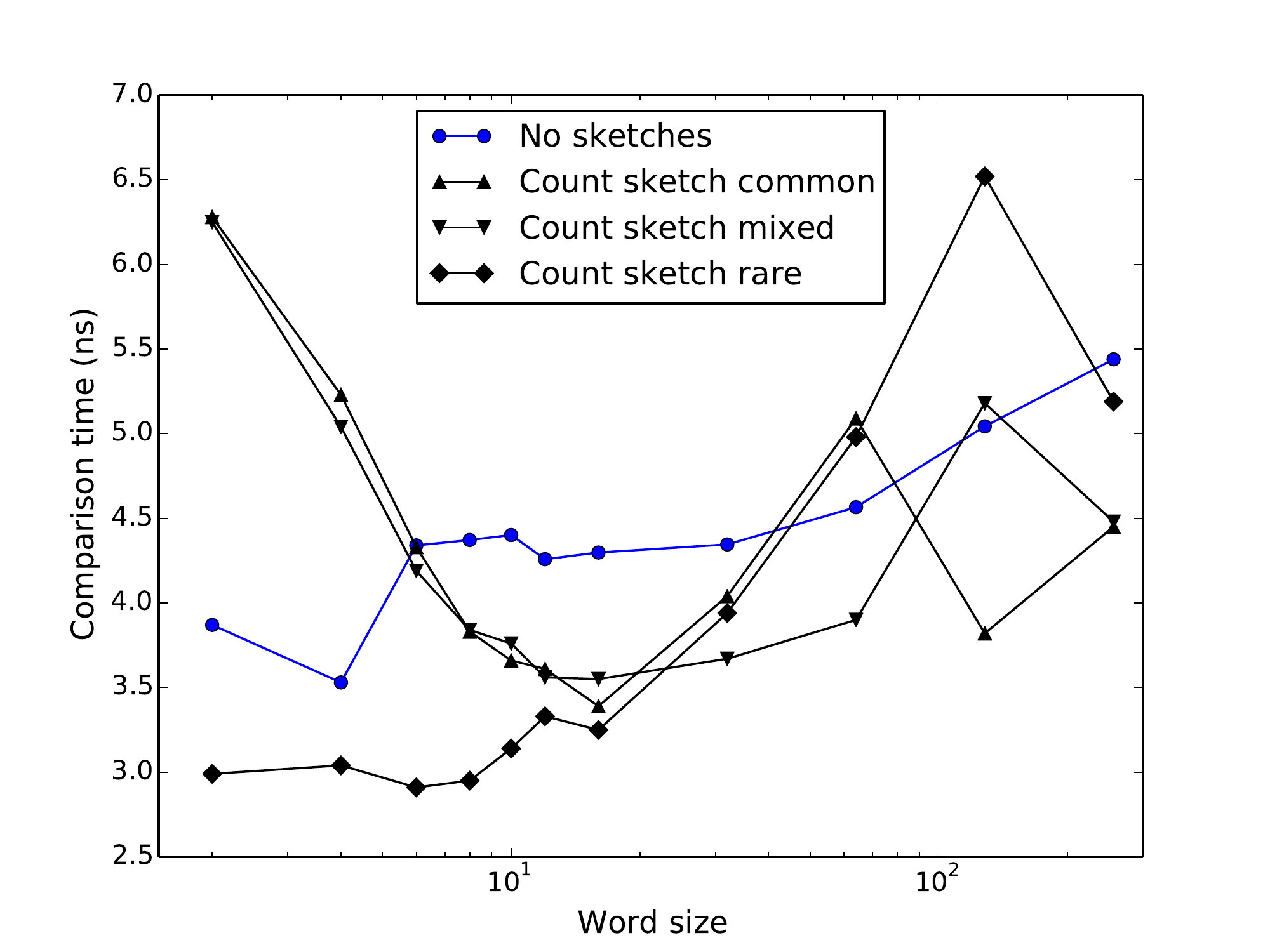}

    \caption[Comparison time vs word size for 1 mismatch using count sketches for words generated over the English alphabet.]{Comparison time vs word size for 1 mismatch using count sketches for words generated over the English alphabet. Each sketch occupies 2 bytes, and time refers to average comparison time between a pair of words. Common sketches use 16 most common letters, rare sketches use 16 least common letters, and mixed sketches use 8 most common and 8 least common letters. Note the logarithmic x-scale.}
    \label{Fig:sketch_word_size_count}
\end{figure}

String sketches which were introduced in Section~\ref{Sec:str_sketches} allow for faster string comparison, since in certain cases we can deduce for two strings $S_1$ and $S_2$ that $D(S_1, S_2) \geqslant k$ for some $k$ without performing an explicit verification.
In our implementation, a sketch comparison requires performing one bitwise operation and one array lookup, i.e.~2 constant operations in total.
We analyze the comparison time between two strings using various sketch types versus an explicit verification.
The sketch is calculated once per query and it is then reused for the comparison with consecutive words, i.e.~we examine the situation where a single query is compared against a dictionary of words.
The dictionary size for which a speedup was reported was around 100 words or more, since in the case of fewer words sketch construction was too slow in relation with the comparisons.
When the sketch comparison was not decisive a verification was performed and it contributed to the elapsed time.
The words were generated over the English alphabet (consult Appendix~\ref{App:letter_freq} in order to see letter frequencies), and each sketch occupied 2 bytes (1-byte sketches were not effective).
Figures~\ref{Fig:sketch_word_size_occ} and \ref{Fig:sketch_word_size_count} contain the results for occurrence and count sketches, respectively.
Consult Appendix~\ref{App:str_sketches} for more information regarding the letter distribution in the alphabet.

\chapter{Conclusions}
\label{Chap:concs}
\lhead{\emph{Conclusions}}

String searching algorithms are ubiquitous in computer science.
They are used for common tasks performed on home PCs such as searching inside text documents or spell checking as well as for industrial projects, e.g.,~genome sequencing.
Strings can be defined very broadly, and they usually contain natural language and biological data (DNA, proteins), but they can also represent various kinds of data such as music or images.
An interesting aspect of string matching is the diversity and complexity of the solutions which have been presented over the years (both theoretical and practical), despite the simplicity of problem formulation (one of the most common ones being ``check if pattern $P$ exists in text $T$'').

We investigated string searching methods which preprocess the input text and construct a data structure called an index.
This allows to reduce the time required for searching, and it is often indispensable when it comes to massive sizes of modern data sets.
The indexes are divided into full-text ones which operate on the whole input text and can answer arbitrary queries, and keyword indexes which store a dictionary of individual $q$-grams (these can corresponds to, e.g.,~words in a natural language dictionary or DNA reads).

Key contributions include the structure called FM-bloated, which is a modification of the FM-index (a compressed, full-text index) that trades space for speed.
Two variants of the FM-bloated were described --- one using $O(n \log^2 n)$ bits of space with $O(m + \log m \log \log n)$ average query time, and one with linear space and $O(m \log \log n)$ average query time, where $n$ is the input text length and $m$ is the pattern length.
We experimentally show that by operating on $q$-grams in addition to individual characters a significant speedup can be achieved (albeit at the cost of very high space requirements, hence the name ``bloated'').

The split index is a keyword index for the $k$-mismatches problem with a focus on the 1-error case.
It performed better than other solutions for the Hamming distance, and times in the order of 1 microsecond were reported for one mismatch for a few-megabyte natural language dictionary on a medium-end PC.
A minor contribution includes string sketches which aim to speed up approximate string comparison at the cost of additional space ($O(1)$ per string).

\section{Future work}

We presented results for the superlinear variant of the FM-bloated index in order to demonstrate its potential and capabilities.
Multiple modifications and implementations of this data structure can be introduced.
Let us recall that we store the count table and occurrence lists for selected $q$-grams in addition to individual characters from the regular FM-index.
This $q$-gram selection process can be fine-tuned --- the more $q$-grams we store, the faster the search should be, but the index size grows as well.
For instance, the linear space version could be augmented with additional 1-, 2-, etc $q$-grams which start at the position of each minimizer, up to an $s$-gram where $s$ is the maximum gap size between two minimizers.
This would eliminate two phases of the search (for prefixes and suffixes, cf. Subsection~\ref{Sec:bloated_lin_space}) where individual characters have to be used for the LF-mapping mechanism.
Moreover, the comparison with other methods could be augmented with an inverted index on $q$-grams, whose properties should be more similar to FM-bloated than those of FM-index variants, especially when it comes to space requirements.

As regards the split index, we describe possible extensions in Subsections~\ref{Sec:split_par} and \ref{Sec:split_inv}.
These include using multiple threads and introducing the functionality of an inverted index on $q$-grams.
Moreover, the algorithm could be possibly extended to handle the Levenshtein distance as well, although this would certainly have a substantial impact on space usage.
Another desired functionality could include a dedicated support for a binary alphabet ($\sigma = 2$).
In such a case, individual characters could be stored with bits, which should have a positive effect on cache usage thanks to further data compaction and possibly an alignment with the cache line size.


\addtocontents{toc}{\vspace{2em}} 

\appendix 
\chapter{Data Sets}
\label{App:data_sets}
\lhead{\emph{Data Sets}}

The following tables present information regarding the data sets that were used in this work.
Table~\ref{Tab:ds_full-text} describes data sets from the popular \ac{pc} corpus~\cite{ferragina2009compressed, pizza}, which were used for full-text indexes (English.30 was extracted from English.50).
Table~\ref{Tab:ds_keyword} describes data sets which were used for keyword indexes.
The English dictionaries come from Linux packages and the webpage by Foster~\cite{foster}, and the list of common misspellings which were used as queries was obtained from the Wikipedia~\cite{typos}.
The DNA dictionaries contain 20-mers which were extracted from the genome of Drosophila melanogaster that was collected from the FlyBase database~\cite{flybase}.
The provided sizes refer to the size of the dictionary after preprocessing --- for keyword indexes, duplicates as well as delimiters (usually newline characters) are removed.
The abbreviation \acs{nl} refers to natural language.

\begin{table}[ht]
\vspace{1em}
\centering
\begin{tabular}{c|ccc}
Name & Source & Type & Size \\
\hline
English.30 & \ac{pc} & \acs{nl} (English) & 30\,MB \\
English.50 & \ac{pc} & \acs{nl} (English) & 50\,MB \\
English.200 & \ac{pc} & \acs{nl} (English) & 200\,MB \\
\end{tabular}
\vspace{4mm}
\caption{A summary of data sets which were used for the experimental evaluation of full-text indexes.}
\label{Tab:ds_full-text}
\end{table}

\begin{table}[ht]
\vspace{1em}
\centering
\begin{tabular}{c|ccc}
Name & Source & Type & Size \\
\hline
iamerican & Linux package & \acs{nl} (English) & 0.79\,MB \\
foster & Foster & \acs{nl} (English) & 2.67\,MB \\
iamerican-insane & Linux package & \acs{nl} (English) & 5.8\,MB \\
misspellings & Wikipedia & \acs{nl} (English) & 42.2\,KB (4,261 words) \\
dmel-tiny & FlyBase & DNA & 6.01\,MB \\
dmel-small & FlyBase & DNA & 135.89\,MB \\
dmel-medium & FlyBase & DNA & 262.78\,MB \\ 
dmel-big & FlyBase & DNA & 627.80\,MB \\

\end{tabular}
\vspace{4mm}
\caption{A summary of data sets which were used for the experimental evaluation of keyword indexes.}
\label{Tab:ds_keyword}
\end{table}

\chapter{Exact Matching Complexity}
\label{App:str_comp}
\lhead{\emph{Exact Matching Complexity}}

\begin{figure}[ht]
    \centering
    \includegraphics[scale=0.55]{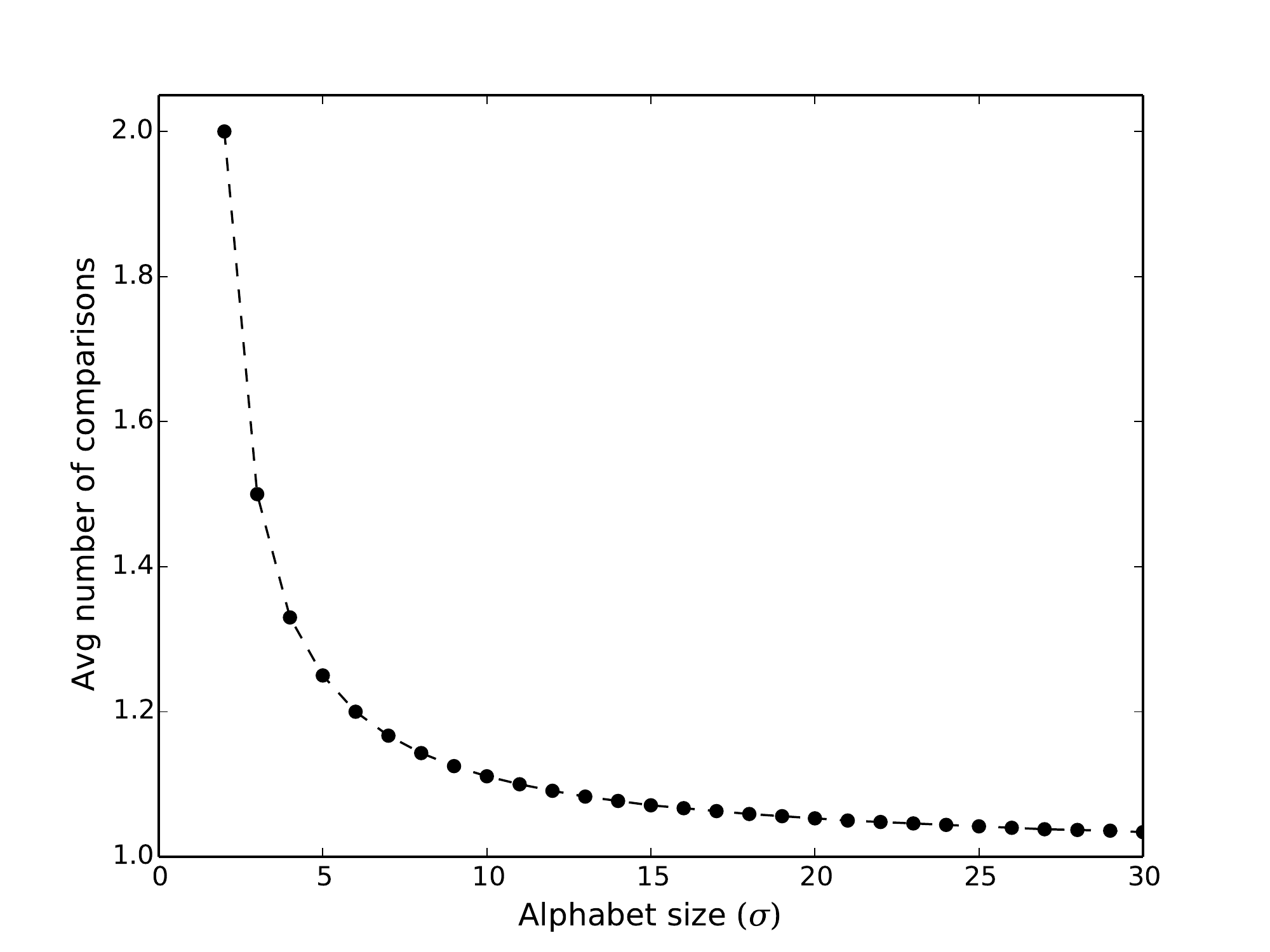}

    \caption[Average number of character comparisons when comparing two random strings from the same alphabet with uniform letter frequency vs the alphabet size.]{Average number of character comparisons when comparing two random strings (for exact matching) from the same alphabet with uniform letter frequency vs the alphabet size $\sigma$.}
    \label{Fig:str_comp}
\end{figure}

In the theoretical analysis we often mention exact string comparison, i.e.~determining whether $S_1$ = $S_2$.
It must hold that $|S_1|=|S_2|$, and the worst-case complexity of this operation is equal to $O(n)$ (all characters have to be compared when the two strings match).
On the other hand, the average complexity depends on the alphabet $\Sigma$.
If, for instance, $\sigma = 2$, we have 1/2 probability that characters $S_1[0]$ and $S_2[0]$ match, 1/4 that characters $S_1[1]$ and $S_2[1]$ match as well, etc, in the case of uniform letter frequencies.
More generally, the probability that there is a match between all characters up to a 0-based position $i$ is equal to $1/\sigma^{i+1}$, and the average number of required comparisons $A_C$ is equal to $1 + 1/(\sigma - 1)$ for any $\sigma \geqslant 2$.
We can derive the following relation: $\lim\limits_{\sigma\to\infty} A_C = 1$, and hence treating the average time required for exact comparison of two random strings from the same alphabet $\Sigma$ as $O(1)$ is justified for any $\sigma$.
In Figure~\ref{Fig:str_comp} we present the relation between the average number of comparisons and the $\sigma$ value.

In the case of real-world $\Sigma$ such as the English language alphabet, context information in the form of $k$-th order entropy should be taken into account.
In a simplified analysis, let us consider the frequencies from Appendix~\ref{App:letter_freq}; the probability that two characters sampled at random match is equal to $0.127^2$ (for \texttt{a})~$+$~$0.091^2$ (for \texttt{t}), etc.
Proceeding in this manner, the probability for the match between the first pair of characters is equal to $6.55\%$, for the first and the second pair $0.43\%$, etc.
As regards an empirical evaluation on the English.200 text, the average number of comparisons between a random pair of strings was equal to approximately 1.075.

\chapter{Split Index Compression}
\label{App:split_comp}
\lhead{\emph{Split Index Compression}}

\begin{figure}[ht]
    \centering
    \includegraphics[scale=0.55]{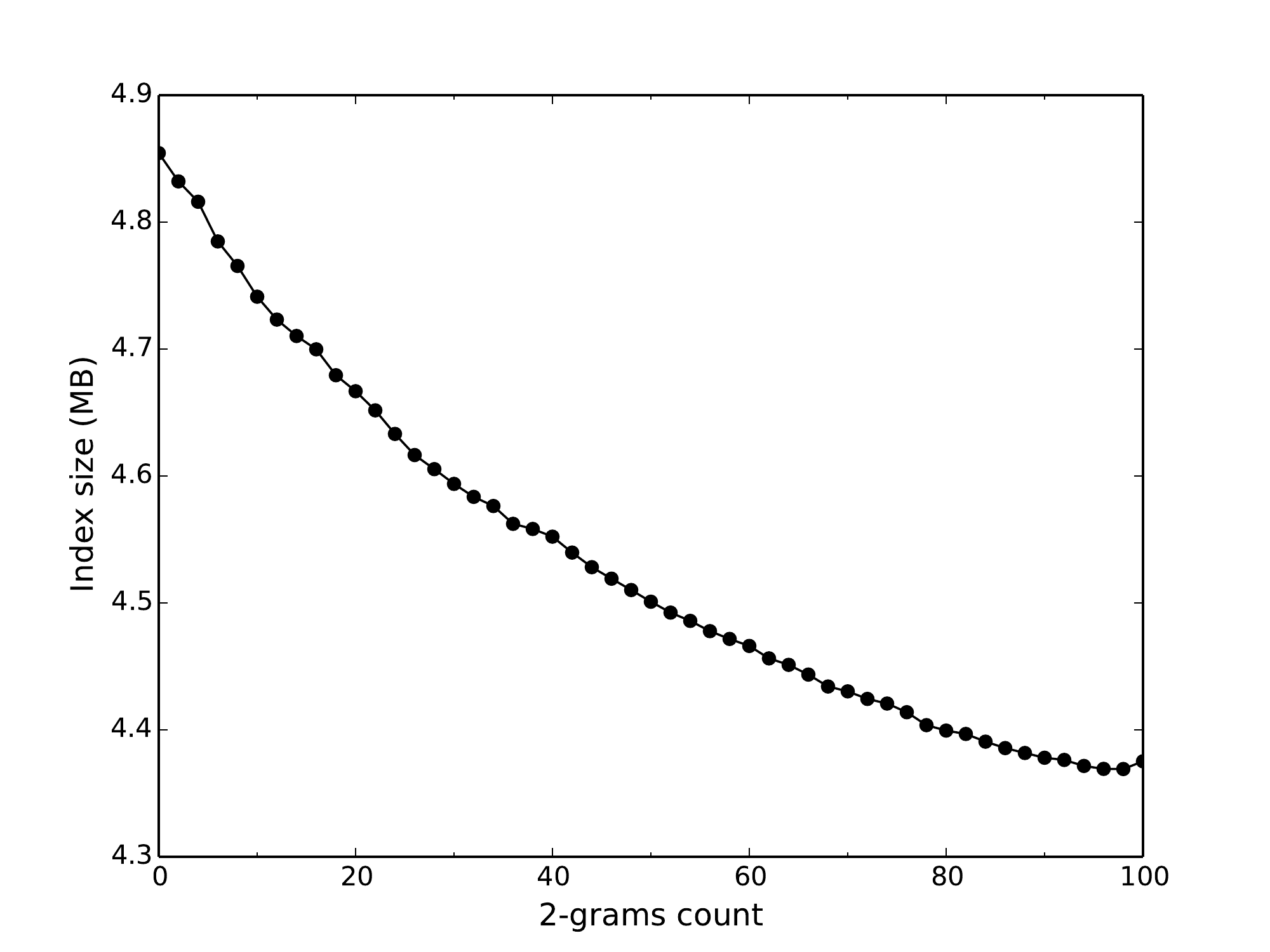}

    \caption[Index size vs the number of 2-grams used for the compression for the English dictionary (split index).]{Index size vs the number of 2-grams used for the compression for the English dictionary. 100 $q$-grams were used, and the remaining $q$-grams were 3-grams.}
    \label{Fig:split_comp_eng_app}
\end{figure}

\begin{figure}[ht]
    \centering
    \includegraphics[scale=0.55]{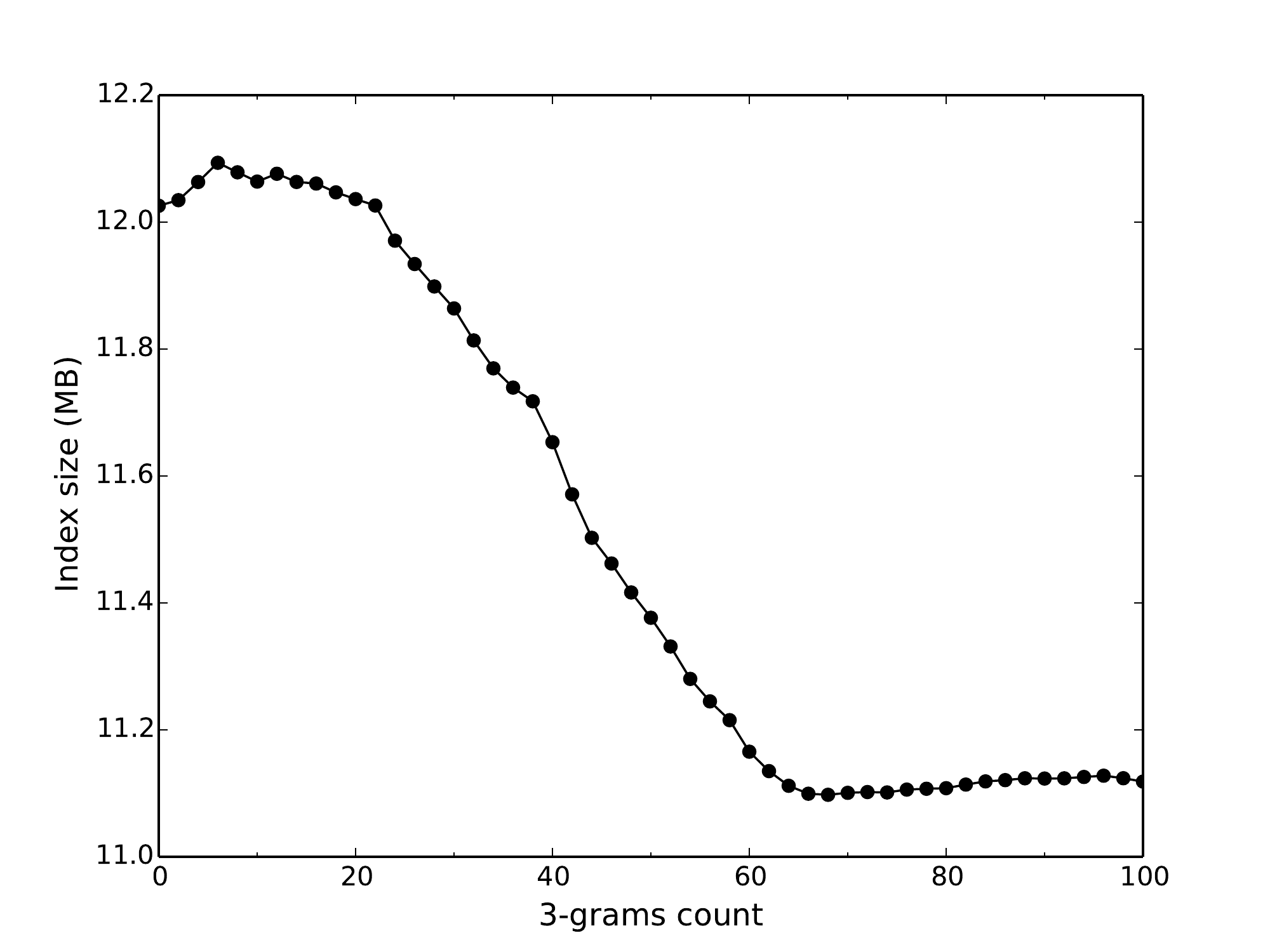}

    \caption[Index size vs the number of 3-grams used for the compression for the DNA dictionary (split index).]{Index size vs the number of 3-grams used for the compression for the DNA dictionary. 100 $q$-grams were used, and the remaining $q$-grams were 4-grams.}
    \label{Fig:split_comp_dna_app}
\end{figure}

This appendix presents additional information regarding the $q$-gram-based compression of the split index (consult Subsection~\ref{Sec:split_index} for the description of this data structure and Section~\ref{Sec:split_index_res} for the experimental results).
In Figures~\ref{Fig:split_comp_eng_app} and~\ref{Fig:split_comp_dna_app} we can see the relation between the index size and the selection of 100 2-grams and 3-grams for the English alphabet (where the 2-grams clearly provided a better compression) and 100 3-grams and 4-grams for the DNA.

\chapter{String Sketches}
\label{App:str_sketches}
\lhead{\emph{String Sketches}}

\begin{figure}[h]
    \centering
    \includegraphics[scale=0.55]{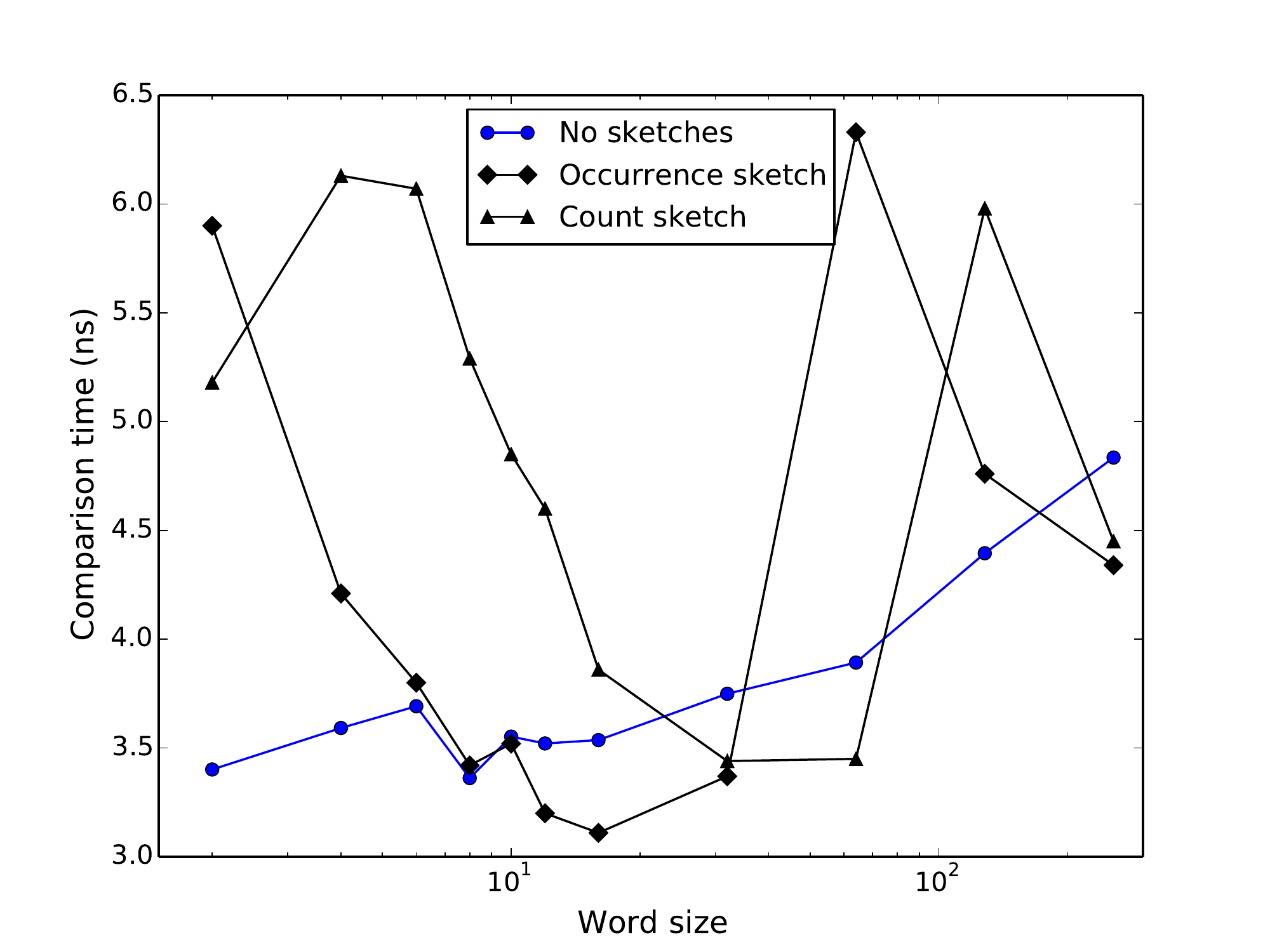}

    \caption[Comparison time vs word size for 1 mismatch using various string sketches generated over the alphabet with uniform letter frequency and $\sigma = 26$.]{Comparison time vs word size for 1 mismatch using various string sketches generated over the alphabet with uniform letter frequency and $\sigma = 26$. Each sketch occupies 2 bytes, and time refers to average comparison time between a pair of words. Note the logarithmic x-scale.}
    \label{Fig:sketch_word_size_rand}
\end{figure}

In Section~\ref{Sec:string_sketches_exp} we discussed the use of string sketches for the English alphabet, where we could take advantage of the varying letter frequency.
Here, we present the results for the alphabet with uniform distribution and $\sigma = 26$.
Instead of selecting the most or the least common letters, the 2-byte sketches contain information regarding 16 (occurrence) or 8 (count) randomly selected letters.
We can see in Figure~\ref{Fig:sketch_word_size_rand} that in this case the sketches do not provide the desired speedup.

\chapter{English Letter Frequency}
\label{App:letter_freq}
\lhead{\emph{English Letter Frequency}}

Frequencies presented in Table~\ref{Tab:letter_freq}~\cite[p.~36]{lewand2000cryptological} were used for the generation of random queries where the letter distribution corresponded to the real-world English use.

\begin{table}[h]
\vspace{1em}
\centering
\begin{tabular}{c|c}
Letter & Frequency \\
\hline
e & 12.702\% \\
t & 9.056\% \\
a & 8.167\% \\
o & 7.507\% \\
i & 6.966\% \\
n & 6.749\% \\
s & 6.327\% \\
h & 6.094\% \\
r & 5.987\% \\
d & 4.253\% \\
l & 4.025\% \\
c & 2.782\% \\
u & 2.758\% \\
\end{tabular}
\begin{tabular}{c|c}
Letter & Frequency \\
\hline
m & 2.406\% \\
w & 2.361\% \\
f & 2.228\% \\
g & 2.015\% \\
y & 1.974\% \\
p & 1.929\% \\
b & 1.492\% \\
v & 0.978\% \\
k & 0.772\% \\
j & 0.153\% \\
x & 0.150\% \\
q & 0.095\% \\
z & 0.074\% \\
\end{tabular}
\vspace{4mm}
\caption{Frequencies of English alphabet letters.}
\label{Tab:letter_freq}
\end{table}

\chapter{Hash Functions}
\label{App:hashes}
\lhead{\emph{Hash Functions}}

Table~\ref{Tab:hash_addr} contains Internet addresses of hash functions which were used to obtain experimental results for the split index (Section~\ref{Sec:split_index_res}).
If the hash function is not listed, it means that our own implementation was used.

\begin{table}[h]
\vspace{1em}
\centering
\begin{tabular}{c|c}
Name & Address\\
\hline
City & \url{https://code.google.com/p/cityhash/} \\
Farm & \url{https://code.google.com/p/farmhash/} \\
FARSH & \url{https://github.com/Bulat-Ziganshin/FARSH} \\
Murmur3 & \url{https://code.google.com/p/smhasher/wiki/MurmurHash3} \\
SpookyV2 & \url{http://burtleburtle.net/bob/hash/spooky.html} \\
SuperFast & \url{http://www.azillionmonkeys.com/qed/hash.html} \\
xxhash & \url{https://code.google.com/p/xxhash/}\\
\end{tabular}
\vspace{4mm}
\caption{A summary of Internet addresses of hash functions.}
\label{Tab:hash_addr}
\end{table}

\backmatter

\label{Bibliography}
\lhead{\emph{Bibliography}}  
\bibliographystyle{alpha}  
\bibliography{Bibliography}  

\chapter{List of Symbols}
\lhead{\emph{List of Symbols}}
\begin{acronym}

\acro{B}[$B$]{block --- either an I/O unit or a fixed-size piece of a bit vector}
\acro{c}[$c$]{character (a string of length 1)}
\acro{C}[$C$]{count table in the FM-index}
\acro{CLine}[$C_L$]{cache line size}
\acro{D}[$D$]{distance metric}
\acro{delta}[$\delta$]{time required for calculating $D(S_1, S_2)$ for two strings over the same alphabet $\Sigma$}
\acro{mD}[$\mathcal{D}$]{dictionary of keywords (for keyword indexes)}
\acro{d}[$d$]{word (string) from a dictionary}
\acro{enc}[$Enc(d)$]{encoded (compressed) word $d$}
\acro{F}[$F$]{first column of the BWT matrix}
\acro{H}[$H$]{hash function}
\acro{Hk}[$H_k(S)$]{$k$-th order entropy of string $S$}
\acro{HT}[$H_T$]{hash table}
\acro{Hw}[$H_W$]{Hamming weight (number of 1s in a bit vector)}
\acro{Ham}[$Ham$]{Hamming distance}
\acro{round}[$\lfloor i \rceil$]{number $i$ rounded to the nearest integer}
\acro{I}[$I$]{index for string matching}
\acro{k}[$k$]{number of errors (in approximate matching)}
\acro{L}[$L$]{last column of the BWT matrix}
\acro{LFsym}[$L_F$]{load factor}
\acro{Lev}[$Lev$]{Levenshtein distance}
\acro{m}[$m$]{pattern size, $m = |P|$}
\acro{n}[$n$]{input size}
\acro{occ}[$occ$]{number of occurrences of the pattern}
\acro{mini}[$M(S)$]{set of minimizers of string $S$}
\acro{P}[$P$]{pattern}
\acro{p}[$p$]{piece of a word in the case of word partitioning}
\acro{pr}[$Pr(e)$]{probability of event $e$}
\acro{QQ}[$\mathcal{Q}$]{collection of $q$-grams}
\acro{S}[$S$]{string}
\acro{Sprime}[$S^\prime$]{string sketch over $S$}
\acro{s}[$s$]{substring}
\acro{mS}[$\mathcal{S}$]{set of substrings (for full-text indexes)}
\acro{saa}[$S_A$]{suffix array}
\acro{alph}[$\Sigma$]{alphabet}
\acro{alphstr}[$\Sigma^*$]{set of all strings over the alphabet $\Sigma$}
\acro{alphSize}[$\sigma$]{alphabet size, $\sigma = |\Sigma|$}
\acro{T}[$T$]{input string (text)}
\acro{TR}[$T_R$]{RRR table}
\acro{TBWT}[$T^{bwt}$]{input text $T$ after applying the BWT}
\acro{w}[$w$]{size of the machine word (typically 32 or 64 bits)}
\acro{v}[$v$]{bit vector}

\end{acronym}

\chapter{List of Abbreviations}
\lhead{\emph{List of Abbreviations}}
\begin{multicols}{2}
\begin{acronym}

\acro{bf}[BF]{Bloom filter}
\acro{bm}[BM]{Boyer--Moore \acroextra{algorithm}}
\acro{bmh}[BMH]{Boyer--Moore--Horspool \acroextra{algorithm}}
\acro{bst}[BST]{binary search tree}
\acro{bwt}[BWT]{Burrows--Wheeler transform}
\acro{csa}[CSA]{compressed suffix array}
\acro{cosa}[CoSA]{compact suffix array}
\acro{dfs}[DFS]{depth-first search}
\acro{dp}[DP]{dynamic programming}
\acro{esa}[ESA]{enhanced suffix array}
\acro{fsm}[FSM]{finite state machine}
\acro{io}[I/O]{input/output}
\acro{kmp}[KMP]{Knuth--Morris--Pratt \acroextra{algorithm}}
\acro{lcp}[LCP]{longest common prefix}
\acro{lcs}[LCS]{longest common subsequence}
\acro{lf}[LF]{load factor}
\acro{mf}[MF]{Mor--Fraenkel \acroextra{algorithm}}
\acro{mphf}[MPHF]{minimal perfect hash function}
\acro{nl}[NL]{natural language}
\acro{nw}[NW]{Needleman--Wunsch \acroextra{algorithm}}
\acro{ocr}[OCR]{optical character recognition}
\acro{pc}[P\&C]{Pizza \& Chili \acroextra{corpus}}
\acro{pies}[PiES]{partitioning into exact searching}
\acro{rk}[RK]{Rabin--Karp \acroextra{algorithm}}
\acro{sa}[SA]{suffix array}
\acro{sc}[SC]{suffix cactus}
\acro{tlb}[TLB]{translation lookaside buffer}
\acro{st}[ST]{suffix tree}
\acro{sw}[SW]{Smith--Waterman \acroextra{algorithm}}
\acro{wt}[WT]{wavelet tree}

\end{acronym}
\end{multicols}

\lhead{\emph{List of Figures}}  
\listoffigures  

\lhead{\emph{List of Tables}}  
\listoftables  

\end{document}